\documentclass[bibyear]{aa}  
\usepackage{txfonts}

\usepackage{xcolor}
\usepackage{graphicx}

\usepackage{afterpage}
\usepackage{soul}
\usepackage{natbib}
\usepackage[bookmarksopen=false,bookmarksnumbered=true,breaklinks=true,colorlinks=true,linkcolor=blue,citecolor=blue]{hyperref}

\usepackage{threeparttable}
\usepackage{longtable}
\usepackage{supertabular}
\usepackage{array}					
\usepackage{booktabs}				
\usepackage{multicol,multirow}
\newcolumntype{C}{>{\begin{math}}c<{\end{math}}}%

\usepackage{comment}

\newcommand{\teff}{\ensuremath{\mathrm{T_{eff}}}}
\newcommand{\logg}{\ensuremath{\log(g)}}

\newcommand{\pc}{\ensuremath{\mathrm{pc}}}
\newcommand{\magg}{\ensuremath{\mathrm{mag}}}

\newcommand{\kelvin}{\ensuremath{\mathrm{K}}}
\newcommand{\mas}{\ensuremath{\mathrm{mas}}}

\newcommand{\yrs}{\ensuremath{\mathrm{years}}}

\newcommand{\dayy}{\ensuremath{\mathrm{day}}}
\newcommand{\days}{\ensuremath{\mathrm{days}}}

\newcommand{\myr}{\ensuremath{\mathrm{Myr}}}
\newcommand{\gyr}{\ensuremath{\mathrm{Gyr}}}
\newcommand{\au}{\ensuremath{\mathrm{au}}}
\newcommand{\mic}{\ensuremath{\mathrm{\mu m}}}

\newcommand{\mj}{\ensuremath{\mathrm{M_{Jup}}}}

\newcommand{\mearth}{\ensuremath{\mathrm{M}_\oplus}}
\newcommand{\rearth}{\ensuremath{\textnormal{R}_{\oplus}}}
\newcommand{\densityearth}{\ensuremath{\rho_{\oplus}}}
\newcommand{\msun}{\ensuremath{\textnormal{M}_\odot}}
\newcommand{\rsun}{\ensuremath{\textnormal{R}_\odot}}

\renewcommand{\degr}{\ensuremath{^\circ}}

\definecolor{lagoon}{rgb}{0.5, 0.7, 0.8}
\definecolor{forest green}{rgb}{0.233, 0.645, 0.233}
\newcommand{\edc}[1]{\textcolor{lagoon}{#1}}
\newcommand{\celia}[1]{\textcolor{lagoon}{ [\textbf{Célia: } #1]}}

\begin{document}
    
   \title{Planetary system architectures with low-mass inner planets
}
\subtitle{Direct imaging\thanks{Based on observations collected at the European Southern Observatory under ESO programmes 099.C-0255(A) 102.C-0489(A), 103.C-0484 (A and B), and 109.23F2.} exploration of mature systems beyond 1 au}
   \authorrunning{Desgrange et al.}
   \author{C.~Desgrange\inst{1}\fnmsep\inst{2}, J.~Milli\inst{1}, G.~Chauvin\inst{3}, Th.~Henning\inst{2}, A.~Luashvili\inst{4}, M.~Read\inst{5}, M.~Wyatt\inst{6}, G.~Kennedy\inst{7}\fnmsep\inst{8}, R.~Burn\inst{2}, M.~Schlecker\inst{9}, F.~Kiefer\inst{10}, V.~D'Orazi\inst{11}\fnmsep\inst{12}, S.~Messina\inst{13}, P.~Rubini\inst{14}, A.-M.~Lagrange\inst{10}, C.~Babusiaux\inst{1}, L.~Matrà\inst{15}, B.~Bitsch\inst{2}, M.~Bonavita\inst{12}\fnmsep\inst{16}, P.~Delorme\inst{1}, E.~Matthews\inst{2}, P.~Palma-Bifani\inst{3}, A.~Vigan\inst{17}
        }

   \institute{
    $^{1}$ Univ. Grenoble Alpes, CNRS, IPAG, F-38000 Grenoble, France;  e-mail: celia.desgrange@univ-grenoble-alpes.fr \\
    $^{2}$ Max Planck Institute for Astronomy, K\"onigstuhl 17, D-69117 Heidelberg, Germany\\ 
    $^{3}$\;Laboratoire Lagrange, UMR7293, Universit\'e  Cote d’Azur, CNRS, Observatoire de la C\^ote d’Azur. Boulevard de l’Observatoire, 06304 Nice, France. \\
    $^{4}$ Laboratoire Univers et Th\'eories, Observatoire de Paris, Universit\'e PSL, CNRS, Universit\'e Paris Cit\'e, 92190 Meudon, France\\
    $^{5}$ Space Research and Planetology Division, Physikalisches Inst., Universit\"at Bern, Switzerland \\
    $^{6}$ Institute of Astronomy, University of Cambridge, Madingley Road, Cambridge CB3 0HA, United Kingdom \\
    $^{7}$ Department of Physics, University of Warwick, Gibbet Hill Road, Coventry, CV4 7AL, UK\\
    $^{8}$ Centre for Exoplanets and Habitability, University of Warwick, Gibbet Hill Road, Coventry CV4 7AL, UK\\
    $^{9}$ Department of Astronomy/Steward Observatory, The University of Arizona, 933 North Cherry Avenue, Tucson, AZ 85721, USA \\
    $^{10}$\;LESIA, Observatoire de Paris, Universit\'e PSL, CNRS, Sorbonne Université, Université de Paris, 5 place Jules Janssen, 92195 Meudon, France \\ 
    $^{11}$ Department of Physics, University of Rome Tor Vergata, Via della Ricerca Scientifica 1, 00133 Rome, Italy\\
    $^{12}$ INAF-Osservatorio Astronomico di Padova, Vicolo dell’ Osservatorio 5, 35122, Padova, Italy\\ 
    $^{13}$ INAF–Osservatorio Astrofisico di Catania, via Santa Sofia, 78 Catania, Italy \\ 
    $^{14}$ Pixyl, 5 Avenue du Grand Sablon, 38700 La Tronche, France\\
    $^{15}$ School of Physics, Trinity College Dublin, The University of Dublin, College Green, Dublin 2, Ireland\\
    $^{16}$ School of Physical Sciences, The Open University, Walton Hall, Milton Keynes, MK7 6AA, UK\\
    $^{17}$ Aix Marseille Univ, CNRS, CNES, LAM, Marseille, France\\
    }
   \date{}

  \abstract
   {The discovery of planets orbiting at less than 1 au from their host star and less massive than Saturn in various exoplanetary systems revolutionized our theories of planetary formation. 
   The fundamental question is whether these close-in low-mass planets could have formed in the inner disk interior to 1 au, or whether they formed further out in the planet-forming disk and migrated inward.
   Exploring the role of additional giant planet(s) in these systems may help us to pinpoint their global formation and evolution.
   } 
   {We searched for additional substellar companions  by using direct imaging in systems known to host close-in small planets. The use of direct imaging complemented by radial velocity and astrometric detection limits enabled us to explore the giant planet and brown dwarf demographics around these hosts to investigate the potential connection between both populations.}
   {We carried out a direct imaging survey with SPHERE at VLT to look for outer giant planets and brown dwarf companions in 27 systems hosting close-in low-mass planets discovered by radial velocity. Our sample is composed of very nearby (< 20 pc) planetary systems, orbiting G-, K-, and M-type mature (0.5--10 Gyr) stellar hosts. We performed homogeneous direct imaging data reduction and analysis to search for and characterize point sources, and derived robust statistical detection limits. 
   The final direct imaging detection performances were globally considered together  with radial velocity and astrometric sensitivity. }
   {Of 337 point-source detections, we do not find any new bound companions. We recovered the emblematic very cool T-type brown dwarf GJ\,229\,B. Our typical sensitivities in direct imaging range from 5 to 30 M$_{\rm Jup}$ beyond 2 au. The non-detection of massive companions is consistent with predictions based on models of planet formation by core accretion. Our pilot study opens the way to a multi-technique approach for the exploration of very nearby exoplanetary systems with future ground-based and space observatories.}
   {}
   

   \titlerunning{Architecture of systems hosting close-in sub-Saturns}
   \maketitle


\bibliographystyle{aa.bst}

\section{Introduction} 

A striking result of the \textit{Kepler} mission is the detection of numerous small planets, mini-Neptunes and super-Earths, separated by the so-called photo-evaporation valley ($R\,\sim\,1.8~\rearth$, referred to as the Fulton gap), and orbiting very close to their host star with a period less than 100 days  \citep{Fulton2017_gap_kepler}.  One fundamental question of particular interest is related to the correlation between the presence of inner super-Earths and outer giant planets from observational and theoretical perspectives.

A positive correlation would imply that outer giant planet(s) could enhance the occurrence of inner low-mass planets, or that conditions that make the growth of giants possible are also favorable for the growth of planets to the inner disk \citep{ChiangLaughlin2013_in_situ_initial_conditions,Schlecker2021_NGPPS_SE_CJ}.
Close-in planets could still have formed in the outer parts of the system, inside or outside the giant planet orbit(s). 
Secular resonances or Kozai interactions with outer giants can promote excitation and growth of inner, low-mass planets to the observable regime \citep{Nagasawa2005_theory_planet_formation_collisions,Munoz2016_theory_planetary_formation_inner_planets_Kozai_migration,Best2023_positive_correlation_transport_secular_resonance}.
If formed farther out, the lower-mass planets could have been scattered to the inner part of the system, and then undergone orbital circularization \citep{TerquemPapaloizou2007_planet_formation_scattering,KennedyKenyon2008_planet_formation_inner_SE}.

Conversely, giant planets can be detrimental to the existence of close-in small planets, since they could halt their migration from the outer to inner regions \citep{Izidoro2015_planet_formation_inner_SE_giant_barrier,Schlecker2021_NGPPS_SE_CJ}. They could cut off the flow of solids toward the inner regions \citep{Morbidelli2016Icar}, and/or stir the velocity distribution of the solids \citep{MustillWyatt2009_stirring}. These negative impacts take place only if the giant planet forms before or at the same time as the small planet.
Finally, formation in the inner disk of low-mass close-in planets would be independent of the presence of a distant giant planet if initial conditions between those regions are uncorrelated \citep{Schlaufman2014_planet_formation_independent_no_correlation_SE_CJ}.

\citet{ZhuWu2018_RV_study_observational_constraints_super_earth_correlation_super_earth_cold_jupiter} and \citet{Bryan2019_RVTransit_study_observational_constraints_super_earth_correlation_super_earth_cold_jupiter} looked for giant planets via the radial velocity (RV) method in samples of $31$ and $65$ systems, respectively, known to host at least one super-Earth by radial velocity and/or transit techniques. Both of these studies found a positive correlation between the presence of inner super-Earths (SE) and outer massive giant planets called cold Jupiters (CJ). They derived conditional probabilities of hosting at least one outer giant planet in systems known to harbor super-Earth(s), which is mathematically represented by the probability P(CJ|SE) and compared it to the absolute probability of a system to host at least one giant planet, noted as P(CJ). These studies have consistent results because \citet{ZhuWu2018_RV_study_observational_constraints_super_earth_correlation_super_earth_cold_jupiter} found a conditional probability  $ P\text{(CJ|SE)}$ of $29\pm18\%$ around Sun-like stars which is greater than the absolute probability of finding at least one giant planet, that is,  $ P\text{(CJ)}=10\%$ \citep{Cumming2008_proba_CJ}, whereas \citet{Bryan2019_RVTransit_study_observational_constraints_super_earth_correlation_super_earth_cold_jupiter} found  $ P\text{(CJ|SE)}  = 34\pm7\%$. They compared it to $P\text{(CJ)}  = 7\pm3\%$ from \citet{Wittenmyer2016_proba_CJ} to be consistent with their definition of cold Jupiter and the spectral type of the sample of host stars considered, FGKM. In addition, they both derived, indirectly, the probability of having super-Earth planets in systems known to host outer giant planets and found that systems possessing outer giant planets are very likely to be accompanied by inner super-Earth(s) as well \citep[$P\text{(SE|CJ)}=90\pm20\%$ from][]{ZhuWu2018_RV_study_observational_constraints_super_earth_correlation_super_earth_cold_jupiter}. 
On the other hand, \citet{Barbato2018_RV_study_observational_constraints_super_earth_correlation_super_earth_cold_jupiter} looked for inner super-Earth planets in a sample of $20$~systems known to host distant giant planets and did not find any. They concluded that the probability of detecting a super-Earth planet in a system known to host a distant giant planet is small ($P\text{(SE|CJ)}  < 9.84\%$). However, this is statistically not at odds with previous observational studies due to observational biases, the limited size of the sample \citep{ZhuWu2018_RV_study_observational_constraints_super_earth_correlation_super_earth_cold_jupiter,Barbato2020}, the eccentricity distribution of the giant planets in those surveys \citep{Bitsch2020_theory_peeble_eccentricity_planets,Bitsch_Izidoro2023}, and the definition of planet categories (i.e., lower and upper bound in planet mass and semi-major axis to define, e.g., a “super-Earth”).

As \citet{ZhuWu2018_RV_study_observational_constraints_super_earth_correlation_super_earth_cold_jupiter} and \citet{Bryan2019_RVTransit_study_observational_constraints_super_earth_correlation_super_earth_cold_jupiter}, \citet{Rosenthal2021_RV_CLS}  recently concluded based on their large California Legacy Survey ($719$~systems)  that systems with inner low-mass planets are more likely to host outer giant ones.
In addition, \citet{Rosenthal2021_RV_CLS} derived the conditional probability P\text{(SE|CJ)} both directly via measurement, and indirectly from Bayesian inference. They found a mismatch of $20\%$ to $30\%$ between the maximum of the conditional probability density measured via direct or indirect measurements. The difference depends on the definition of an outer giant planet, in particular the range of semi-major axis and masses  ($3$--$7~\au$ and $95$--$4\,130~\mearth$, i.e., $0.3$--$13~\mj$, or $0.23$--$10~\au$ and $30$--$6\,000~\mearth$, i.e.,  $0.09$--$18.9~\mj$). 

On the other hand, \citet{Bonomo2023_no_correlation_SE_CJ}  argued for the absence of an excess of cold Jupiters in small-planet systems based on their RV survey on planet-transiting systems. 
However, \citet{Bitsch_Izidoro2023} highlighted that the survey from \citet{Bonomo2023_no_correlation_SE_CJ} might have been biased in the first place, as the eccentricities of their transiting low-mass planets are mainly low, hinting to the presence of only a few giant planets (see Section~\ref{sec:discussion_influence_ecc}). In other ways, \citet{Zhu2023_metallicity} indicated that the metallicity dimension could resolve the discrepancy between \citet{Bonomo2023_no_correlation_SE_CJ} and previous works, stating that the positive correlation of close-in super-Earths and distant cold Jupiters is verified around metal-rich hosts.

Despite those investigations to name a few, the origin of the architectures of exoplanetary systems is still an open question. The link between close-in small mass planets and distant giant planets, hence the conclusions on their formation and evolution, depend on the definition of the planet categories \citep[e.g.,][]{Rosenthal2021_RV_CLS, Schlecker2021_NGPPS_SE_CJ}, and on observational biases \citep[e.g.,][]{Barbato2018_RV_study_observational_constraints_super_earth_correlation_super_earth_cold_jupiter,ZhuWu2018_RV_study_observational_constraints_super_earth_correlation_super_earth_cold_jupiter,Bryan2019_RVTransit_study_observational_constraints_super_earth_correlation_super_earth_cold_jupiter}. To date, observational studies on this correlation mainly used RV and transit detection methods, which are limited somehow to the inner regions of exoplanetary systems.

\smallskip

In this paper, we look for outer giant planets or brown dwarfs with direct imaging in $27$ nearby exoplanetary systems known to harbour close-in  ($\lesssim1~\au$) low-mass ($\lesssim95~\mearth$) planet(s) by the RV detection method. We complement our direct imaging observations (VLT/SPHERE) with archival data from RVs (HARPS) and proper motion coupled to the excess noise (\textit{Gaia} and \textit{Hipparcos}) to illustrate the global parameter space explored in these exoplanetary systems.
In Section \ref{sec:sample}, we introduce our {direct imaging survey, and the properties of the targeted systems. In Section \ref{sec:obs}, we present the high-contrast imaging observations and archival data, while in Section \ref{sec:data_reduction}, we describe the data reduction. In Section \ref{sec:results_obs}, we report our observational results, including the directly imaged point-source detections, and the status of these planetary or substellar companion candidates based on two criteria: the color magnitude diagram (CMD), and the  
proper motion diagram (PMD). We indicate the detection limits for each system by combining the direct imaging, radial velocity and astrometry detection methods. In Section \ref{sec:discussion}, we discuss our results. We compare our direct-imaging results with previous studies, including theoretical predictions in Section \ref{sec:comparison_obs}. 
In Section \ref{sec:discussion_influence}, we consider constraints from intrinsic parameters of the planetary systems in the presence of additional planets. 
In Section \ref{sec:conclusion}, we conclude on how this survey and in-depth study is particularly relevant to demonstrate to what extent current observational facilities used in synergy can challenge the current global vision of planetary formation.

\section{Sample and target properties \label{sec:sample}}

We describe here how we selected the systems of our sample, and their known architectures and properties. 

\subsection{Sample selection}

Our direct imaging survey aims to explore the link between close-in low-mass  planets and outer giant planets, as it could inform us on how the close-in low-mass planets formed. 
Therefore, we built a sample of $27$ exoplanetary systems suspected to harbour at least one close-in ($\lesssim1~\au$)  low-mass ($\lesssim95~\mearth$, i.e., sub-Saturn mass, see Fig.~\ref{fig:intro_exoplanets}) planet revealed by various RV spectrographs (see Table~\ref{tab:target_properties}), and observable from the southern hemisphere at the Very Large Telescope (VLT). To maximize the sensitivity of high-contrast imaging observations, we restricted the sample to stars within $20~\pc$. The sample includes stars covering a relatively broad range of stellar masses, from $0.3~\msun$ to $1\,\msun$ corresponding to spectral types from M, K and G types (see Fig.\,\ref{fig:target_properties}).  

\begin{figure}[h]
    \includegraphics[width=\linewidth]{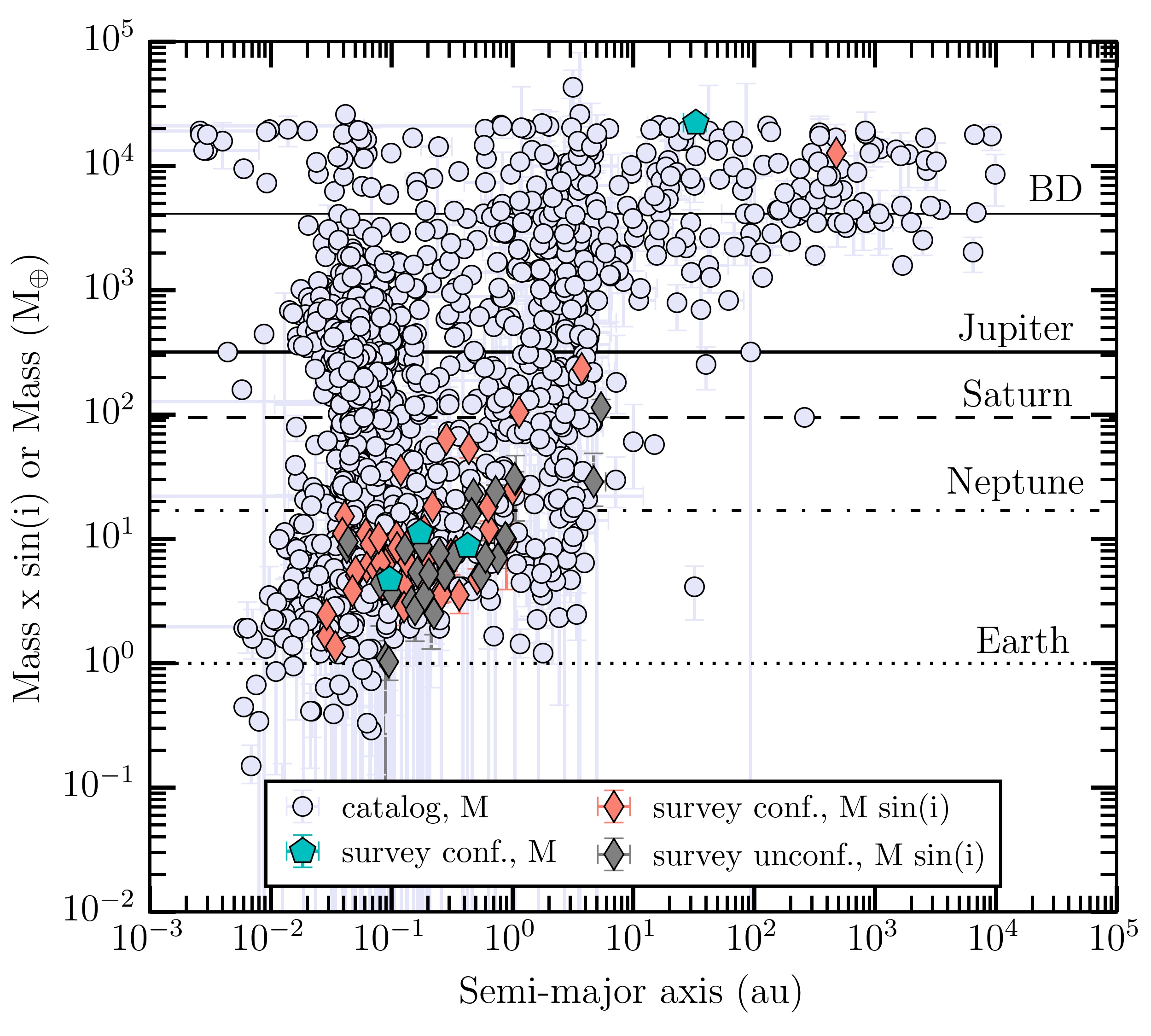}
    \caption{Masses and semi-major axes of the planets in our sample. Minimal masses of both confirmed (“conf.”) and unconfirmed (“unconf.”) planets are shown in orange and gray diamonds, respectively, and absolute masses of the confirmed companions GJ\,229\,B and in HD\,136352 are represented in blue pentagons, in the context of all the exoplanets known so far\textsuperscript{\scriptsize\,a} (“catalog”).
    }
    \label{fig:intro_exoplanets}
    \small \textsuperscript{\scriptsize a} from the catalog {\tiny \url{http://exoplanet.eu/catalog/}}
\end{figure}


 \begin{figure}[h!]  \centering
    \includegraphics[width=\linewidth]{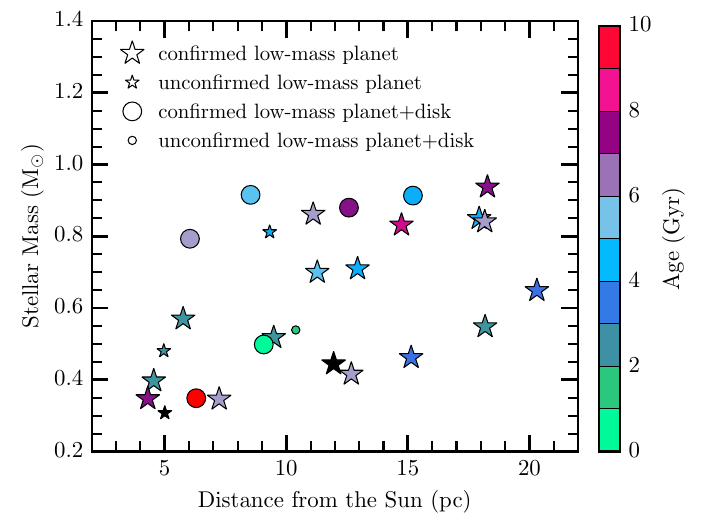}  
    \caption{Target properties of our sample: star (circle) markers correspond to systems without (with) a debris disk discovered. 
    The four small markers corresponds to the four systems for which only suggested or debatted low-mass planets exist based on the literature (see Appendix~\ref{app:biblio_indiv_systems}). 
    }
    \label{fig:target_properties}
\end{figure}

The challenge of this survey is to directly image planets around relatively mature and quiet stars where high-precision RV monitoring detected low-mass planets. Indeed old systems ($>100~\myr$) are rarely observed in direct imaging to search for planets. With time, giant exoplanets cool down and their emitted light decreases, making their detection more difficult given current XAO planet imager performance.  In addition,  we note that RV-based low-mass planets come with a bias, because the presence of a signal from a giant planet may often lead to a stop of the observations on this system (see the discussion in Section~\ref{sec:comparison_obs}).

We derived homogeneously the masses for each star in the sample by following the method from \citet{GaiaArenou2022_stellar_mass}, see their Appendix~E. This uses the magnitude in the G band from the \textit{Gaia} Data Release 3, as well as the age and the metallicity of the stars homogeneously re-analyzed in the following sections. This is the first time that the mass  of GJ~649 is estimated. As for the other stars of the sample, the masses derived are consistent with the previous ones published at $1\,\sigma$ or $2\,\sigma$ when considering only the \textit{Gaia} uncertainty, as the uncertainty value from the literature is not always available.

The detailed target properties (distance, spectral type, stellar host mass and radius, duplicity or multiplicity, number of planets and presence of a debris disk) are reported in Table\,\ref{tab:target_properties}. We derived homogeneously the age for each star of our sample and the metallicity for each GK star. Both methods are described in Appendix~\ref{app:age} and~\ref{app:metallicity}, respectively, and associated results in Tables~\ref{tab-age} and~\ref{tab:stellar_feh}, and in Figs.~\ref{fig:hist_age} and~\ref{fig:hist_metallicity}. We note that three out of the twelve GK stars from our sample have super-solar metallicity (HD\,102365, HD\,154088 and HD\,3651), two consistent with solar-like metallicity (HD\,69830, 61\,Vir) and for the fifteen M-stars, no reliable metallicity constraints  (see Appendix~\ref{app:metallicity}). This could impact our results concerning the GK-stars, because the presence of giant planets if formed via core-accretion \citep{Pollack1996_CA} is correlated with super-solar metallicity \citep[see e.g.,][and our discussion in Section~\ref{sec:discussion_influence_metallicity}]{FischerValenti2003_obs_metallicity}. Stellar metallicity was not a criterion back to the time of the sample design, already restricted to $27$ systems owing to required visibility from the VLT, presence of suggested close-in low-mass planets and the very nearby location of the systems (within $20~\pc$) to maximize high-contrast imaging sensitivity. 
Below, we describe the architectures of the systems from our sample. 

\begin{table*}[h]
    \centering
    \caption{Target properties, including both stellar and system properties.}
     \setlength{\tabcolsep}{3pt} 
    \begin{tabular}{lCcCCCCCCccc}

\hline\hline\noalign{\smallskip}
System & \textnormal{Distance} & Spectral Type & \textnormal{Stellar mass}
& \textnormal{Stellar radius} &  \multicolumn{2}{C}{\textnormal{Super-Earth}} &  \multicolumn{2}{C}{\textnormal{Sub-Saturn }} & Ref. & \textnormal{Debris}  & \textnormal{Binary} \\ 
&(\pc) && ( \msun) &  (\rsun) & \textnormal{conf.} & \textnormal{unconf.} & \textnormal{conf.} & \textnormal{unconf.} &  & \textnormal{disk} & system \\ 
\hline\hline\noalign{\smallskip}
GJ\,163 & 15.14 & M3.5 & 0.46\pm0.05 & $--$ & 2 & 2 & 3 & 2 &  1, ${a}$ & -- & -- \\
GJ\,176 & 9.49 & M2.5 & 0.52\pm0.05 & $--$ & 1 & 0 & 1 & 1 & 2, ${b}$ & -- & -- \\ 
GJ\,180 & 11.95 & M2 & 0.44\pm0.05 & $--$ & 2 & 1 & 2 & 1 & 3, ${ac}$ & -- & -- \\
BD-061339 & 20.31 & M0V & 0.65\pm0.05 & $--$ & 0 & 1 & 1 & 2 & 4, $ad$, 5, ${ad}$ & -- & -- \\
GJ\,229 & 5.76 & M2 & 0.57\pm0.05 & $--$ & 2 & 0 & 2 & 0 & 3, ${ac}$ & -- & yes \\
GJ\,422 & 12.68 & M3.5 & 0.42\pm0.05 & $--$ & 1 & 0 & 1 & 0 & 6, ${ac}$ & -- & -- \\
GJ\,433 & 9.08 & M1.5 & 0.50\pm0.05 & $--$ & 2 & 0 & 2 & 1 & 7, ${a}$ & yes & -- \\
GJ\,581 & 6.30 & M3 & 0.35\pm0.05 & 0.30\pm0.01 & 3 & 3 & 3 & 3 & 8, ${a}$ & yes & -- \\
Wolf\,1061 & 4.31 & M3V-M4.5V & 0.35\pm0.05 & 0.33\pm0.01 & 2 & 1 & 2 & 1 & 9, ${a}$  & -- & -- \\
GJ\,649 & 10.39 & M1.5 & 0.54\pm0.05 & 0.54\pm0.02 & 0 & 1 & 0 & 1 & 10, ${e}$  & yes & -- \\
GJ\,667C & 7.24 & M1.5 & 0.35\pm0.05 & 0.30\pm0.02 & 2 & 5 & 2 & 5 & 11, ${ade}$  & -- & yes \\
GJ\,674 & 4.55 & M2.5V & 0.40\pm0.05 & $--$ & 1 & 0 & 1 & 0 & 12, ${a}$  & -- & -- \\
GJ\,682 & 5.01 & M3.5 & 0.31\pm0.05 & $--$ & 0 & 2 & 0 & 2 & 3, ${ac}$  & -- & -- \\
GJ\,832 & 4.97 & M1.5 & 0.48\pm0.05 & 0.50\pm0.02 & 0 & 1 & 0 & 1 & 13, ${adf}$  & -- & -- \\
GJ\,3998 & 18.18 & M1.5 & 0.55\pm0.05 & $--$ & 2 & 0 & 2 & 0 & 14, ${g}$ & -- & -- \\
HD\,20794 & 6.04 & G8V & 0.79\pm0.05 & $--$ & 3 & 3 & 3 & 3 & 15, ${a}$ & yes & -- \\
HD\,3651 & 11.11 & K0V & 0.86\pm0.06 & $--$ & 0 & 0 & 1 & 0 & 16, ${h}$ & -- & yes \\
HD\,38858 & 15.21 & G2V & 0.91\pm0.06 & $--$ & 1 & 0 & 1 & 1 & 17, ${a}$ & yes & -- \\
HD\,40307 & 12.93 & K2.5V & 0.71\pm0.05 & $--$ & 4 & 2 & 4 & 2 & 18, ${a}$ & -- & -- \\
HD\,69830 & 12.58 & G8 & 0.88\pm0.06 & 0.91\pm0.02 & 3 & 0 & 3 & 0 & 19, ${a}$ & yes & -- \\
HD\,85512 & 11.28 & K6V & 0.70\pm0.05 & $--$ & 1 & 0 & 1 & 0 & 15, ${a}$ & -- & -- \\
HD\,99492 & 18.16 & K2V & 0.84\pm0.05 & $--$ & 0 & 0 & 1 & 0 & 20, ${e}$ & -- & yes \\
HD\,102365 & 9.32 & G3V-G5V & 0.81\pm0.05 & $--$ & 0 & 1 & 0 & 1 & 21, ${f}$ & -- & yes \\
61\,Vir & 8.53 & G7V & 0.92\pm0.05 & $--$ & 2 & 0 & 2 & 1 & 22, ${f}$ & yes & -- \\
HD\,136352 & 14.74 & G2V & 0.83\pm0.05 & $--$ & 3 & 0 & 3 & 0 & 23, ${a}$  & -- & -- \\
HD\,154088 & 18.27 & K0IV-V & 0.94\pm0.05 & $--$ & 1 & 0 & 1 & 0 & 24, ${a}$  & -- & -- \\
HD\,189567 & 17.93 & G2V & 0.85\pm0.05 & $--$ & 2 & 0 & 2 & 0 & 24, ${a}$  & -- & -- \\
\noalign{\smallskip}\hline\hline
\end{tabular}
\label{tab:target_properties}
\tablefoot{For each system, we indicate the presence of confirmed (“conf.”) or debated (“unconf.”) super-Earth and sub-Saturn mass planets, and if it is known to host a debris disk (see the references in Appendix~\ref{app:biblio_indiv_systems}).  The system is marked as binary if it is bound to a star or brown dwarf with a mass $\gtrsim 20~\mj$ (references in Appendix~\ref{app:biblio_indiv_systems}). We give the references for the first planet discovered (even if the planet was debated afterwards) for each system in the column “Ref.” and the RV instrument(s) enabling the discovery. We review the literature on the presence of additional planet(s) and their (un)confirmed status in Appendix~\ref{app:biblio_indiv_systems}, with corresponding references therein. The stellar masses come from this work. The distances are taken from the \textit{Gaia} Early Data Release 3, with distance uncertainties given in Appendix~\ref{app:biblio_indiv_systems}.

\tablebib{ Discovery papers: 1 = \citet{Bonfils2013_GJ163_5planets}, 2 = \citet{Endl2008_GJ176_1planet_Neptune_wrong}, 3 = \citet{Tuomi2014_GJ180_GJ229_GJ422_GJ433_GJ682}, 4 = \citet{LoCurto2013_BD061339_2planets}, 5 = \citet{Arriagada2013_BD061339_2planets}, 6 = \citet{Feng2020_GJ180_GJ229_GJ422_GJ433}, 7 = \citet{Delfosse2013_GJ667_GJ433_RV_charac_planets}, 8 = \citet{Bonfils2005_GJ581_1planet}, 9 = \citet{Wright2016_Wolf1061_3planets_discovery_RV}, 10 = \citet{Johnson2010_GJ649_1planet}, 11 = \citet{Anglada-Escude2012_GJ667C_RV_3planets}, 12 = \citet{Bonfils2007_GJ674_1planet}, 13 = \citet{Wittenmyer2014_GJ832_2planets_cwrong}, 14 = \citet{Affer2016_GJ3998_2planets_mass}, 15 = \citet{Pepe2011_HD20794_HD85512}, 16 = \citet{Fischer2003_HD3651_1planet_subSaturn_discovery}, 17 = \citet{Mayor2011_GJ38858_HD136352_HD154088_HD189567}, 18 = \citet{Mayor2009_HD40307_3planets}, 19 = \citet{Lovis2006_HD69830_3planets}, 20 = \citet{Marcy2005_HD99492_planetb}, 21 = \citet{Tinney2011_HD102365_planetb}, 22 = \citet{Vogt2010_61Vir_RV_3exoplanets_discovery}, 23 = \citet{Udry2019_HD136352_3planets_RV}, 24 = \citet{Unger2021_HD154088_HD189567}. Instruments used for the discovery paper: $a$ = ESO/HARPS, $b$ = Hobby-Eberly Telescope \citep[HET,][]{Ramsey1998_RV_HET} with the High-Resolution Spectrograph \citep[HRS,][]{Tull1998_RV_HET_HRS}, $c$ = Ultraviolet and Visual Echelle Spectrograph \citep[UVES,][]{Dekker2000_RV_UVES}, $d$ = Magellan Clay telescope Planet Finder Spectrograph \citep{Crane2006_RV_PFS,Crane2008_RV_PFS,Crane2010_RV_PFS}, $e$ = Keck Observatory High-Resolution Echelle spectrometer \citep[HIRES,][]{Vogt1994_RV_HIRES}), $f$ = Anglo-Australian Telescope (AAT) with the University College London Echelle Spectrograph \citep[UCLES,][]{Diego1990_RV_UCLES},  $g$ = Telescopi Nazionale Galileo (TNG) HARPS-N \citep{Cosentino2012_RV_HARPSN}, $h$ = Lick Hamilton spectrometer \citep{Vogt1987_RV_LICK}.}}
\end{table*}

\subsection{Prior knowledge of the planetary system architectures}

\begin{figure*}[h]
    \centering
    \includegraphics[width=0.495\linewidth]{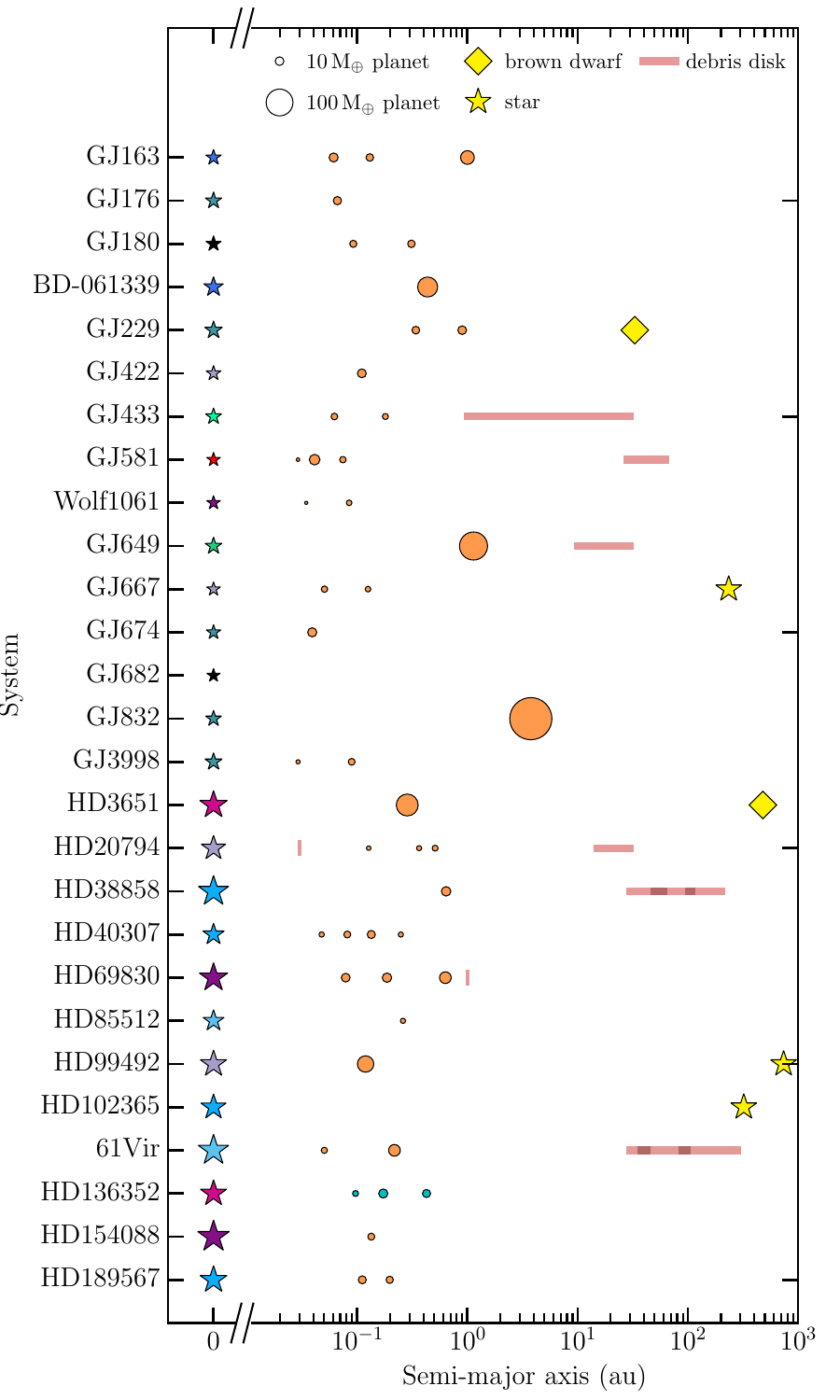} \hfill \includegraphics[width=0.495\linewidth]{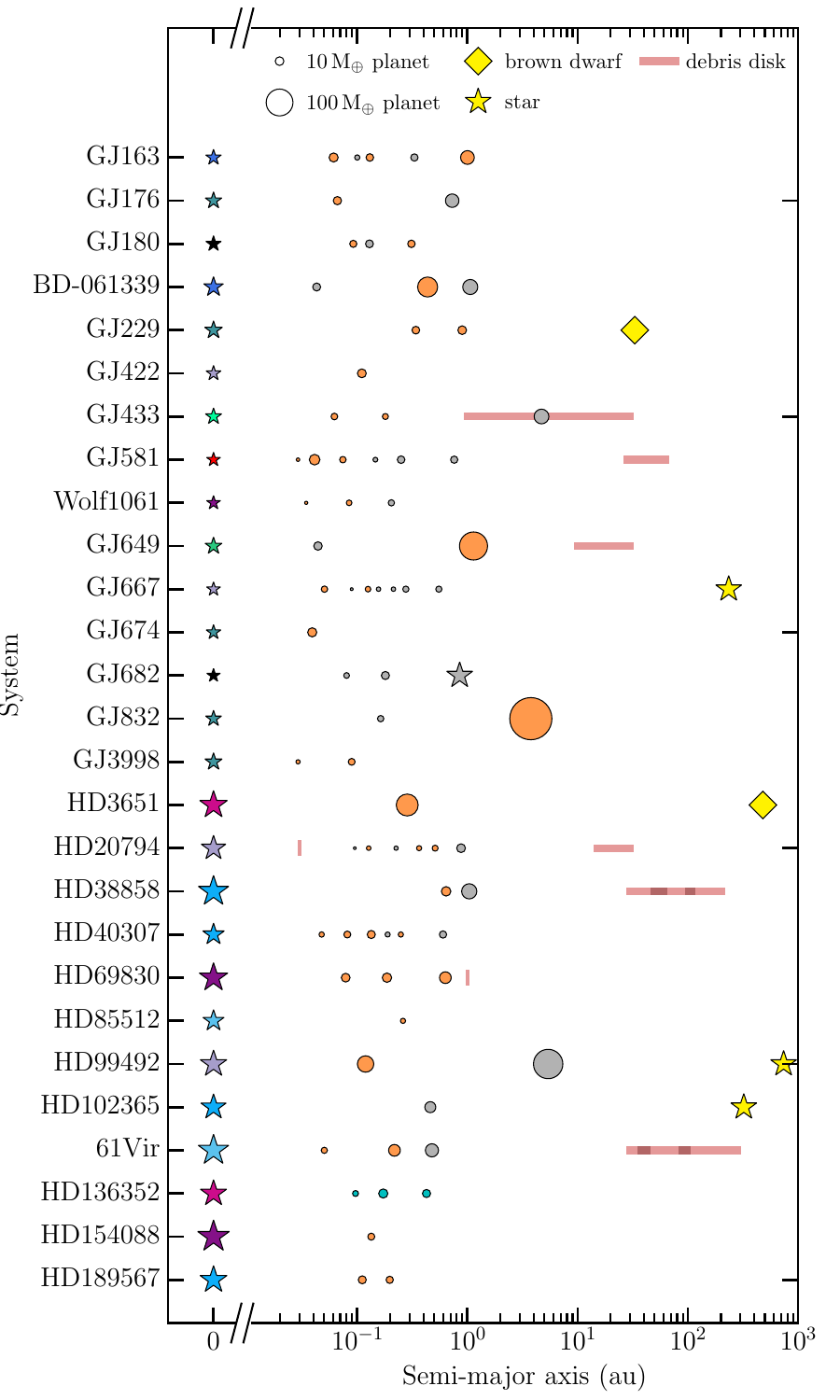}
    \caption{Overview of the $27$ systems in our sample. The planets are represented in orange if the minimal mass is used to scale the circle radius, in blue (HD\,136352) if the absolute mass is used instead. On the \textit{right} image, we added the unconfirmed planets in gray. We indicate the expected locations of the known debris disk(s) with red horizontal rectangles, with two possible structures for 61\,Vir and HD\,38858, either a single-belt or a two-belt (darker region only) architecture. For the host star, the color code is the same than in Fig.~\ref{fig:target_properties}, and the bigger the star marker is, the higher the stellar mass. The references for each system (star, planet, debris disk and/or binary companion) are indicated in Appendix~\ref{app:biblio_indiv_systems}. 
    } 
    \label{fig:overview_system_linear}
\end{figure*}

We point out that the systems in our sample show a diversity of architectures, see Fig. \ref{fig:overview_system_linear}. Some systems are known to host a debris disk (61~Vir, GJ~433, GJ~581, GJ~649, HD~20794, HD~38858 and HD~69830), others to be part of binary (GJ~229, HD~3651, HD~99492, and HD~102365) or even multiple-star (GJ~667~C) system.
Table~\ref{tab:target_properties} gives details on the architecture of each system in our sample, whereas Table~\ref{tab:summary_target_prop} summarizes the number of systems with a specific architecture. 
We provide a review on the state of the art of each system in the Appendix \ref{app:biblio_indiv_systems}.

\begin{table}[htbp]
\centering \small
\caption{Summary of the characteristics of the systems in our sample. 
}\label{tab:summary_target_prop}
\begin{tabular}{lccc}
\hline\hline\noalign{\smallskip}
Planetary systems considered & Total & Debris & Binary  \\
  &  &  disk &  system \\
\noalign{\smallskip}\hline\hline\noalign{\smallskip}
All the systems & 27 & 7 & 5 \\
Systems with confirmed sub-Saturn & 23 & 6 & 4 \\
Systems with confirmed super-Earth & 19 & 6 & 1 \\
\noalign{\smallskip}\hline\hline
\end{tabular}
\end{table}

Regarding the planets known in our sample, we note that some are debated in the literature, and thus we label them as unconfirmed. In the $27$~systems of the survey, four systems have only unconfirmed low-mass planets, while
the $23$~other systems are known to host such confirmed planets (see Fig.~\ref{fig:hist_multiplicity_Sub_Saturns}).
We define in this work low (minimal) mass planets as being less massive than $95~\mearth$ (i.e., sub-Saturn-mass planets). Some of the known planets in our sample correspond to the so called super-Earth  (rocky) or sub-Neptune planets (with a gaseous envelope). We consider (e.g., in Tables~\ref{tab:target_properties} and~\ref{tab:summary_target_prop}) as super-Earth a planet with a (minimal) mass less than $20~\mearth$, and stress that this definition differs in studies \citep[e.g.,][]{Izidoro2015_planet_formation_inner_SE_giant_barrier,Schlecker2021_NGPPS_SE_CJ,Mayor2011_GJ38858_HD136352_HD154088_HD189567,Bryan2019_RVTransit_study_observational_constraints_super_earth_correlation_super_earth_cold_jupiter,Burn2021_NGPPS}.  

\begin{figure}[h!]
    \centering
    \includegraphics[width=\linewidth]{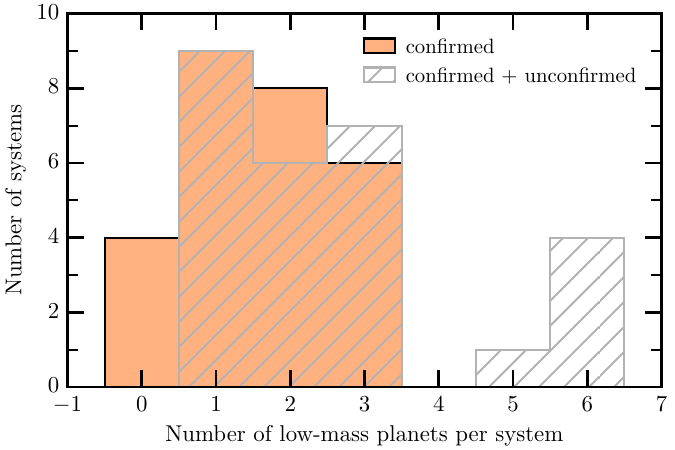}
    \caption{Number of close-in ($\lesssim1~\au$) low-mass ($\lesssim95~\mearth$)  planets per system in our sample. In orange is indicated the multiplicity based on the confirmed planets, while in gray, it is based on both confirmed and unconfirmed planets. Bins are centered on their corresponding multiplicity value, i.e., four systems do not have confirmed low-mass planets (GJ\,649, GJ\,682, GJ\,832, HD\,102365), even though at least one has been suggested in the literature (see details in Appendix~\ref{app:biblio_indiv_systems}). 
    }
    \label{fig:hist_multiplicity_Sub_Saturns}
\end{figure}

For a handful of companions in our sample, the absolute masses are known via transit or astrometry.
The transiting planets are the three known planets in the system HD~136352  \citep[transits seen with TESS and CHEOPS;][]{Kane2020_HD136352_transit_age_mass,Lovos2022_HD136352_Transit_TESS,Delrez2021_HD136352_Transit_CHEOPS}. In particular, the planet HD~136352~b has a radius of $1.66\pm0.05~\rearth$ (see 
Appendix \ref{app:biblio_indiv_systems}}), i.e., close to the photo-evaporation valley observed at about $1.8~\rearth$ for close-in planets. The photo-evaporation valley is believed to represent the separation between super-Earth and sub-Neptune-like planets \citep{Fulton2017_gap_kepler}. 
%
%
In addition, by combining RV measurements with the astrometry from \textit{Gaia} and \textit{Hipparcos}, the absolute mass of the brown dwarf GJ~229~B was constrained to  $71.4\pm0.6~\mj$ \citep{Brandt2021_GJ229B_dynamic_mass}. 

\section{Observations \label{sec:obs}}

We first present the $47$ VLT/SPHERE observations from our survey, then the archival data used (HARPS, \textit{Gaia} and \textit{Hipparcos}).

\subsection{VLT/SPHERE \label{sec:new_obs_DI}}


Our survey was conducted between 2017 and 2022. The observations were acquired with the VLT/SPHERE high-contrast instrument \citep{Beuzit2019_sphere}. All first-epoch observations were obtained with the IRDIFS mode, while additional observations to confirm the status of the point-source detections were acquired either with the IRDIFS or IRDIFS-EXT modes. Both of the modes combine simultaneously the IRDIS \citep{Dohlen2008} and IFS instruments \citep{Claudi2008}. The IRDIFS mode uses IRDIS in dual-band imaging \citep[DBI,][]{Vigan2010dbi} with the H2H3 filter doublet ($\lambda_{H2}=1.593\pm0.055~\mathrm{\mu m}$, $\lambda_{H3}=1.667\pm0.056~\mathrm{\mu m}$), and IFS in the YJ ($0.95$--$1.35\,\mu$m) spectral bands, while the  IRDIFS-EXT mode operates for IRDIS with the K1K2 filter ($\lambda_{K1}=2.103\pm0.102~\mathrm{\mu m}$, $\lambda_{K2}=2.255\pm0.109~\mathrm{\mu m}$) and for IFS with the YJH ($0.97$--$1.66\,\mu$m) bands.
We used the pupil-tracking mode for each observational sequence. This strategy of observation  enables to process the data with the  angular \citep[ADI,][]{Marois2006_ADI} and/or spectral differential imaging techniques \citep[SDI,][]{Racine1999_speckle_noise,Sparks2002}, with the combination of ADI and SDI noted ASDI, to achieve higher contrast at sub-arcsecond separations.

Table \ref{tab:obslog} reports the individual setup of the $47$ observations, as well as the observational conditions; those distributions are shown in Fig. \ref{fig:hist_obslog}.
The seeing ($\epsilon$) and the atmospheric coherence time parameters ($\tau_0$) were obtained from  the Atmospheric Site Monitor on the platform of the Paranal observatory, whereas the Strehl ratio (Sr) is measured by the SPHERE eXtreme AO \citep[SAXO,][]{Petit2014} real-time computer \citep[SPARTA,][]{Fedrigo2006sparta}. 

Most of the systems were observed once, while a few up to three or even five times, due to companion candidates or because the observations were repeated as the observational conditions degraded, and hence the requirements were not matched.



\subsection{Archival data (HARPS, \textit{Gaia}, and \textit{Hipparcos}) \label{sec:arxiv_data_RV_PMa}}

In synergy with direct imaging observations, we use radial velocity and astrometric measurements  to better constrain the presence of giant companions.

All the systems of the survey have low-mass planets discovered by radial velocity, and except HD\,3651, all the systems of the survey were observed at least with HARPS. 
In this work, we use the RV data to set constraints on the detection limits in terms of planet mass and location in each system of the survey.  We only use RV data from HARPS to avoid potential biases between instruments, though we note a possible bias between before and after the change of fiber in May 2015 \citep{LoCurto2013_BD061339_2planets}. A more complete analysis using all available RV measurements would be optimal, but this is beyond the scope of this paper focused on the direct imaging exploration of these mature systems. We note that in-depth studies exploiting various RV spectrographs have already been published in the literature on the systems from our sample, see Appendix \ref{app:biblio_indiv_systems}.

In addition, \textit{Gaia} and \textit{Hipparcos} data measurements exist for all the targets except GJ\,667C. We use both the proper motion anomaly, \citep{Kervella2019_PMa_inc,Kervella2022_PMa} and excess noise (Kiefer et al. in prep.)
 to look for the evidence of the presence of a binary companion, and to set constraints on the presence of planets for each system. The proper motion anomaly is computed by using long-term proper motion, i.e., the difference between \textit{Hipparcos} and \textit{Gaia} measurements (over a baseline of 2015.5~$-$~1991.25~=~24.25 years) and near-instantaneous proper motions. Regarding the excess noise, this represents the difference between the dispersion of the estimated \textit{Gaia} measurements at the estimated times of observations and the instrumental and intrinsic \textit{Gaia} noise (Kiefer et al. in prep).
We note that from the \textit{Gaia} Data Release 3, none of our targets is classified as a non-single star, either via astrometry or spectroscopy.

\section{Data reduction \label{sec:data_reduction}}

In the following sections, we describe the data reduction regarding the high-contrast imaging observations, more briefly the one for the radial velocity measurements.

\subsection{Direct imaging \label{sec:data_reduction_DI}}

The SPHERE observations were first pre-processed by the SPHERE Data Center \citep{delorme2017_red1step}, using the SPHERE Data Reduction and Handling pipeline \citep{Pavlov2008_red1step}. The pre-processing includes correction for bad pixels, dark current, flat non-uniformity, the sky background for both IRDIS and IFS, and for the wavelength and cross-talk between spectral channels in the calibration for IFS. 
The centering of the coronagraphic images is carried out by means of the four satellite spots which determine the precise position of the star behind the coronagraphic mask.
The astrometry of both IRDIS and IFS on sky is calibrated with past regular observations of a star crowded field (47 Tuc) as described in \citet{Maire2016,Maire2021}. It includes measurements of the detector plate scale, true north and  distortion. 

Second, we independently applied on all the observations two post-processing pipelines named ANDROMEDA \citep{Cantalloube2015_andromeda} and SpeCal \citep{galicher2018_red2step} to remove the starlight residuals in the coronagraphic images via the ADI or ASDI techniques, and to reveal putative planetary companions. 
ANDROMEDA is a forward-modelling approach using the maximum likelihood estimator \citep{mugnier2009}. It performs a simple pair-wise subtraction before tracking for the specific planetary signature that appears after the subtraction. Jointly for all pair of subtracted images, it fits the flux and position of each planetary signature. 
The SpeCal pipeline includes a set of different baseline algorithms such as classical ADI (cADI, \citealt{Marois2006_ADI}), locally optimized combination of images (LOCI/TLOCI, \citealt{Lafreniere2007_tloci}) and principal component analysis (PCA, \citealt{Amara2012_pca,Soummer2012_pca}).

 \begin{figure*}[h!]
    \centering 
    \includegraphics[width=0.495\linewidth]{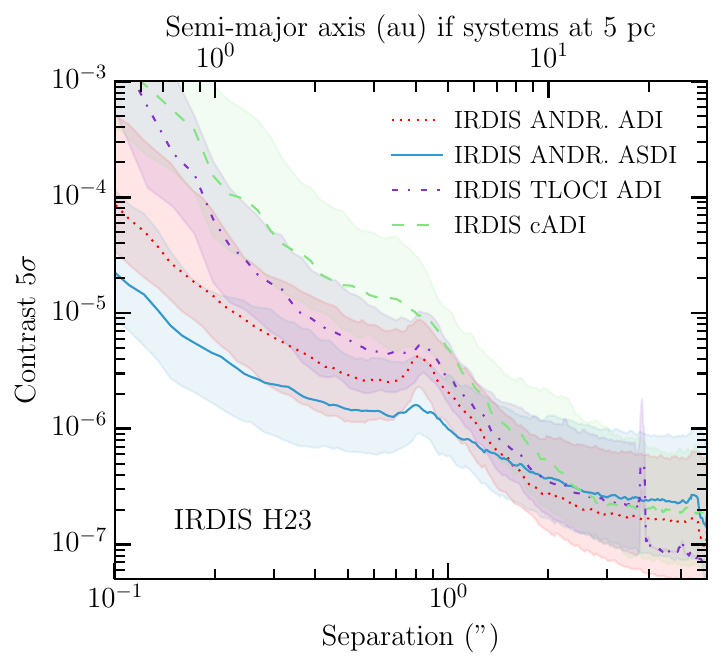} \includegraphics[width=0.495\linewidth]{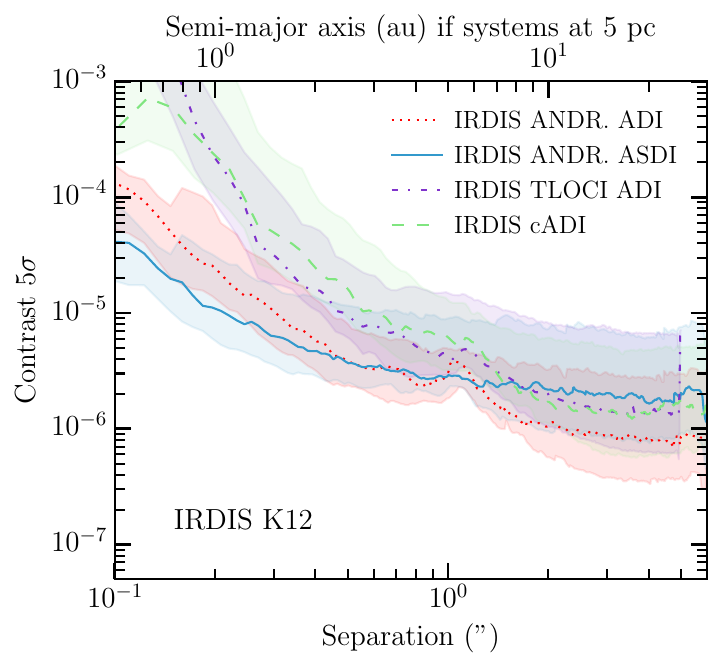} \vspace{3mm}
    
    \includegraphics[width=0.495\linewidth]{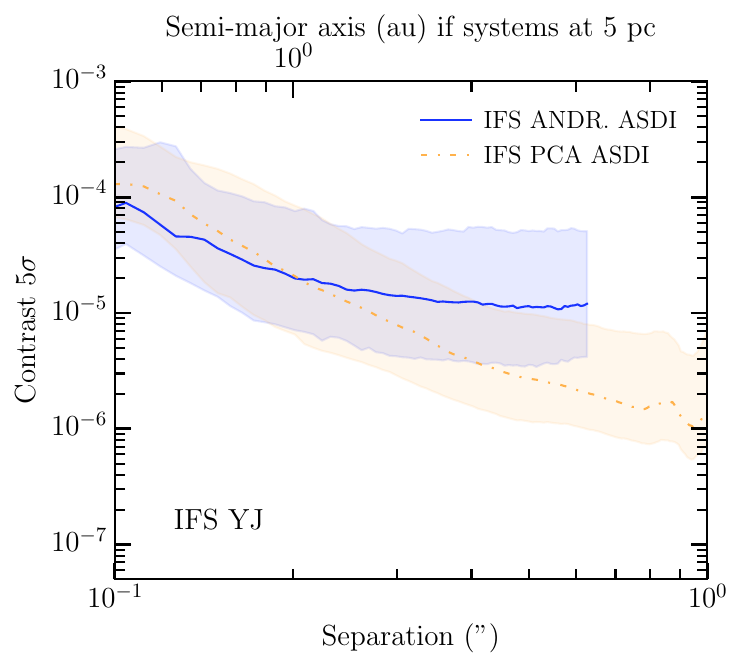} \includegraphics[width=0.495\linewidth]{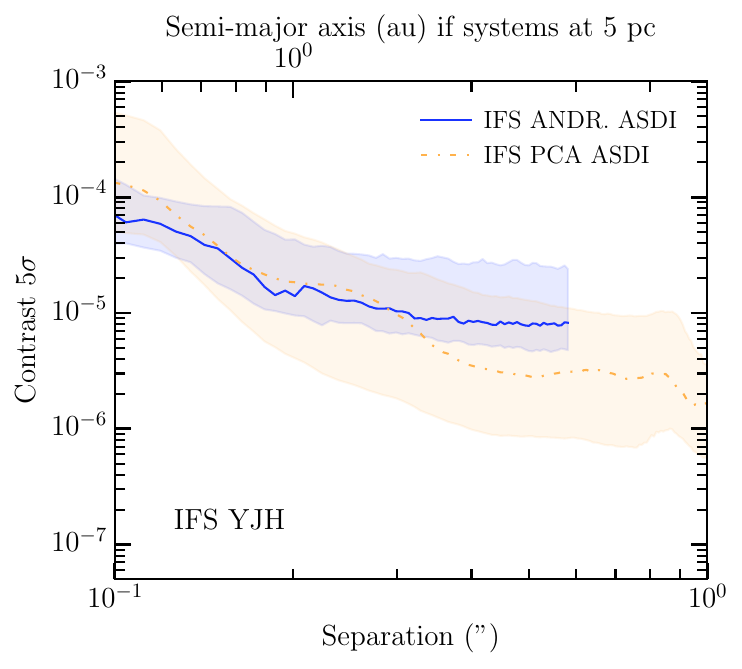}
    \caption{Sensitivity of our observations in terms of contrast with respect to the host star for the whole survey. \textit{Top panel:} Detection limits from SPHERE-IRDIS in the H23  (\textit{on the left}) and K12 (\textit{on the right}) bands for ANDROMEDA-ADI, ANDROMEDA-ASDI, SpeCal-TLOCI ADI and SpeCal-cADI post-processing. 
    \textit{Bottom panel:} Detection limits from SPHERE-IFS in YJ (\textit{on the left}) and in YJH (\textit{on the right}) for ANDROMEDA-ASDI and SpeCal-PCA-ASDI reductions. The detection limits are given as a function of the separation to the star, also expressed as semi-major axis for systems located at $5~\pc$, following semi-major axis (au) $=$ separation (") $\times$  distance~(pc).}
    
    \label{fig:detlims_contrast_curve}
\end{figure*}

The detection limits  of the different post-processing techniques are shown in  Fig. \ref{fig:detlims_contrast_curve}. Regarding IRDIS observations, ANDROMEDA-ASDI and ADI perform better than SpeCal-cADI and SpeCal-TLOCI-ADI at close separations, while for IFS data, SpeCal-PCA-ASDI \citep{Mesa2015} performs better than ANDROMEDA-ASDI or ADI.
For the sake of completeness, we show the IRDIS and IFS reduced images in terms of contrast relative to the host star in Figs.  \ref{fig:mosaic_super-earths_survey_IRDIS_cADI}, \ref{fig:mosaic_super-earths_survey_IRDIS_TLOCI} and \ref{fig:mosaic_super-earths_survey_IFS_PCAPad}, for the mentioned SpeCal algorithms, and in Figs. \ref{fig:mosaic_super-earths_survey_IRDIS_ANDROMEDA} and \ref{fig:mosaic_super-earths_survey_IFS_ANDROMEDA} the signal-to-noise ratio (S/N) maps for ANDROMEDA-ADI, respectively.

Companion candidates are identified based on the following methodology: detection of close ($\lesssim 2$") and faint ($\lesssim 10^{-5}$) point-like sources are estimated from the more sophisticated processing ANDROMEDA (both ADI and ASDI versions), whereas easier point-source detections (farther out or brighter) are considered from the baseline processing cADI. A comparison between ANDROMEDA and cADI detections is shown in Fig. \ref{fig:hist_plot_all_candidates}, with a few additional point-source detections in ANDROMEDA reduced images.

\subsection{Radial velocities \label{sec:data_reduction_RV}}

The HARPS RV data were processed following the ESO pipeline to obtain time series of the RV signals and uncertainties. The RV measurements were mainly used to estimate the RV detection limits described in Section~\ref{sec:results_obs_detlims}. 
By using the \texttt{DACE} tool \citep{Buchschacher2015_DACE}, we removed from the time series the Keplerian signal associated with the planets already discovered and confirmed by RV based on the literature values, except for GJ~649 and GJ~3998 as only a limited number of HARPS measurements are available. This step enables to remove the impact of known confirmed planets on determination of the detection limits.

The change of fiber in HARPS in May 2015 may cause some differences in the RV signal measured between before and after the operation  \citep{LoCurto2015_HARPS_change_fiber}. This can result in an offset between both measurements of about $20$--$50~\mathrm{m \,s^{-1}}$. We added the offset as a free parameter when relevant. If only a negligible number of measurements have been acquired after the fiber operation, we did not consider them in our RV time series.

\begin{figure*}[p]\centering
    \includegraphics[width=\linewidth]{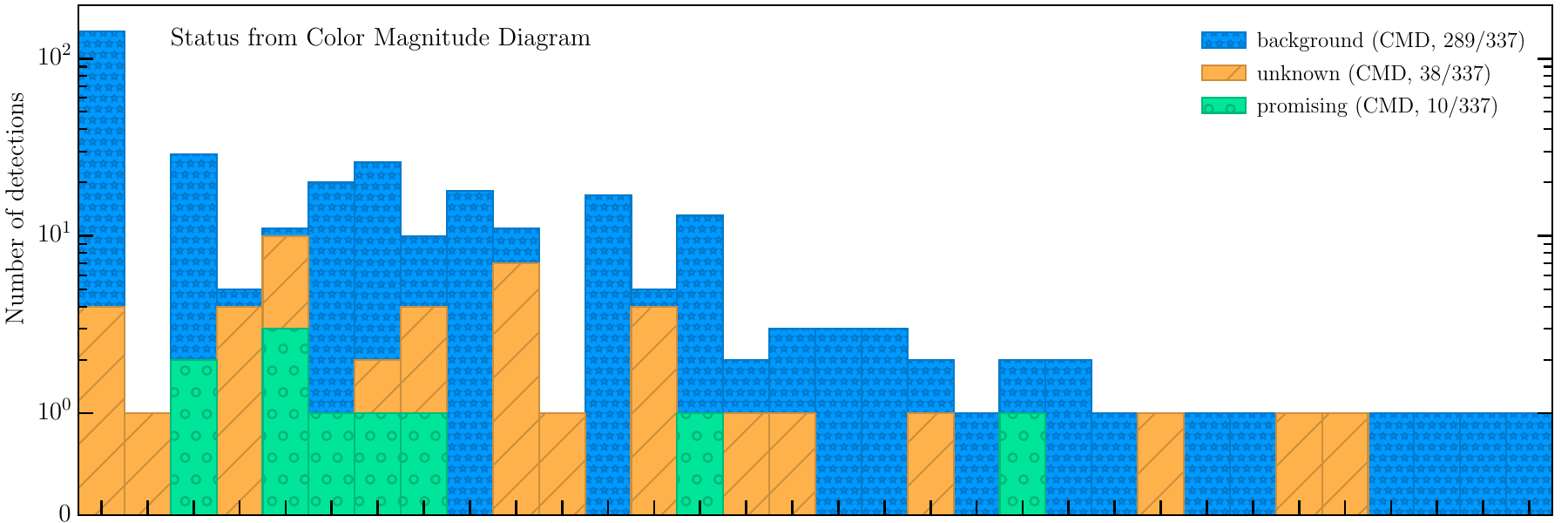} \vspace{-0.2cm}
    
    \includegraphics[width=\linewidth]{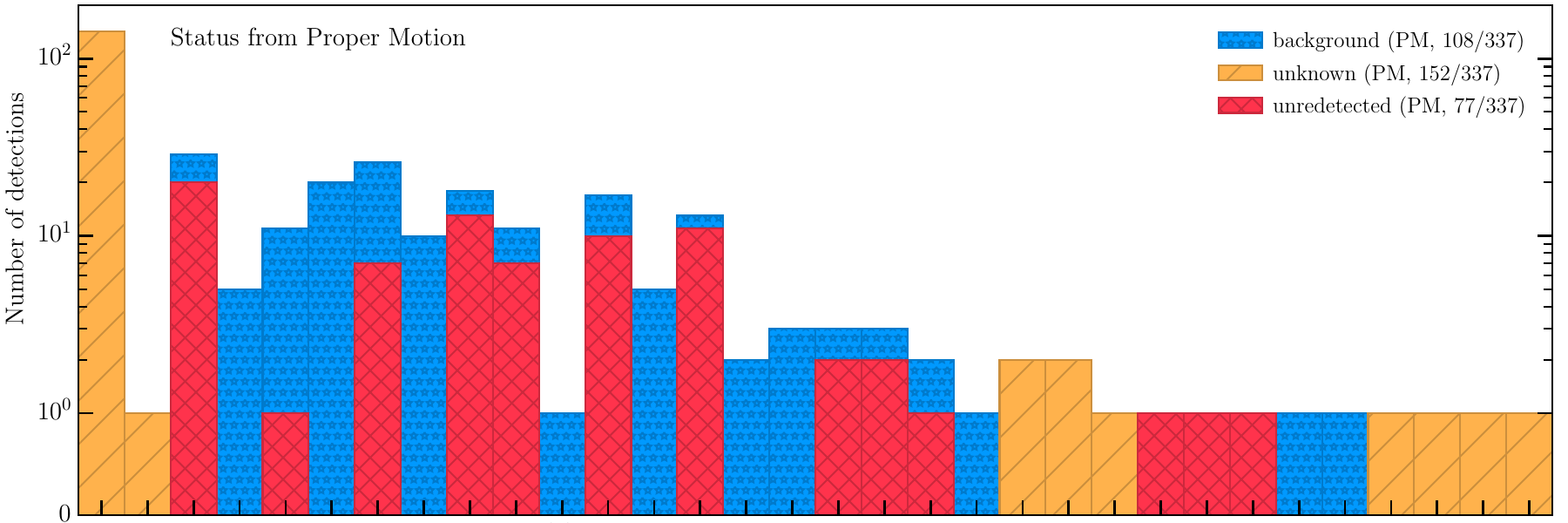} \vspace{-0.2cm}
    
    \includegraphics[width=\linewidth]{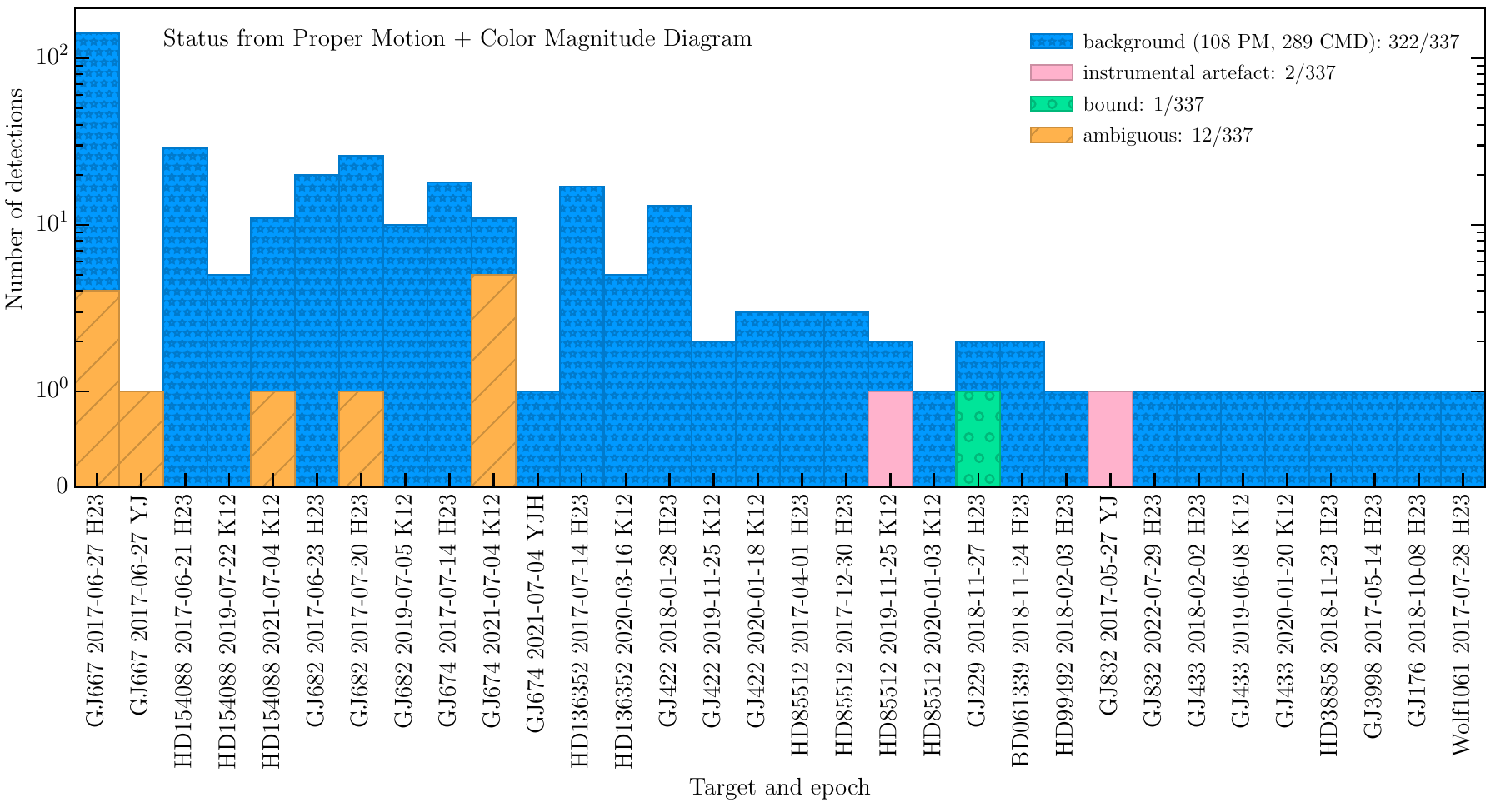}
    
    \caption{Status of all the point-source detections with SPHERE-IRDIS and SPHERE-IFS ranked by system, epoch of observation and filter. \textit{Top~panel:} based only on the Color Magnitude Diagram criterion. \textit{Middle panel:} based only on the Proper Motion Diagram criterion. \textit{Bottom panel:} Final status of the point-source detections based on both criteria. There is one already known bound companion, corresponding to the brown dwarf GJ~229~B, and none new. \vspace{0cm} 
    }
    \label{fig:hist_status_all_candidates}
\end{figure*}

\begin{figure*}[h!]
    \begin{center}
    \includegraphics[width=0.49\linewidth]{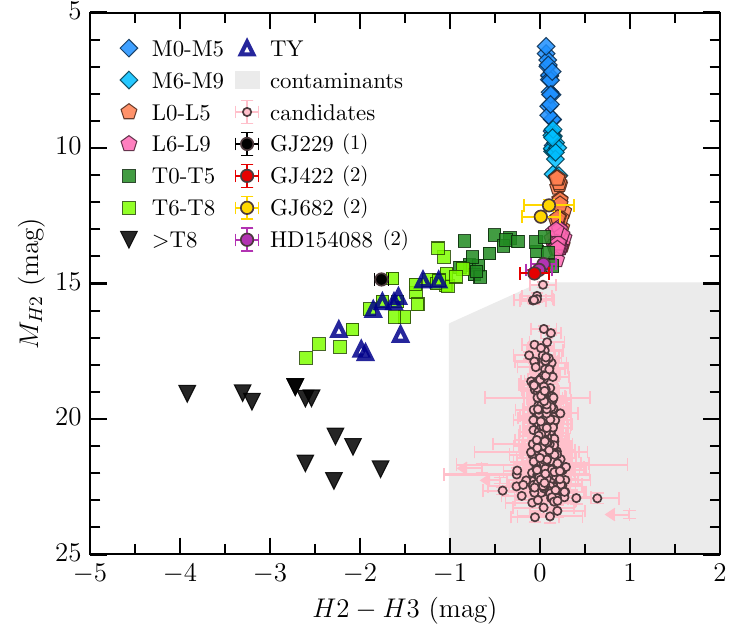} \hfill \includegraphics[width=0.49\linewidth]{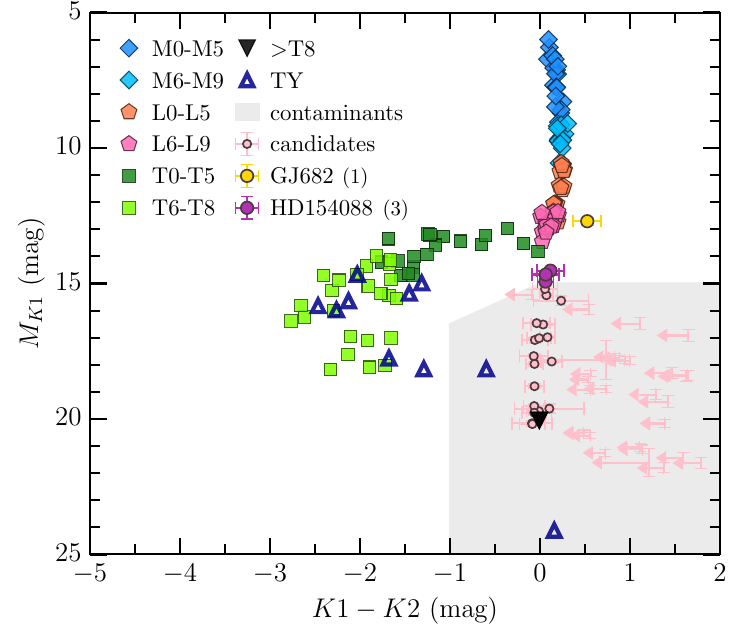}

    \includegraphics[width=0.49\linewidth]{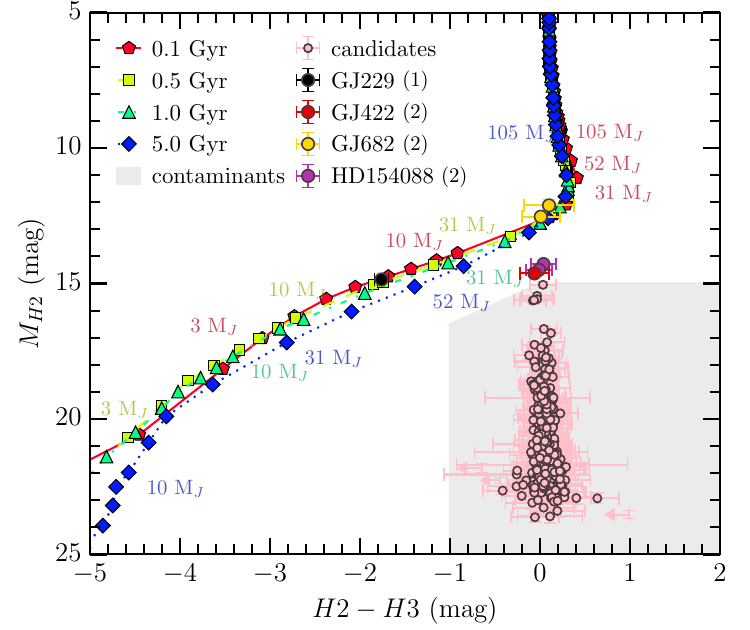}  \hfill 
    \includegraphics[width=0.49\linewidth]{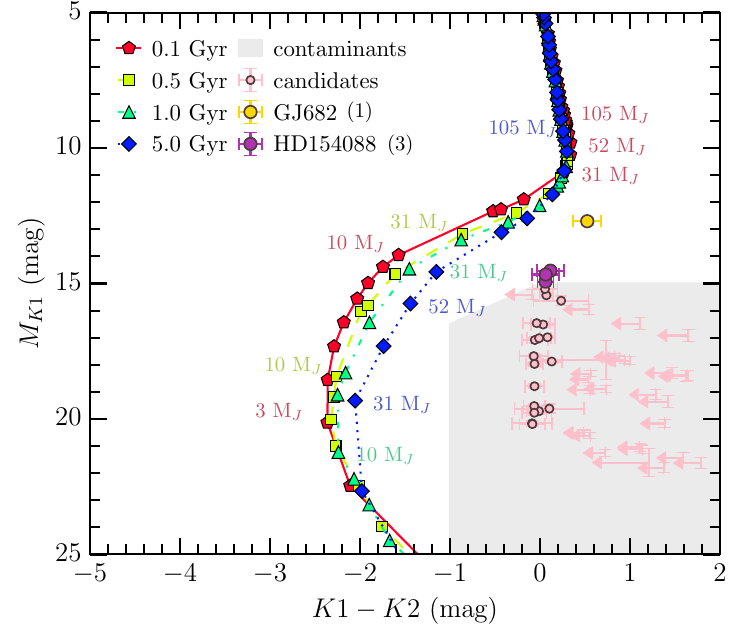}
     
    \caption{Color Magnitude Diagram in the band H23 (\textit{left}) and K12 (\textit{right}) for all the point sources detected in the H2 and/or K1 band over the whole survey. \textit{Top~panel:} The sequence of known giant planets  and brown dwarfs of different spectral types is shown\textsuperscript{\scriptsize\,a}.
    \textit{Bottom panel:} Isochrones from the evolutionary models \texttt{COND} \citep{Baraffe2003_COND} are plotted for different object masses.}
    \end{center} \vspace{-0.35cm}
    \small \textsuperscript{\scriptsize a} The photometry from MLTY brown dwarfs is taken from {\tiny \url{https://cass.ucsd.edu/~ajb/browndwarfs/spexprism/index.html}}
    \label{fig:CMD_H23_K12_empiric_and_theory}
\end{figure*}

\section{Results \label{sec:results_obs}}

We describe the classification of our detections of giant planet or brown dwarf candidates in Section \ref{sec:results_obs_DI}. In Section \ref{sec:results_obs_DI_candidates}, we report the bound and ambiguous point-source detections, and the ones likely caused by instrumental artifacts. In Section \ref{sec:results_gaiapmex}, we derive constraints on the potential companions consistent with measurements from \textit{Gaia} and \textit{Hipparcos} for each system of our sample. Last in Section \ref{sec:results_obs_detlims}, we derive the probability of detection in terms of mass and semi-major axis for each system providing the detection sensitivities of our VLT/SPHERE high-contrast imaging observations, the \textit{Gaia} and \textit{Hipparcos} data, and the HARPS radial velocity measurements.

\subsection{Overview of the direct imaging results \label{sec:results_obs_DI}}

Over $47$ SPHERE observations, $337$ point sources have been detected by SpeCal (bright or large-separation companion candidates) and ANDROMEDA (faint and close candidates). 
Table \ref{tab:detections_per_target_summarize} lists the number of companion candidates for each observation both for the IRDIS and IFS instruments, as well as their detection status (background contaminant, companion, instrumental artifact or ambiguous), also shown in Fig. \ref{fig:hist_status_all_candidates}.  

\begin{table*}
    \centering \small
     \caption{Overview of the SPHERE-IRDIS and SPHERE-IFS point-source detections per observation, as well as the status of these detections.
    }
    \begin{tabular}{lccc||C||CC|CC||CCCCC} 
    \hline\hline\noalign{\smallskip}
\multicolumn{4}{c||}{Observation information} & \textnormal{Detections} & \multicolumn{8}{c}{Status of the detections} \\
\noalign{\smallskip}\hline\noalign{\smallskip}
\multicolumn{4}{c||}{ } & \textnormal{} & \multicolumn{2}{c|}{(CMD)} & \multicolumn{2}{c||}{(PM)}  & \multicolumn{4}{c}{(CMD and/or PM)} \\

Target & \text{Date} & Instru. & Filter &  \text{Total}  & \text{Bkg.}  &  \text{Promising } & \text{Bkg.} &  \text{Not re-det.}  &  \text{Bkg.} & \text{Inst. Art.} & \text{Bound} & \text{Ambi.}  \\
    \noalign{\smallskip}\hline\hline\noalign{\smallskip}\noalign{\smallskip}
GJ\,176 & 2018-10-08 & IRDIS & H23 & 1 & 1 & 0 & 0 & 0 & 1 & 0 & 0 & 0 \\
BD-061339 & 2018-11-24 & IRDIS & H23 & 2 & 2 & 0 & 0 & 0 & 2 & 0 & 0 & 0 \\
GJ\,229 & 2018-11-27 & IRDIS & H23 & 2 & 1 & 1 & 0 & 0 & 1 & 0 & 1 & 0 \\
GJ\,422 & 2018-01-28 & IRDIS & H23 & 13 & 12 & 1 & 2 & 11 & 13 & 0 & 0 & 0 \\
GJ\,422 & 2019-11-25 & IRDIS & K12 & 2 & 1 & 0 & 2 & 0 & 2 & 0 & 0 & 0 \\
GJ\,422 & 2020-01-18 & IRDIS & K12 & 3 & 2 & 0 & 3 & 0 & 3 & 0 & 0 & 0 \\
GJ\,433 & 2018-02-02 & IRDIS & H23 & 1 & 1 & 0 & 1 & 0 & 1 & 0 & 0 & 0 \\ 
GJ\,433 & 2019-06-08 & IRDIS & K12 & 1 & 0 & 0 & 1 & 0 & 1 & 0 & 0 & 0 \\ 
GJ\,433 & 2020-01-20 & IRDIS & K12 & 1 & 0 & 0 & 1 & 0 & 1 & 0 & 0 & 0 \\
Wolf\,1061 & 2017-07-28 & IRDIS & H23 & 1 & 1 & 0 & 0 & 0 & 1 & 0 & 0 & 0 \\
GJ\,667 & 2017-06-27 & IRDIS & H23 & 142 & 138 & 0 & 0 & 0 & 138 & 0 & 0 & 4 \\
GJ\,667 & 2017-06-27 & IFS & YJ & 1 & 0 & 0 & 0 & 0 & 0 & 0 & 0 & 1 \\
GJ\,674 & 2017-07-14 & IRDIS & H23 & 18 & 18 & 0 & 5 & 13 & 18 & 0 & 0 & 0 \\
GJ\,674 & 2021-07-04 & IRDIS & K12 & 11 & 4 & 0 & 4 & 7 & 6 & 0 & 0 & 5 \\
GJ\,674 & 2021-07-04 & IFS & YJH & 1 & 0 & 0 & 1 &  0 &  1 & 0 & 0 & 0 \\
GJ\,682 & 2017-06-23 & IRDIS & H23 & 20 & 19 & 1 & 20 & 0 & 20 & 0 & 0 & 0 \\
GJ\,682 & 2017-07-20 & IRDIS & H23 & 26 & 24 & 1 & 19 & 7 & 25 & 0 & 0 & 1 \\
GJ\,682 & 2019-07-05 & IRDIS & K12 & 10 & 6 & 1 & 10 & 0 & 10 & 0 & 0 & 0 \\
GJ\,832 & 2017-05-27 & IFS & YJ & 1 & 0 & 0 & 0 & 1 & 0 & 1 & 0 & 0 \\
GJ\,832 & 2022-07-29 & IRDIS & H23 & 1 & 1 & 0 & 0 & 1 & 1 & 0 & 0 & 0 \\
GJ\,3998 & 2017-05-14 & IRDIS & H23 & 1 & 1 & 0 & 0 & 0 & 1 & 0 & 0 & 0 \\
HD\,38858 & 2018-11-23 & IRDIS & H23 & 1 & 1 & 0 & 0 & 0 & 1 & 0 & 0 & 0 \\
HD\,85512 & 2017-04-01 & IRDIS & H23 & 3 & 3 & 0 & 1 & 2 & 3 & 0 & 0 & 0 \\
HD\,85512 & 2017-12-30 & IRDIS & H23 & 3 & 3 & 0 & 1 & 2 & 3 & 0 & 0 & 0 \\
HD\,85512 & 2019-11-25 & IRDIS & K12 & 2 & 1 & 0 & 1 & 1 & 1 & 1 & 0 & 0 \\
HD\,85512 & 2020-01-03 & IRDIS & K12 & 1 & 1 & 0 & 1 & 0 & 1 & 0 & 0 & 0 \\
HD\,99492 & 2018-02-03 & IRDIS & H23 & 1 & 1 & 0 & 0 & 0 & 1 & 0 & 0 & 0 \\
HD\,136352 & 2017-07-14 & IRDIS & H23 & 17 & 17 & 0 & 7 & 10 & 17 & 0 & 0 & 0 \\
HD\,136352 & 2020-03-16 & IRDIS & K12 & 5 & 1 & 0 & 5 & 0 & 5 & 0 & 0 & 0 \\
HD\,154088 & 2017-06-21 & IRDIS & H23 & 29 & 27 & 2 & 9 & 20 & 29 & 0 & 0 & 0 \\
HD\,154088 & 2019-07-22 & IRDIS & K12 & 5 & 1 & 0 & 5 & 0 & 5 & 0 & 0 & 0 \\
HD\,154088 & 2021-07-04 & IRDIS & K12 & 11 & 1 & 3 & 10 & 1 & 10 & 0 & 0 & 1 \\
\noalign{\smallskip}\hline\hline\noalign{\smallskip}\noalign{\smallskip}
    \end{tabular}
    \label{tab:detections_per_target_summarize}
    \tablefoot{The observations with non-detections are not listed. More details on the classification of the companion candidates are given in Section~\ref{sec:results_obs_DI} and~\ref{sec:results_obs_DI_candidates}. 
     Notes: “Bkg.” stands for a background star; “Not re-det.” for not re-detected; “Inst. Art.” for instrumental artifact, “Ambi.” for ambiguous, “CMD” for Color Magnitude Diagram; “PM” for Proper Motion. }
\end{table*}

\begin{table*}
    \centering \small
    \caption{Relative astrometry and photometry with respect to their host star of the point-source detections in our SPHERE survey labeled as promising, ambiguous or instrumental artifacts.
    } 
    \begin{tabular}{lccCCCCCCcc} 
    \hline\hline\noalign{\smallskip}
Target & \text{Epoch} & Instr.-Filt.  & \Delta \alpha & \Delta \delta & \text{Separation} & \text{PA} & \text{Contrast} & \text{S/N} & \text{Status} & ID    \\
 & &  & \text{(mas)} & \text{(mas)}  &  \text{(mas)} &  \text{($\degr$)}  &  &  &  & \\
    \noalign{\smallskip}\hline\hline\noalign{\smallskip}\noalign{\smallskip}
GJ\,229 & 2018-11-27 & IRDIS-H2 & 228\pm12 & -5204\pm3 & 5209\pm12 & 177.5\pm0.8 & (2.0\pm0.1)\times10^{-4} & 88 & bound & -- \\ 
 \noalign{\smallskip}\hline\noalign{\smallskip}
GJ\,667 & 2017-06-27 & IFS-YJ & -535\pm3 & -159\pm3 & 558\pm3 & 253.5\pm0.3 &  (3.1\pm0.4)\times10^{-5} & 7 & ambi.  & 1  \\ 
GJ\,667 & 2017-06-27 & IRDIS-H2 & -2526\pm15 & -1437\pm10 & 2906\pm15 & 240.4\pm0.2 & (8.0\pm0.1)\times10^{-7} & 13 & ambi.  & 2 \\ 
GJ\,667 & 2017-06-27 & IRDIS-H2 & -3462\pm8 & 587\pm8 & 3511\pm8 & 279.6\pm0.1 & (4.3\pm0.8)\times10^{-7} & 6 & ambi.  & 3  \\ 
GJ\,667 & 2017-06-27 & IRDIS-H2 & 3760\pm40 & 3004\pm12 & 4813\pm40 & 51.4\pm0.2 & (6.8\pm1.0)\times10^{-7} & 8 & ambi.  & 4 \\ 
GJ\,667 & 2017-06-27 & IRDIS-H2 & 5885\pm8 & -424\pm8 & 5901\pm8 & 94.1\pm0.1 & (1.2\pm0.2)\times10^{-6} & 7 & ambi.  & 5 \\ 
 \noalign{\smallskip}\hline\noalign{\smallskip}
GJ\,674 & 2021-07-04 & IRDIS-K1 & -1550\pm27 & -2677\pm6 & 3093\pm27 & 210.1\pm0.2 & (9.4\pm1.7)\times10^{-7}  & 6 & ambi.  & 6  \\ 
GJ\,674 & 2021-07-04 & IRDIS-K1 & 986\pm10 & -5092\pm10 & 5187\pm10 & 169.0\pm0.6 & (2.4\pm0.2)\times10^{-6}  & 12 & ambi.  & 7 \\ 
GJ\,674 & 2021-07-04 & IRDIS-K1 & -2382\pm9 & -2825\pm9 & 3695\pm9 & 220.1\pm0.2 & (2.6\pm0.3)\times10^{-6}  & 12 & ambi.  & 8 \\ 
GJ\,674 & 2021-07-04 & IRDIS-K1 & -3206\pm11 & -4354\pm11 & 5407\pm11 & 216.4\pm0.2 & (2.6\pm0.3)\times10^{-6}  & 12 & ambi.  & 9  \\ 
GJ\,674 & 2021-07-04 & IRDIS-K1 & 226\pm7 & -2601\pm7 & 2611\pm7 & 175\pm2 & (1.6\pm0.2)\times10^{-6} & 9 & ambi.  & 10 \\ 
 \noalign{\smallskip}\hline\noalign{\smallskip}
 GJ\,682 & 2017-07-20 & IRDIS-H2 & 3571\pm5 & 1667\pm6 & 3941\pm5 & 65.0\pm0.1 & (3.6\pm0.5)\times10^{-7}  & 8 & ambi.  & 11  \\ 
 \noalign{\smallskip}\hline\noalign{\smallskip}
 GJ\,832 & 2017-05-27 & IFS-YJ & 487\pm3 & 194\pm3 & 524\pm3 & 68.3\pm0.3  & (2.7\pm0.9)\times10^{-6} & 4 & inst. art.  & --  \\
 \noalign{\smallskip}\hline\noalign{\smallskip}
 HD\,85512 & 2019-11-25 & IRDIS-K1 & 911\pm7 & 11\pm4 & 912\pm6 & 89.3\pm0.2 & (4.5\pm2.2)\times10^{-6} & 6 & inst. art.  & --  \\ 
 \noalign{\smallskip}\hline\noalign{\smallskip}
 HD\,154088 & 2021-07-04 & IRDIS-K1 & 3525\pm8 & -4408\pm7 & 5644\pm8 & 141.3\pm0.1 & (1.0\pm0.2)\times10^{-6} & 6 & ambi.  & 12  \\ 
    \noalign{\smallskip}\hline\hline\noalign{\smallskip}
    \end{tabular}
    \label{tab:detections_promising}
    \tablefoot{ More information on the “Status” column is given in Section~\ref{sec:results_obs_DI} and~\ref{sec:results_obs_DI_candidates}.}
\end{table*}

\begin{figure}[h!]
    \centering
    \includegraphics[width=\linewidth]{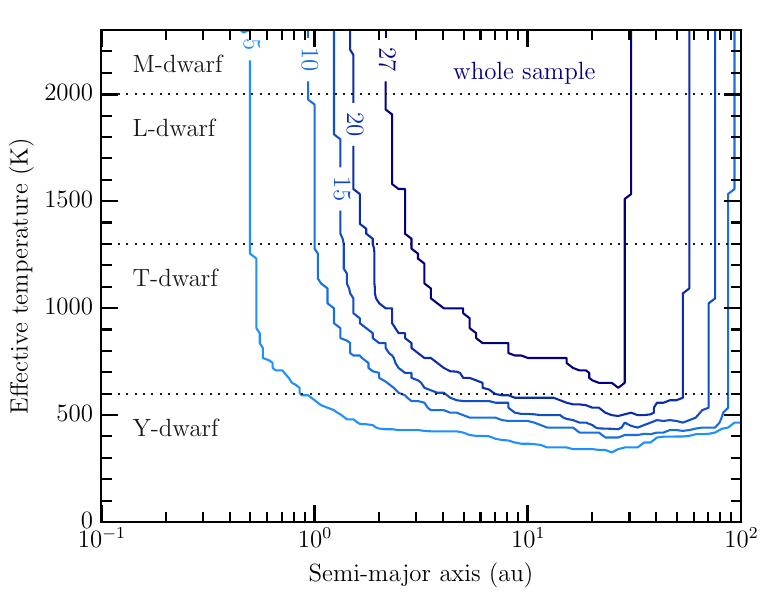}
    \caption{
    Sensitivity for our direct imaging survey.  The isocontours give the number of stars that have the sensitivity to detect companions at a probability of $50\%$ for this given semi-major axis and effective temperature.
    The dotted lines represent the indicative effective temperatures of the transitions between Y and T-dwarfs ($\sim600\,\kelvin$), T and L-dwarfs ($\sim1300\,\kelvin$), L and M-dwarfs ($\sim2000\,\kelvin$). }
    \label{fig:detlim_Teff_HD85512}
\end{figure}

We used two independent criteria to find out about the nature of the companion candidates.
The first criterion consists in extracting the photometry in two bands, H2H3 (or K1K2), and comparing it with empirical and theoretical predictions on a CMD, see Fig \ref{fig:CMD_H23_K12_empiric_and_theory}. The empirical expectation is based on a sequence of old known brown dwarfs and substellar objects for different spectral types (symbols). The theoretical predictions are from the \texttt{COND} evolutionary model  \citep[lines,][]{Baraffe2003_COND}. This model describes the evolution of the internal structure of a planet with time, and gives constraints on the physical properties of the planet, such as its luminosity, mass, effective temperature, surface gravity, and radius. The promising point-source detections are defined as having similar photometric magnitudes than the empirical sequence of known substellar objects, while the other detections are likely to be background contaminants. In particular, one can define the area of background contaminants as the one corresponding to faint objects (M$_\text{H2}\gtrsim15~\magg$) and without color ($\text{H2-H3}\lesssim-1~\magg$). Reciprocally, the brighter point-source detections or with a significant color are considered as promising. For promising detections, we asked for additional epochs in the K12 dualband, such as for the systems GJ\,422, GJ\,682, and HD\,154088. 

    
    The same principle applies for K12 observations, except some difficulties arise to define a background contaminant zone. Indeed, our direct imaging observations are sensitive to companions with an effective temperature down to $400\,\mathrm{K}$ (see Fig. \ref{fig:detlim_Teff_HD85512}) corresponding to giant planets of TY spectral type. Those objects are very faint in the K1 band, and some of them with a color K1-K2 close to zero, hence they could be confused with background stars (see Fig. \ref{fig:CMD_H23_K12_empiric_and_theory}).  In addition, another issue arose, as a dozen point-source detections are only detected in the K1 band (and not in the K2 band). On the one hand, this could be interpreted positively as the K2 band corresponds to methane absorption.  Thus, it may advocate for the presence of an exoplanet with methane in its atmosphere. On the other hand, if the point-source detection is faint in the K1 band, the absence of detection in the K2 band could suggest that the detection is rather an instrumental artifact than a real object. 
    We note that both the GJ\,682 and HD\,154088 systems possess promising point-source detections based on the CMD in the K12 bands, coinciding with the promising detections in the H23 bands.  
    In summary, we could flag $289$ companion candidates as background contaminants, $10$ as promising, and $38$ remain as unknown as only detected with the IFS, or in one of the IRDIS dual band, H2 or K1 (see Fig. \ref{fig:hist_status_all_candidates}, \textit{top panel}).
    

        
    The second criterion is the relative motion of the point-source detections with respect to their host stars when comparing at least two epochs. If a detection is a background star, its expected location in the second epoch can be derived by knowing the proper and parallactic motion of its host star. Conversely, if a companion candidate is detected at about the same location relative to the star between two epochs, it suggests it is bound to the star, by assuming its orbital motion is small. Of the $27$ systems of our sample, $11$ have been observed at least twice, either as having promising companion candidates or because the observational conditions did not satisfy the requirements asked for. We have $108$ point-source detections flagged as background contaminants based on this absolute criterion, summarized in Fig. \ref{fig:hist_status_all_candidates}. Proper motion diagrams are shown in Appendix (Figs. \ref{fig:PMD_GJ422_2+3epochs}--\ref{fig:PMD_HD154088_2+3epochs}).
   For $77$ companion candidates, new observations were acquired at a second epoch, but the companion candidates were not re-detected at the expected location for a bound object and/or a stationary background contaminant. For some cases, the predicted location for a background contaminant was out of the field of view given the high proper motion of the star (see next Section). The $152$ “unknown” point-source detections correspond to companion candidates in system observed only once.

In total, we were able to flag $322$/$337$ point-source detections as background contaminants, with $289$ being from the CMD only and $108$ from proper motion. We discuss the other $15$ remaining companion candidates in the following section.

\subsection{Substellar companion candidates \label{sec:results_obs_DI_candidates}}


Of the $337$~point-source detections with SPHERE-IRDIS and SPHERE-IFS, one is already known as the companion GJ\,229\,B \citep{Nakajima1995_GJ229_1planet_BD}, two are identified as instrumental artifacts (GJ\,832, HD\,85512), and the twelve others remain with an ambiguous nature, as we lack sufficient information to conclude definitively. Table~\ref{tab:detections_promising} lists these $15$ point-source detections, labeled as “bound”, “instrumental artifact” “ambiguous”, respectively, along with their relative astrometry and contrast with respect to their host star.

\begin{figure*}[h!] \centering
    \includegraphics[height=5.cm]{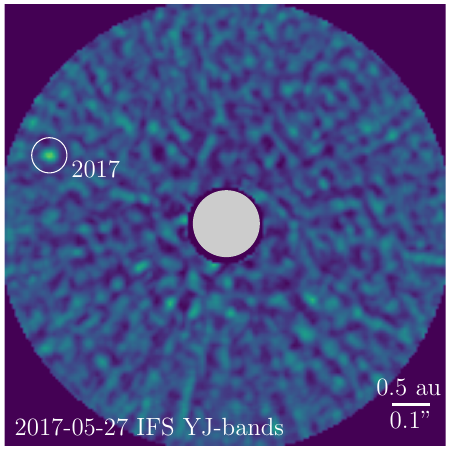} \hspace{0.5cm}
    \includegraphics[height=5.cm]{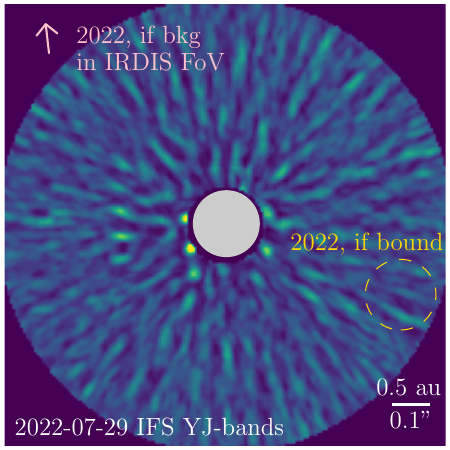}
    \hspace{0.3cm}
    \includegraphics[height=5.cm]{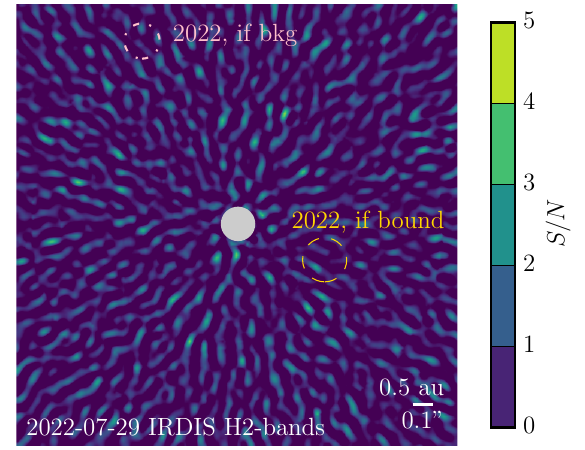}
    \caption{Companion candidate in the system GJ\,832: detected in the epoch 2017-05-27 (in the white circle) but not in 2022-07-29, neither at the expected location if bound (yellow dashed circle) or if a background contaminant (“bkg”, in the pink dashed circle visible only in the IRDIS field of view) in spite of better conditions of observation and higher detection sensitivity (see text for more information).\label{fig:GJ832_proposal_verdict} \vspace{0.1cm}}
\end{figure*}


We list below the reasons why ambiguous companion candidates are labeled as such. 
We note that the ambiguous companion candidates are all detected in systems with a crowded field of view (see Table~\ref{tab:detections_per_target_summarize}).
\begin{itemize} 
    \item candidate 1 (GJ\,667): detected only with the IFS in the combined YJ bands in 2017-06-27, so we could not use the CMD criterion. There is no additional epoch of observation on the system GJ\,667.
    \item candidate 2 (GJ\,667): detected only in the band H2 (and not in the band H3) in 2017-06-27, hence we cannot conclude to its color H2-H3. There is no second epoch on the system GJ\,667.
     
    \item candidates 3, 4, and 5 (GJ\,667): detected only in the band H2 in 2017-06-27 (and not in the band H3), hence we cannot use their color H2-H3 to conclude. There is no additional epoch of observation on the system GJ\,667. We note these candidates are at separations $\gtrsim 3$", and what should look like point sources are in practice elongated azimuthally for separations $\gtrsim 3$" for the observation on GJ\,667. This is caused by a too fast rotation of the field of view  with respect to the frame exposure time. The greater the separation, the stronger the effect is. This could bias our flux estimation of the companion candidates~3,~4 and~5.
    
   
    
    \item candidates 6, 7, 8, 9 and 10 (GJ\,674): candidates detected only in the K1 band in 2021-07-04. If the companion candidates are background stars, there are not expected to be in the field of view ($11$"$\times 11$" for IRDIS) of our other epoch of observation, 2017-07-14, acquired in H23, considering their location and the fact that the star GJ\,674\,A moves of $0.57$"/yr in right-ascension, and $-0.88$"/yr in declination \citep{Gaia2020_EDR3_Vizier_motion}.
    
    \item candidate 11 (GJ\,682): faint companion candidate detected only in the H2 band in 2017-07-20. The companion candidate is in the field of view of the previous epoch of observation 2017-06-23, but the epoch was acquired under poorer observation conditions which lead to poorer detection sensitivity. The companion candidate is out of the field of view of the third epoch of observation, 2019-07-05, as the proper motion of the star GJ\,682\,A is high
    \citep[$0.71$"/yr in right-ascension, $-0.94$"/yr in declination,][]{Gaia2020_EDR3_Vizier_motion}
    
    \item candidate 12 (HD\,154088): companion candidate detected both in K1 and K2 bands in 2021-07-04 with a promising CMD, but not detected in the epoch of observation 2019-07-22 in K12 dual band (acquired with poorer condition of observations) or in the other dual band H23 in 2017-06-21. Due to the proper motion of the star HD\,154088\,A  \citep[$0.83$"/yr in right-ascension, $-0.27$"/yr in declination,][]{Gaia2020_EDR3_Vizier_motion}, if the companion candidate is a background star, it is expected to be out of the field of view.
\end{itemize}

Regarding the point-source detection in GJ\,832 that we classified as an instrumental artifact, this was detected in the IFS data (epoch 2017-05-27) at a separation of $524~\mas$ (i.e., $2.7~\au$), which could have been consistent with the giant planet already known from RV measurements ($m\times\sin(i)=0.74\pm0.06~\mj$). 
However, if the imaged companion candidate was indeed the RV planet, the system  GJ\,832 should have been much younger (about $100~\myr$), than the age estimation of $2.4\pm0.3~\gyr$ (see Table \ref{tab-age}). Indeed, at $100~\myr$, the imaged companion candidate could correspond to a planet of $1~\mj$ based on the evolutionary models \texttt{COND}, and be consistent with a RV-planet seen with an inclination of about $50\deg$. 
Unfortunately, we did not detect the companion candidate again in the latter observation which was acquired under better observing conditions (2022-07-29), and reached better detection sensitivity. There was no detection either at the expected location of a background object
, or at its expected location if bound and coinciding with GJ\,832\,b (see Fig.\ref{fig:GJ832_proposal_verdict}). This location was expected in the epoch 2022-07-29 on the opposite side of the star to its location in the epoch 2017-05-27, as the period of GJ\,832\,b  is about ten years from RV measurements, and the observations were taken five years apart. 
Therefore, this point-source detection at low $S/N$ ($S/N\sim4$) is likely to be an instrumental artifact. The same conclusion applies to the point-source detection in HD\,85512 (epoch 2019-11-25). This companion candidate was not re-detected in the later epoch acquired with better observational conditions and in the same filter, either as a bound or background object, thereby we conclude this to be an artifact.

\subsection{Results from proper motion anomaly and excess noise (Gaia-Hipparcos) \label{sec:results_gaiapmex}}


The \textit{Gaia} and \textit{Hipparcos} data were analyzed with the \texttt{GaiaPMEX} tool (Kiefer et al. in prep) to determine the possible mass and semi-major axis (sma, hereafter) of candidate companions within the selected systems using the proper motion anomaly (PMa, hereafter) from the~\citet{Kervella2022_PMa} catalog and the \texttt{astrometric\_excess\_noise} (AEN, hereafter; for a definition see~\citet{Lindegren2018} and \citet{Kiefer2019_Gaia}) from the Gaia DR3~\citep{Gaia2021_EDR3}.

\begin{table}
    \centering
    \caption{Distribution of parameters sampled at each tested bin of the mass -- semi major axis (“sma”) grid with the \texttt{GaiaPMEX} tool.}
    \label{tab:distro}
    \begin{tabular}{lcc}
        \noalign{\smallskip}\hline \hline  \noalign{\smallskip}
        Parameter &  Type & Bounds or law \\
        \noalign{\smallskip}\hline \hline  \noalign{\smallskip}
        $\log M_c$ & uniform &  $\log M_c\pm\Delta\log M_c$  \\
        $\log \text{sma}$ & uniform &  $\log \text{sma}\pm\Delta\log \text{sma}$ \\ 
        $e$ & uniform &  [$0$, $0.9$] \\
        $\omega$ & uniform &  [$0$, $\pi$] \\
        $\Omega$ & uniform &  [$0$, $2\pi$] \\
        $\phi$ & uniform &  [$0$, $1$] \\
        $I_c$ & uniform or $\sin i$ & [$0$, $\pi/2$] \\
        $\varpi$ & normal & $\mathcal N$(PLX,\,$\sigma_{\rm PLX}^2$)  \\
        $M_\star$ & normal & $\mathcal N$($M_\star$,\,$\sigma_{M_\star}^2$)  \\
        \noalign{\smallskip}\hline \hline  \noalign{\smallskip}
    \end{tabular}
    \tablefoot{“$M_c$” stands for the mass of the companion, “PLX” for the parallax, other parameters are defined in Section~\ref{sec:results_gaiapmex}.}
\end{table}

\begin{figure*}[h!]
    \centering 
    \includegraphics[width=0.45\linewidth]{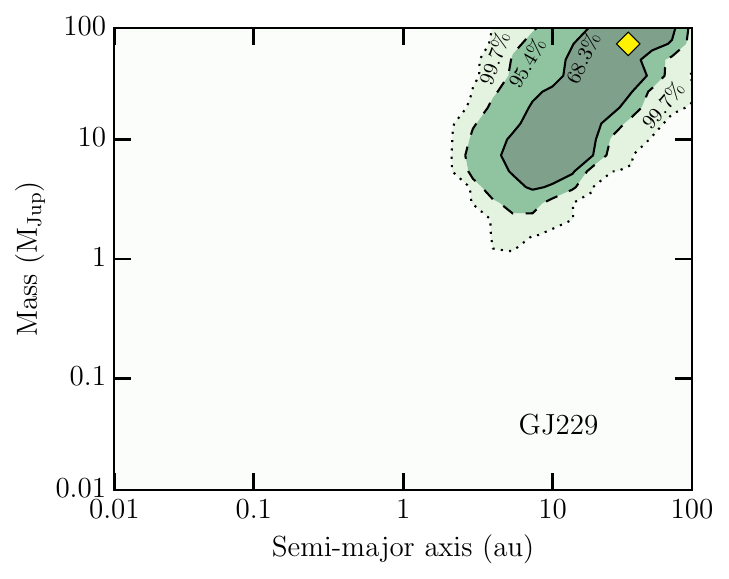} \qquad  \includegraphics[width=0.45\linewidth]{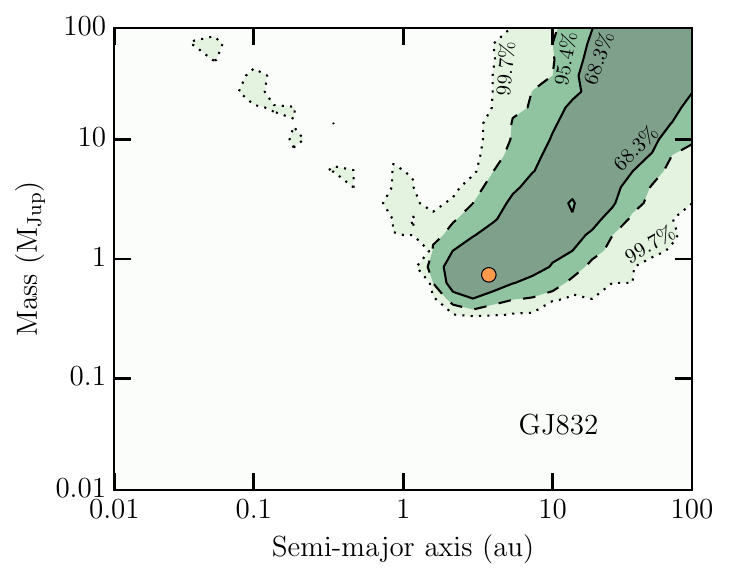} 
    \caption{Constraints derived on the presence of a substellar companion in  the systems GJ\,229 (\textit{left}) and  GJ\,832 (\textit{right}), by using the tool \texttt{Gaia\_PMEX} based on the proper motion anomaly and excess noise from \textit{Gaia} and \textit{Hipparcos}. The known massive companions are represented with the yellow diamond for GJ\,229\,B and the orange circle for GJ\,832\,b, consistent with the \texttt{Gaia\_PMEX} constraints.
    }
    \label{fig:Gaia_PMEX_GJ229_GJ832}
\end{figure*}

In a nutshell, the \texttt{GaiaPMEX} tool models the AEN and PMa on a grid of possible mass and semi-major axis of a hypothetical companion orbiting the targeted star. It will be described in depth in Kiefer et al. (in prep.)
. For each (mass,\,sma)-node of the grid, all parameters, including $e$ the eccentricity, $\omega$ the periastron longitude, $\Omega$ the longitude of ascending node, $\phi$ the phase, $I$ the orbital inclination, $M_\star$ the stellar mass, and $\varpi$ the parallax, are drawn randomly from prior-distributions shown in Table~~\ref{tab:distro}. Astrometric and spacecraft attitude noises~\citep{Lindegren2021}, and other non-astrophysical jitters that participate to the generation of an AEN and a PMa, are estimated for the targeted stars given their $G$-mag, $G_{bp}-G_{rp}$ color, RA and DEC (Kiefer et al. in prep.). Table~\ref{tab:paramPMEX} summarizes the parameters used for running \texttt{GaiaPMEX} for GJ\,832 and GJ\,229 the only two that led to a positive detection (see Section~\ref{sec:results_gaiapmex}). Based on those parameters, the Gaia DR3 observations and the Hipparcos position at epoch J1991.25 of the targeted stars are simulated 1000~times per node, leading to theoretical distributions of AEN and PMa at all nodes. From those distributions, the likelihoods of the observed AEN and PMa are obtained and then further transformed into a posterior probability of the given companion mass and semi-major axis, through a Bayesian framework that will be thoroughly explained in Kiefer et al. (in prep.). Resulting posterior maps for the GJ\,832 and GJ\,229 companions are shown in Fig.~\ref{fig:Gaia_PMEX_GJ229_GJ832}, with the presence of a companion at a significance of $8.8\,\sigma$ and $51\,\sigma$, respectively, both consistent with objects already known: the radial velocity discovered Jovian planet GJ\,832\,b \citep{Bailey2009_GJ832_1planet} and the imaged brown dwarf GJ\,229\,B \citep{Nakajima1995_GJ229_1planet_BD}. The other systems from our survey do not show strong proper motion anomaly and/or excess noise ($< 2\,\sigma$).


\subsection{Detection limits from direct imaging, radial velocity and proper motion anomaly and excess noise \label{sec:results_obs_detlims}}

\begin{figure*}[p]
    \centering
    \includegraphics[width=\linewidth]{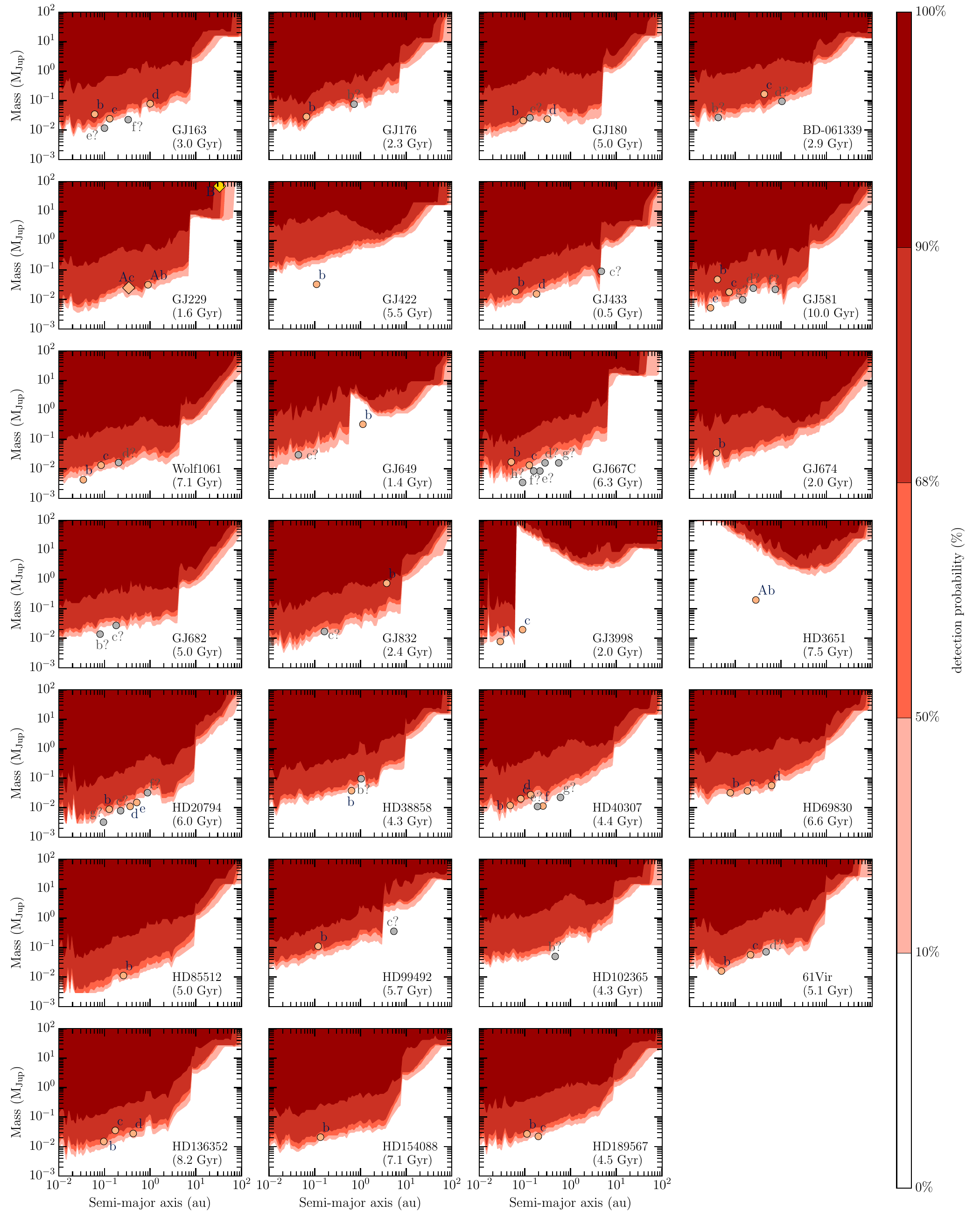}
    \caption{Detection limits from RV (HARPS) + DI (SPHERE-IRDIS and SPHERE-IFS) + astrometric (\textit{Gaia} and \textit{Hipparcos}) data for the whole survey. Already known planets are overplotted in a filled orange (gray) circle if confirmed (unconfirmed). 
    Although some planets seem beyond the reach of RV data, they were all detected with RV data (except the brown dwarf GJ\,229\,B) by using different RV instruments, whereas the limits plotted here only considered HARPS measurements. 
    }
    \label{fig:mosaic_detlims_cond2003}
\end{figure*}

Although we did not detect a new companion bound to its host star, we can set constraints on the presence of outer companions in the different systems.

Regarding our direct imaging observations, we used the evolutionary models \texttt{COND} \citep{Baraffe2003_COND}, to convert our ANDROMEDA detection limit maps given in contrast relative to the host star into masses, both for SPHERE-IRDIS and SPHERE-IFS observations. This conversion depends in particular on the age of the system, as the older the system is, the fainter the companion in the near infrared, and thus the poorer is the mass sensitivity of direct imaging observations. However, mature systems are generally required for RV observations to be sensitive to low-mass planets, as the host stars must be stable enough.

We used our \texttt{MESS3} tool that can combine the available RV data with the high contrast imaging data and with the \textit{Gaia}-\textit{Hipparcos} data to derive the detection limits associated to each target (see Fig.~\ref{fig:mosaic_detlims_cond2003}). The \texttt{MESS3} tool is an extension of \texttt{MESS2}, described in \citet{Lannier2017} based on RV and high contrast imaging data. 
\texttt{MESS3}, as previous \texttt{MESS} versions, generates synthetic planets on a grid of semi-major axes and masses, with orbital parameters drawn from prior distributions, and it tests their detectability based on observations given as inputs. Direct imaging inputs are the detection limit maps expressed in terms of mass via evolutionary models. RV inputs are the RV time series that will be combined in \texttt{MESS3} with the local power analysis (LPA) approach described in \cite{meunier2012} to assess planet detectability. Long-term trends do not count as detections, because only planets with periods lower than two times the time baseline are considered \citep{Lannier2017}.
 As for \textit{Gaia} and  \textit{Hipparcos} inputs, these are detection limit maps  based on the proper motion anomaly and \textit{Gaia} astrometric excess noise.


%

\begin{figure*}[h!]
    \centering  
    \includegraphics[width=0.49\linewidth]{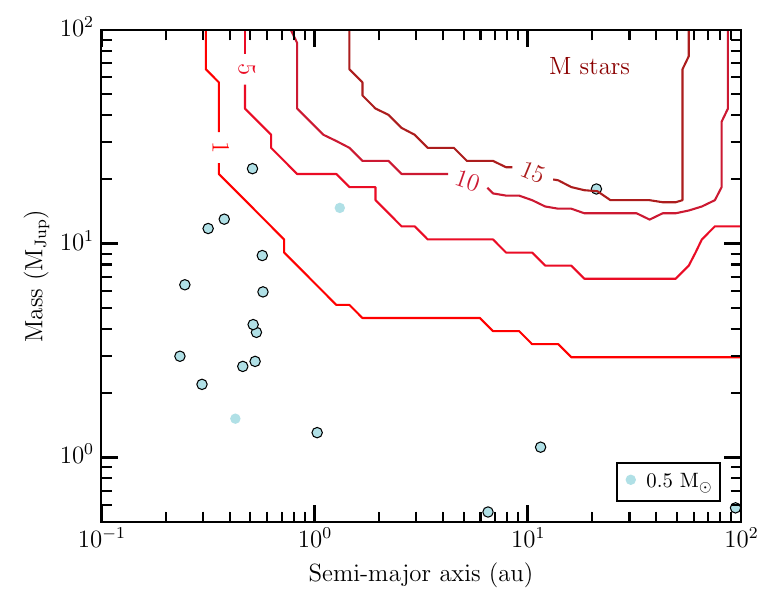}   \hfill
    \includegraphics[width=0.49\linewidth]{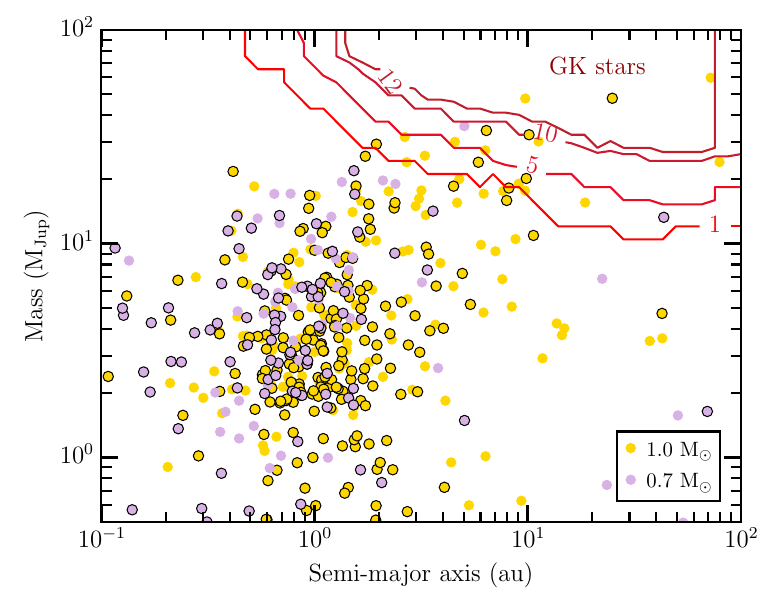}   
    \caption{Detection sensitivity of companions for our direct imaging SPHERE observations of 15 M-stars (\textit{left}) and 12 GK-stars (\textit{right}).  The isocontours give the number of stars from our survey that have the sensitivity to detect companions at a probability of $50\%$.
    The theoretical population of companions formed via the Bern core-accretion models considering $1000$~systems are indicated in circles of colors associated to stellar masses \citep{Burn2021_NGPPS}. These symbols have a black outline if the systems also host close-in ($\leq1~\au$) low-mass ($\leq95~\mearth$) planets.
    }
    \label{fig:Fig4_Vigan_detlim_obs_vs_theory}
\end{figure*}

From Figures~\ref{fig:mosaic_detlims_cond2003} and~\ref{fig:detlim_pedagogical}, we have the following results: RV measurements are sensitive between $0.01$ and $1~\au$ to planets down to a few Earth-masses, between $1$ and $5~\au$ to giant planets down to about $30$ Earth-masses (RV). $\textit{Gaia}$ and $\textit{Hipparcos}$ set constraints mainly between $0.1$ and $30~\au$, down to about one Jupiter mass. Eventually, direct imaging observations provide constraints beyond $1~\au$ down to mostly companions of $20~\mj$ (see Fig.~\ref{fig:Fig4_Vigan_detlim_obs_vs_theory}), and in the most favorable systems,  of a few Jupiter masses beyond a few astronomical units (GJ\,433)  or dozen of astronomical units (e.g.,~GJ\,229, GJ\,649, GJ\,674, GJ\,832), see Fig.~\ref{fig:mosaic_detlims_cond2003_DIonly}.


\section{Discussion \label{sec:discussion}}

We discuss our results using a generic approach in Section~\ref{sec:comparison_obs} by considering studies on other surveys and planet population models. In Section~\ref{sec:discussion_influence}, we use an individual approach by looking into the intrinsic parameters of the planetary systems from our sample.


\subsection{Comparison to theories and surveys \label{sec:comparison_obs}}

Our results do not contradict previous observational studies as we are sensitive to a different companion category, typically beyond $1~\au$ and only massive objects (at least a few Jupiter-masses and mostly above $20~\mj$, see Fig.~\ref{fig:Fig4_Vigan_detlim_obs_vs_theory}). This survey is complementary, as RV measurements to date can only detect planets up to about a dozen of astronomical units, and only suggest trends for any planets further out.

Twelve of our systems are part of the survey from \citet{Bryan2019_RVTransit_study_observational_constraints_super_earth_correlation_super_earth_cold_jupiter}, and five from  \citet{ZhuWu2018_RV_study_observational_constraints_super_earth_correlation_super_earth_cold_jupiter}, who both did not find additional distant giant planets in their analysis of RV measurements for these systems. 
We note that past RV studies have often been biased positively or negatively for systems hosting Jupiter-mass planets. In some RV surveys, the primary goal was to find Earth-like objects in the so-called habitable zone of their host star. Thus, if a massive giant planet was identified in the measurements, the RV-monitoring of the system would be stopped, because it would have been much more difficult to find an Earth-like planet. On the contrary, some other RV surveys would have kept the system and observed it more extensively. In the former case, the systems with small planets discovered so far with RV may be biased toward systems without giant planets easily accessible to RV measurements, and the contrary would happen in the latter case. Taking into account these biases in RV-monitored systems is challenging.




From a theoretical perspective, models of planetary population synthesis give predictions for the architectures of planetary systems based on current knowledge and in the limits of computational capacities. The New Generation Planetary Population Synthesis \citep[NGPPS,][]{Emsenhuber2021_NGPPS_intro}, latest version of the Bern models, aim to be a comprehensive global end-to-end model of planetary system formation and evolution in the core-accretion framework. By coupling N-body interactions, internal structure calculations, and long-term thermodynamical evolution model, with the initial system parameters (e.g., stellar mass and metallicity, disk mass), the Bern models predict the architecture of planetary systems, and in particular physical and observable quantities for the planets at different ages.
%
Based on these models, \citet{Schlecker2021_NGPPS_SE_CJ} suggested a weak positive correlation between the presence of super-Earths and cold-Jupiters around Sun-like stars, but this correlation depends of the range of masses and periods defining each planet category, as also highlighted in \citet{Rosenthal2021_RV_CLS} from their large RV survey. A more robust prediction of their study is the link between the composition of close-in super-Earths and the presence of an outer Jupiter-mass planet.  If a distant giant planet is present, the inner super-Earth is denser, volatile-depleted. The formation of each category is driven by the disk mass, with both super-Earths and cold Jupiters forming preferentially in an intermediate mass disk. 

Additional parameters influence the presence of a given type of planet, such as the stellar metallicity and stellar mass, or can give constraints on the presence of planets, as the multiplicity of planets investigated in \citet{Burn2021_NGPPS}, the existence of debris disks \citep[e.g.,][]{Marino2017_61Vir_ALMA_general_recent_view,Pearce2022_DD_obs} or the stellar duplicity or multiplicity \citep[e.g.,][]{ZuckerMazeh2002_binary_theory_planetary_formation_correlation_mass_period_deficit_inner_massive_planets,Eggenberg2004_binary_theory_planetary_formation_migration_inner_giant_planet,Fontanive2019_binary_systems_fraction}. By using New Generation Planetary Population Synthesis (NGPPS) simulations,
\citet{Burn2021_NGPPS} found that super-solar stellar metallicity ([Fe/H]~$>0.1$) favors the presence of 
sub giant ($30$--$100~\mearth$) and giant planets ($>100~\mearth$), in agreement with previous theoretical \citep{IdaLin2004_theory_metallicity} and observational \citep{Valenti_Fischer2005_61Vir_mass_HD3651_age} studies. 
In addition, the more massive the star, the greater the fraction of systems with planets more massive than $2~\mearth$. A striking result is the absence of formation of (sub) giant planet around $0.1$ and $0.3~\msun$-stars in the NGPPS simulations. Yet, \citet{Morales2019_Science_GJ3512} discovered via RV with CARMENES observations two giant planets of minimal mass $0.46\pm0.02~\mj$ and  $0.20\pm0.02~\mj$ 
around a $0.12~\msun$ star in the system GJ\,3512, and those exoplanets are confirmed by \citet{LopezSantiago2020_GJ3512}. Another giant exoplanet of minimal mass $0.21\pm0.02~\mj$ orbiting a $0.15~\mj$ star was recently discovered \citep{Quirrenbach2022_TZAri_giant_planet_star_0.15Msun}. This highlights the discrepancies still to be solved between theoretical predictions and observations, or the existence of alternative formation mechanisms \citep{Schlecker2022_RV_NGPPS_core_accretion_Mdwarfs_RV2022}.

By comparing the detection probabilities from our direct imaging observations obtained with \texttt{MESS3} (Fig.~\ref{fig:Fig4_Vigan_detlim_obs_vs_theory}) to massive companions obtained in systems hosting close-in low-mass planets from Bern population models, we see that only a few of those companions are reachable by our observations. In more detail, we use the grids from \citet{Burn2021_NGPPS} synthesized for a population of $1\,000$ planetary systems around a host star of  $1.0~\msun$, $0.7~\msun$, $0.5~\msun$, and $0.3~\msun$, to match our sample. The Bern grids directly give the mass and semi-major axis of the planets of the $1\,000$  systems at different ages. 
We consider the surviving planets at the age of $5~\gyr$ that are displayed in circles of different colors in Fig.~\ref{fig:Fig4_Vigan_detlim_obs_vs_theory}, and we highlight the ones in systems hosting as well close-in ($\leq1~\au$) low-mass ($\leq95~\mearth$) planets by circles with black outline. 
The more massive the star is, the greater the number of massive companions orbiting within a few astronomical units, as underlined in \citet{Burn2021_NGPPS}. We can see that the best sensitivities in our survey are reached for M stars, i.e., low-mass stars (see Fig.~\ref{fig:mosaic_detlims_cond2003_DIonly}, e.g., GJ\,433, GJ\,674).

We computed the number of planets that these model predict should have been detected given our detection performance from the direct imaging SPHERE observations. For each system, we used the grid with the closest stellar mass, and at the corresponding age.   
This gives a probability of detecting at least one planet at a confidence level of $50\%$ of $0.8\%$ for the most favorable system (the solar-mass star HD\,154088) from our sample, while the detection probability is $0\%$ for the least favorable systems, corresponding mainly to low-mass stars, because Bern models do not form giant planets for such stars \citep{Burn2021_NGPPS}.
Overall, this results in a probability of $3.7\%$ to detect at least one massive distant planet in one of our systems at a confidence level of $50\%$.
A more faithful comparison to our sample would be to look at theoretical predictions only for the planetary systems with close-in low-mass planets ($\lesssim1~\au$, $\lesssim95~\mearth$). In this case, the detection performance for each system increases at best to $0.9\%$ at a confidence level of $50\%$. On the whole, this results in a probability of $4.5\%$ to detect at least one massive distant planet in one of our systems at a confidence level of $50\%$. The probabilities slightly increase with respect to planetary systems for which we do not assume that already close-in low-mass planet(s) exist. This is due  to the weak positive correlation of inner low-mass planets and outer giant planets in the Bern models \citep{Schlecker2021_NGPPS_SE_CJ}.
Overall, the non-detection of a distant massive giant planet in our high-contrast imaging observations is compatible with predictions based on core-accretion models of planet formation. Nonetheless, we note that the formation of large planets and brown dwarfs could be enhanced by other formation pathways, such as gravitational disk instability \citep[e.g.,][]{Boss1997_GI} or gravo-turbulent fragmentation \citep[e.g.,][]{PadoanNordlund2004_gravoturbulent_fragmentation_clouds}, not discussed in this work.

Based on the sensitivity of our SPHERE observations to mostly brown dwarfs in the mature systems of our sample, and the biases of RV surveys disregarding systems with massive companions when looking for Earth-like planets, looking into the correlation between the presence of close-in small planets and outer massive companions is highly challenging. A detailed statistical study taking into account the observational biases, as well as  the formation and evolution of planets and brown dwarfs from different pathways applied to different stellar types (Solar-like to M stars), from population or parametric models \citep[e.g.,][]{Vigan2021_shine_demographics} is beyond the scope of this paper, and could be addressed in a future work.

\subsection{Constraints from intrinsic parameters of the planetary systems  \label{sec:discussion_influence}}

In this section, we peer into the architecture of the planetary systems of our sample to get an insight on where we could expect additional companions. We first focus on the stellar parameters such as the stellar multiplicity, metallicity and mass, then on the presence of a debris disk, eventually on the multiplicity and eccentricity of the known planet(s). 

From a larger point of view, we note that the stellar neighborhood can play a role to shape planetary systems, for instance with flyby events \citep{Kalas1996_flyby} consisting in an encounter between a given system and a perturber such as a star. In our sample, \citet{Bertini2023_flybys_DD_HD20794_HD38858_HD69830} reported that the systems HD\,20794, HD\,38858, and HD\,69830 have experienced (or will) one to six flybys. The consequence of flyby events can be similar to the presence of a binary star and is described below.

\subsubsection{Influence of the stellar duplicity or multiplicity \label{sec:discussion_influence_binary}}

Since about $50\%$ of Sun-like stars are part of binaries or higher-order multiple system, binaries cannot be neglected to have a complete picture of planet formation and evolution \citep[e.g.,][]{Heintz1969_binary_stats,AbtLevy1976_multiplicity_stats,DuquennoyMayor1991_mutiplicity_influence,Raghavan2010_multiplicity_stats,Tokovinin2014_multiplicity_stats}.

One of the key parameters is the separation of the two binary-components. Close binaries ($\lesssim50$--$100~\au$) inhibit planet formation \citep[e.g.,][]{Kraus2012_tight_binaries_inhibit_planet_formation,Kraus2016_tight_binaries_inhibit_planet_formation,Wang2014_tight_binaries_inhibit_planetary_formation,Bonavita2007_binaries_frequency_planets,Bonavita2020_binaries_frequency_planets}. In the Milky Way,  \citet{Kraus2016_tight_binaries_inhibit_planet_formation} reported that $20\%$ of all the solar-type stars cannot host planetary systems owing to the influence of a binary companion. 
A relatively wide binary (about $100$--$1000~\au$) may impact the evolution of the planet via the high-eccentricity migration, as it may facilitate the planet's migration via dynamical interactions, by increasing its eccentricity until it undergoes circularization from the primary star (as could do flybys). This migration may be caused by Kozai-Lidov process \citep[e.g.,][]{FabryckyTremain2007_HEM_Kozai,Naoz2012binary_HEM_Kozai,Dong2014_HEM_process}, but not only \citep{Ngo2016_binary_HEM_noonlyKozai}. 
On the contrary, very wide binaries ($\gtrsim1000~\au$) may not have consequences on planetary formation and evolution, as exoplanets in these systems have the same properties as the ones around single stars \citep{FontaniveBardalezGagliuffi2021_compilation_binary_systems,Bonavita2007_binaries_frequency_planets,Bonavita2020_binaries_frequency_planets}. 

In this context, the outer wide ($200$--$1000~\au$) binary dwarf or stellar companion in the systems HD\,3651, HD\,99492, and HD\,102365 from our sample should not prevent the formation of planets, nor have a significant impact on the formation of giant exoplanets beyond~$\gtrsim 0.5~\au$ \citep{FontaniveBardalezGagliuffi2021_compilation_binary_systems}, but could influence their evolution. 
In particular, the binary dwarf HD\,3651\,B could explain the high eccentricity \citep[$0.63\pm0.04$,][]{Fischer2003_HD3651_1planet_subSaturn_discovery} of the close-in giant planet HD\,3651\,A\,b if induced via the high-eccentricity migration process mentioned previously. 
On the other hand, in spite of the proximity of the binary companion GJ\,229\,B to its host star ($33.3^{+0.4}_{-0.3}~\au$), the small mass ratio ($0.1$, see dedicated paragraph in Appendix \ref{app:biblio_indiv_systems}) relative to GJ\,229\,A  could explain the formation of small inner planets in the system, especially if the brown dwarf forming during the disk phase was on a circular orbit and inclined with respect to the disk \citep{Cadman2022_binary_simues}.

Regarding GJ\,667\,AB+C, the hierarchical triple stellar system in our sample, this has one central binary AB (semi-major axis of $13~\au$, total mass of $ 1.27~\msun$), and a third star C of $0.35~\msun$, located at a projected distance of $235~\au$ \citep{Delfosse2013_GJ667_GJ433_RV_charac_planets,Soderhjelm1999_GJ667}. The close binary GJ\,667\,AB can be treated dynamically as a  single object and thus previous conclusions for binary systems hold for this system as well. Planet formation is not inhibited in triple systems, nor in GJ\,667 or in others \citep[e.g.,][]{Mugrauer2007a_triple_system,Mugrauer2007b_triple_system}.

\subsubsection{Influence of the stellar metallicity and mass \label{sec:discussion_influence_metallicity}}

It is now well established observationally that the occurrence rate of a giant planet is strongly dependent on the stellar metallicity \citep[e.g.,][]{FischerValenti2003_obs_metallicity,Santos2004_metallicity}. The host star metallicity also increases with increasing planet mass or radius   \citep[e.g.,][]{Narang2018_metallicity_correlation_kepler}. These trends
are often interpreted as a direct signature of the core accretion \citep[][]{Pollack1996_CA}
formation paradigm explained by a more massive and frequent core assembly at higher metallicities \citep[e.g.,][]{IdaLin2004_theory_metallicity,Bitsch2015_metallicity_pebble_accretion,NduguBitsch2018_pebble_accretion_cluster_environment_metallicity,Emsenhuber2021_NGPPS_paperII,Schlecker2021_NGPPS_SE_CJ}. This correlation however reverses for planet masses greater than $4$--$5~\mj$, as the averaged host star metallicity decreases with increasing planetary mass \citep{Santos2017_correlation_metallicity_giant_planet,Narang2018_metallicity_correlation_kepler,Schlaufman2018_metallicity_CA_GI,Swastik2021_metallicity_DIP}. They suggest that this turn-over  traces different formation pathways for super-Jupiter and brown dwarf companions. These companions with masses larger than $4$--$5~\mj$  would be formed by gravitational instability or even gravo-turbulent fragmentation. In our context, detection limits are mainly limited to these massive companions (see Fig.~\ref{fig:Fig4_Vigan_detlim_obs_vs_theory}) and the metallicity of the stellar host probably has little impact on our chances of maximizing the discovery rate. 

From radial velocity and transit surveys, it is also known that the giant planet occurrence rate increases approximately linearly with stellar mass from low-mass stars up to intermediate-mass stars \citep{Johnson2010_planet_formation_depending_stellar_mass,Ghezzi2018_planet_formation_depending_stellar_mass_metallicity}. The trend is also predicted by core accretion theory due to a more massive initial reservoir of solids available to their formation \citep{KennedyKenyon2008_planet_formation_giant,Liu2019_peeble,Schlecker2022_RV_NGPPS_core_accretion_Mdwarfs_RV2022}. For the low-mass stars in our sample, however, we expect a very low occurrence, almost null, of giant planets formed by core accretion (see Fig.~\ref{fig:Fig4_Vigan_detlim_obs_vs_theory}). For solar-type stars, despite an increasing rate of giant planets formed by core accretion, our sensitivities degrade significantly and remain mainly sensitive to the population of super-Jupiters or brown dwarfs,  here again preferably formed by gravitational instability or gravo-turbulent fragmentation and
for which a universal distribution of the companion mass ratio, which is roughly flat \citep[e.g.,][]{ReggianiMeyer2013}, is expected.

\subsubsection{Constraints from a debris disk \label{sec:comparison_theory_debris_disk}}

The presence of a debris belt is helpful to constrain the location and mass of putative planets in the system. Those planets may sculpt the belts, and/or stir them via secular interactions, in particular their edges.

In our sample, seven systems are known to harbour a debris disk: GJ\,433, GJ\,581, GJ\,649, HD\,20794, HD\,38858, HD\,69830, 61\,Vir.
The system HD\,69830 is a specific case 
because there is no significant level of cool dust detected \citep[e.g.,][]{Marshall2014_HD69830_DD_Herschel}, unlike the others. The existence of the hot belt in the mature HD\,69830 system  raised particular interests in the literature (see Appendix \ref{app:biblio_indiv_systems}), since it cannot be replenished, and may be explained by recent collisions between planetesimals \citep{Wyatt2007_HD69830_DD,Wyatt2010_HD69830_DD,HengTremaine2010_long_lived_planetesimal_discs}.  
A comparison of the four debris disks around GK stars (HD\,69830, 61\,Vir, HD\,20794, and HD\,38858) from our sample can be found in \citet{Kennedy2015_HD38858_HD20794_61Vir_HD69380_submm_debris_belt_kuiper_analog}, while the observations of the three debris disks around M stars (GJ\,433, GJ\,581, and GJ\,649) are presented in \citet{Kennedy2018_GJ433_GJ649_DD}.

\citet{Pearce2022_DD_obs} provided constraints on the location and mass required for planets in four of our studied systems (GJ\,581, GJ\,649, HD\,69830, and $61$\,Vir), by assuming the belts are sculpted or stirred by one or several planets. The outermost planet HD\,69830\,c ($a=0.63~\au$, $m\sin i=18~\mearth$) could have stirred and sculpted its disk, while  GJ\,649\,b $a=1.1~\au$, $m\sin i=0.33~\mj$) could have stirred its disk, and maybe sculpted it. Although, the outermost planets 61\,Vir\,c (yet unconfirmed: $a=0.49~\au$, $m\sin i=23~\mearth$), and GJ\,581\,d ($a=0.074~\au$, $m\sin i=5.7~\mearth$) have in principle a minimal mass sufficient to sculpt their debris disk, they are located too far and inside of the belts for this sculpting to have been effective over the system ages. As a matter of fact, \citet{Marino2017_61Vir_ALMA_general_recent_view} showed that a $>10~\mearth$ stirring planet in 61\,Vir should be located between $10$ and $20~\au$. 
In short, no giant planets are required to explain the single/multi-belt architecture in those four systems, even though such planets are not ruled out.


\subsubsection{Influence of the planet multiplicity or eccentricity \label{sec:discussion_influence_ecc}}

From theoretical predictions, the multiplicity of small inner planets may be reduced by the presence of outer giant planet(s) in a given system \citep
{Schlecker2021_NGPPS_SE_CJ,Bitsch_Izidoro2023}, also reported in a combined theoretical and observational analysis of the \textit{Kepler} results from \citet{Hansen2017_kepler_multiplicity_reduced_CJ}, or tentatively from observations by \citet{ZhuWu2018_RV_study_observational_constraints_super_earth_correlation_super_earth_cold_jupiter}. 

Figure \ref{fig:hist_multiplicity_Sub_Saturns} represents the distribution of the multiplicity of the known low-mass planets in our sample. 
All the systems have less than $3$ low-mass planets confirmed
, and a few up to $6$ planets but with some controversy in the literature (see our review in Appendix \ref{app:biblio_indiv_systems}).
In particular, the systems GJ\,176, GJ\,422, GJ\,674, HD\,154088 only host one close-in small mass planet so far, while in BD-061339 or HD\,99492 a second companion is suggested but debated. Hence, this low-multiplicity may hint toward the presence of a yet to be discovered distant giant planet. 
.

\begin{figure}[h!]
    \centering
    \includegraphics[width=\linewidth]{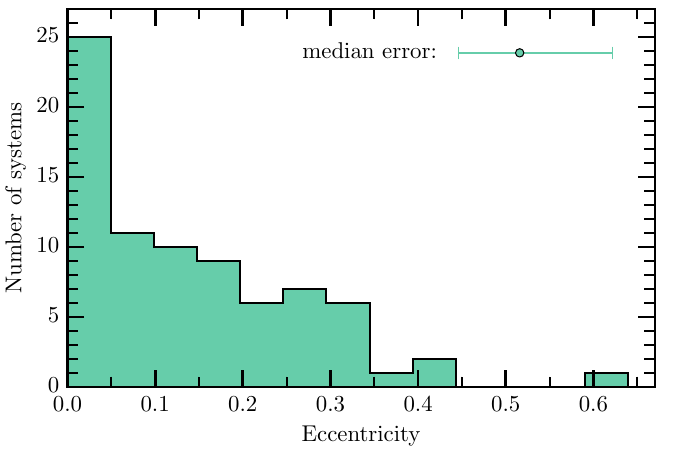}
    \caption{Distribution of the eccentricity of the close-in ($\lesssim1~\au$) low-mass ($\lesssim95~\mearth$) planets known in our sample. We indicate the median error in the upper right corner.   
    }
    \label{fig:hist_eccentricity}
\end{figure}

Furthermore, the eccentricity of known planets can help to constrain the presence of additional planets, as an eccentric orbit may result from scattering with another planet \citep[e.g.,][]{Raymond2009_theory_obs_eccentricities_exoplanets_scattering}. Thus, an eccentric close-in small planet may hint at the presence of an outer more massive companion \citep{Bitsch2020_theory_peeble_eccentricity_planets,Bitsch_Izidoro2023}. In our sample, most planets have low to moderate eccentricities (see Fig.
~\ref{fig:hist_eccentricity}). The more eccentric confirmed low-mass planets are GJ\,163\,d ($e=0.41\pm0.07$,  $m\sin i=25~\mearth$) and GJ\,667\,C\,c ($e=0.34\pm0.10$,  $m\sin i=4.25~\mearth$). Even though both planets are part of multi-planetary systems, since GJ\,163 hosts between $3$ and $5$~planets (of minimal masses $\leq 25~\mearth$), while GJ\,667\,C between $2$ and $7$ (of minimal masses $\leq 7~\mearth$), see Appendix~\ref{app:biblio_indiv_systems} and Fig.~\ref{fig:overview_system_linear}
, scattering events only between these small planets cannot explain the large observed eccentricity of GJ\,163\,d and GJ\,667\,C\,c. This would require the presence of an outer giant planet \citep{Bitsch_Izidoro2023}.

Conversely, if eccentric giant planets are not sufficiently dynamically separated from inner systems of low-mass planets, they may disrupt the latter. Both planetesimal and pebble accretion based formation models robustly predict this disruption, which becomes more likely the more eccentric, massive, and closer in the giant planet is \citep{Schlecker2021_NGPPS_SE_CJ,Bitsch2020_theory_peeble_eccentricity_planets}.
Thus, we do not expect inner low-mass planets in the system HD\,3651 ($e=0.64\pm0.04$, $m\sin i=0.20~\mj$, $a=0.28~\au$) even though the star has a slightly super-solar metallicity  ($0.08\pm0.06$) and is relatively massive ($0.86\pm0.06~\msun$). In addition, this should also be unlikely in GJ\,649 ($e=0.30\pm0.08$, $m\sin i=0.33~\mj$, $a=1.14~\au$) or  BD-061339 ($e=0.31\pm0.11$, $m\sin i=0.17~\mj$, $a=0.44~\au$).

However, we note that there are observational biases in the estimation of planet eccentricities from RV measurements.  \citet{Zakamska2011_bias_eccentricity_planets_RV} reported that the eccentricity of single-planets may be overestimated, and that the conclusions of their analysis may be applied to multi-planet systems too. In addition, the  RV signal of two circular planets can be confused with the RV signal of one eccentric planet
\citep[e.g.,][]{Wittenmeyer2007_HD3651_other_2planets,Wittenmyer2013_GJ649_2planets_cunconfirmed,Anglada-Escude2010_eccentricity_RV_signal_confusion}. These biases depend on the quality of the RV measurements, their number and sampling.
In any case, measuring the eccentricity at a given time period only gives partial information as the eccentricities of planets in multi-planet systems can oscillate through time owing to secular perturbations \citep[e.g.,][]{Zakamska2011_bias_eccentricity_planets_RV,ReadWyatt2016_HD38858_61Vir_secular_interactions}.

Regarding the systems in our sample, an in-depth study of secular perturbations can help to provide constraints on the presence of additional planets. For instance, \citet{ReadWyatt2016_HD38858_61Vir_secular_interactions} showed that for the single planetary system HD~38858, the presence of an outer eccentric planet~c is unlikely as secular perturbations would have excited the planet~b which is known to have a circular orbit. If such an additional planet exists, it could point out to the presence of a third inner planet~d, close to the HD~38858~b, that would stabilize it. Regarding two-planet systems, \citet{ReadWyatt2016_HD38858_61Vir_secular_interactions} highlighted that it is complex to constrain deeply the presence of an outer eccentric planet based on secular interactions only, such as for the system 61~Vir, and that instead, the structure of an outer debris disk is more restrictive.

\section{Conclusion and perspectives \label{sec:conclusion}}

We carried out a survey on $27$ very nearby systems. Although all these systems are suspected to host at least one low (minimal) mass  ($\lesssim95~\mearth$) planet orbiting on a small orbit ($<1~\au$), they depict diverse planetary architectures: seven harbor a debris disk, five are part of a binary or triple hierarchical stellar system, and two host a brown dwarf companion beyond 10\,au.
In this work, we homogeneously derived target properties (age, stellar mass, stellar metallicity), processed data reduction and analyzed them. We also provide a complete review of the literature on each system in Appendix~\ref{app:biblio_indiv_systems}.

Our main findings are as follows:



\begin{enumerate}
    \item In the $47$ epochs of observation in the near infrared with VLT/SPHERE-IRDIS and IFS  published for the first time in this paper, we detected $337$ point sources. We did not discover new bound companions even though $12$ remain pending, as information is missing to conclude. We detected the already known brown dwarf GJ~229~B. We classified two companion candidates as instrumental artifacts, whereas the $328$ others are classed as background stars based on analysis of their proper motion or location in a Color Magnitude Diagram.
    \item By combining direct imaging (SPHERE), radial velocity (HARPS), and proper motion coupled to excess noise (\textit{Gaia} and \textit{Hipparcos}) observations, the existence of planets in each system is constrained down to a few Earth-masses between $0.01$ and $1~\au$, down to about $30~\mearth$ between $1$ and $10~\au$, and down to about a dozen of Jupiter-masses beyond $10~\au$. Regarding the high-contrast imaging observations alone, we are sensitive down to $3$ to $30~\mj$ beyond $0.5$ or $2~\au$ depending on the system.
    \item Given these detection limits, our non-detection of new giant planets is consistent with state-of-the-art planet formation models following the core accretion paradigm \citep{Burn2021_NGPPS}. 
    \item If additional planets yet undiscovered exist in some systems of our sample, by examining the properties and architecture of each system, and previous work from the literature, 
    we identified several particularly promising targets for follow-up. We could preferentially expect additional planets in:
    \begin{itemize}
        \item GJ\,163 and GJ\,667\,C: both systems have eccentric low-mass planets ($0.41\pm0.07$ and $0.34\pm0.10$, respectively). 
        The multiplicity of GJ\,667\,C should not have prevented the formation of massive planets as the two components GJ\,667\,AB and  GJ\,667\,C are distant enough from each other ($\gtrsim 235~\au$).
        \item 61\,Vir and  GJ\,581: by assuming the inner edge of the debris disk is sculpted by a planet, this would require a planet of $\gtrsim19~\mearth$ if located at $30\pm10~\au$ (61\,Vir) or $\gtrsim3.4~\mearth$ if located at $14\pm6~\au$ (GJ\,581). It would need to be more massive if the planet is closer-in \citep{Pearce2022_DD_obs}.
    \end{itemize}
\end{enumerate}

Very nearby planetary systems offer unique possibilities to couple direct imaging, radial velocity and astrometry from \textit{Gaia} and \textit{Hipparcos}. This is true not only across a system, but also for the coupled characterization of a single exoplanet. In the coming years, \textit{Gaia} would hint for giant exoplanets with orbital periods up to about $10~\yrs$, and direct imaging and interferometric instruments, such as for instance the future upgrades VLT/SPHERE+  \citep{Boccaletti2020_SPHERE+} and VLTI/GRAVITY+ \citep{Eisenhauer2019_gravity+}. They will be able to explore the population of massive giant planets down to the snowline expanding the current overlap between these techniques (but again mainly for the youngest systems). This is of utmost importance as different detection techniques provide diverse physical information on the exoplanet, and particularly a measurement of the dynamical mass, the luminosity and the atmospheric properties, which can help to constrain theoretical models of formation and evolution. 

With this work, we have not answered our initial question set in the introduction concerning the correlation between close-in low-mass planets and giant outer planets.
We however showed that, with the current state-of-the-art instrumentation, the high-contrast imaging detection limits are mainly sensitive to companions down to $20~\mj$ around mature stars. The question remains open from a direct imaging perspective, though radial velocity and transit studies seem to suggest a tentative positive correlation, that would require larger samples to be fully established.  For the decades to come, the \textit{James Webb Space Telescope} with the MIRI instrument or the forthcoming high-contrast imagers such as METIS (2030+) or PCS (2035+) on the ELT will pave the way for the imaging in thermal and/or reflected light, and the demographics characterization of more mature and smaller planets around very nearby stars.

\begin{acknowledgements}
C.D. and J.M. thank Lucile Mignon, Xavier Bonfils and Xavier Delfosse for constructive discussions on radial velocity data. C.D., J.M. and G.C. thank Clémence Fontanive for discussions on exoplanet demographics and multiple systems. C.D. and J.M. thank Pierre Kervella for his help  on \textit{Gaia} data.  Last, C.D. is grateful to Mickaël Bonnefoy, Nadège Meunier, Nicolas Cuello, Melissa Hobson, and Aurélia Leclerc for fruitful discussions on this paper.
This work has made use of the SPHERE Data Centre, jointly operated by OSUG/IPAG (Grenoble), PYTHEAS/LAM/CESAM (Marseille), OCA/Lagrange (Nice), Observatoire de Paris/LESIA (Paris), and Observatoire de Lyon. This work has made use of data from the European Space Agency (ESA) space mission Gaia (\url{https://www.cosmos.esa.int/gaia}), processed by the Gaia Data Processing and Analysis Consortium (DPAC).
This work has been supported by a grant from Labex OSUG (Investissements d’avenir – ANR10 LABX56). This project has received funding from the European Research Council (ERC) under the European Union's Horizon 2020 research and innovation programme (COBREX; grant agreement n° 885593)
C.D., T.H., R.B. acknowledge support from the European Research Council under the European Union’s Horizon 2020 research and innovation program under grant agreement No. 832428-Origins. 
G.M.K. is supported by the Royal Society as a Royal Society University Research Fellow.
The results reported herein benefited from collaborations and/or information exchange within the program “Alien Earths”
(supported by the National Aeronautics and Space Administration under agreement No. 80NSSC21K0593) for NASA’s Nexus for Exoplanet System Science
(NExSS) research coordination network sponsored by NASA’s Science Mission Directorate.
B.B., thanks the European Research Council (ERC Starting Grant 757448-PAMDORA) for their financial support.

\end{acknowledgements}

\bibliography{biblio.bib}

\newpage

\begin{appendix}

\twocolumn

\section{State of the art of each system \label{app:biblio_indiv_systems}}
In this appendix, we briefly describe each system of the survey, and state-of-the-art studies on it.

\paragraph{GJ\,163}

The system GJ\,163 is located at $15.135\pm0.004~\pc$ \citep{Gaia2021_EDR3}.
The host star is a M3.5 dwarf \citep{Hawley1996_GJ163_GJ422_GJ433_GJ667_GJ674_GJ682_GJ832_spectral_types_HD102365_binary}. We estimated the mass of the star to $0.46\pm0.05~\msun$ and its age via gyrochronology to $3\pm7~\gyr$. Both are consistent with the previous estimates from \citet{Bonfils2013_GJ163_5planets} and \citet{Tuomi2013_GJ163_4planets}.  The question of GJ\,163's membership of the old disk or halo populations has been discussed in the literature \citep[e.g.,][]{Leggett1992_GJ163_GJ649_GJ674_colors_membership_pop,Bonfils2013_GJ163_5planets,Tuomi2013_GJ163_4planets} based on its relatively high UVW galactic velocities, and is still debated. GJ\,163 could be either a rather young star accelerated to the typical velocity of an old dynamical population, or belong to the metal-rich tail of an old population of stars. \\
\citet{Bonfils2013_GJ163_5planets} discovered three planets, GJ\,163~b, GJ\,163~c and GJ\,163~d with minimum masses of $11~\mearth$, $7.7~\mearth$ and $25~\mearth$ and semi-major axes of $0.061~\au$, $0.13~\au$, and $1.0~\au$, respectively. Two additional planets may be present, GJ\,163~e ($11~\mearth$, $0.10~\au$) and GJ\,163~f ($7.2~\mearth$, $1.0~\au$). \citet{Tuomi2013_GJ163_4planets} confirmed the presence of three  planets~b,~c and~d, and have tentative evidence for the planet~f. Thus, in our work, we consider the planets~b,~c and~d as confirmed and the planets~e and~f as unconfirmed.


\paragraph{GJ\,176}
The system GJ\,176 is located at $9.485\pm0.002~\pc$ \citep{Gaia2021_EDR3}. Its host star is a M2.5 dwarf \citep{Joy1974_GJ176_GJ229_GJ581_GJ649_BD06_Wolf_spectral_types}.  We estimated the stellar mass to $0.52\pm0.05~\msun$,  consistent with the previous estimate from \citet{AstudilloDefru2017_GJ176_GJ581_GJ674_activity_mass}. However, our age estimate based on gyrochronology gives a younger age of $2.3\pm0.6~\gyr$ compared with $8.8^{+2.5}_{-2.8}~\gyr$ from \citet{Veyette2018_GJ176_GJ581_GJ649_Wolf1061_age_chemo_kinematic}. \citet{vonBraun2014_61Vir_GJ176_GJ649_stellar_radius} derived a stellar radius of $0.45\pm0.02~\rsun$. \citet{Peacock2019_GJ176_GJ832_EUV} investigated the extreme ultraviolet (EUV) emission of the star, that plays a key role in the stability and photochemistry of exoplanet atmospheres.  \\
%
\citet{Endl2008_GJ176_1planet_Neptune_wrong} discovered one Neptune-like planet with a minimum mass of $24.1\pm3.1~\mearth$ at a period of  $10.2~\days$. However, this planet has been called into question by \citet{Butler2009_GJ176_rv_nondetection_Endl} and \citet{Forveille2009_GJ176_1planet_superearth}. \citet{Forveille2009_GJ176_1planet_superearth} found the planet~b to be rather a super-Earth-like planet ($8.4~\mearth$, $8.8~\days$) by using HARPS RV measurements. 
\citet{Robertson2015_GJ176_stellar_activity_1planet} and \citet{Trifonov2018_GJ176_GJ581_rv_1planet} confirmed the existence of the planet~b by studying the stellar activity for the former or  by adding CARMENES measurements for the latter. They reported a minimal mass of $9.06^{+1.54}_{-0.70}~\mearth$ at a semi-major axis of $0.066\pm0.001~\au$ (period of $8.8~\days$).

\paragraph{GJ~180}

The system GJ~180 is located at $11.949\pm0.002~\pc$ \citep{Gaia2021_EDR3}.
The host star is a M2 dwarf \citep{Gaidos2014_GJ180_GJ3998_Wolf1061_spectral_type}, and has a stellar mass estimated in our work of $0.44\pm0.05~\msun$ which is consistent with the previous estimate from \citet{Zechmeister2009_GJ180_GJ229_GJ422_GJ433_GJ682_Mdwarfs}.  \\ 
The system hosts two or three known planets discovered by radial velocity measurements. The planets~b and~c were found by \citet{Tuomi2014_GJ180_GJ229_GJ422_GJ433_GJ682}, while \citet{Feng2020_GJ180_GJ229_GJ422_GJ433} confirmed the existence of b ($m \sin i=6.75^{+1.37}_{-1.78}~\mearth$, $a = 0.092\pm0.008~\au$, $P=17.132\pm0.008~\days$), found no evidence significant enough for c ($8.3_{-5.3}^{+3.5}~\mearth$, $0.129\pm0.017~\au$, $24.33\pm0.07~\days$ from \citet{Tuomi2014_GJ180_GJ229_GJ422_GJ433_GJ682}), and discovered an additional planet~d ($7.49^{+2.66}_{-2.33}~\mearth$, $0.31\pm0.03~\au$, $106.34\pm0.03~\days$).

\paragraph{BD-06 1339 (GJ~221)}
The system BD-06~1339 (also known as GJ~221) is the furthest from the Sun in our sample, at $20.305\pm0.008~\pc$ \citep{Gaia2021_EDR3}. We estimated an age of $2.9\pm1.2~\gyr$ by using chromospheric activity, a stellar mass of $0.65\pm0.05~\msun$, and a sub-solar stellar metallicity [Fe/H] of $-0.20\pm0.10$, all in agreement with \citet{LoCurto2013_BD061339_2planets} and/or \citet{Arriagada2013_BD061339_2planets}.
The star is a K7 dwarf \citep{Stephenson1986_BD061339_spectral_type}.  \\
It hosts one confirmed planet named~c (minimal mass of $53~\mearth$ at a semi-major axis of $0.44~\au$, period of $126~\days$), and putatively two additional planets,~b ($8.5~\mearth$, $0.04~\au$, $3.9~\days$) and~d ($30~\mearth$, $1.1~\au$, $1496~\days$). The planets~b and~c were found simultaneously by \citet{Arriagada2013_BD061339_2planets,LoCurto2013_BD061339_2planets}. \citet{Tuomi2014_BD061339_planet_d} indicated the presence of the third planet~d. However, this planet~d is ignored in the later paper from \citet{Simpson2022_BD061339_stellar_activity_false_alarm}, who recently reported that the planet~b is likely to be a false alarm caused by stellar activity. The only confirmed planet in this system is the one known as BD-06~1339~c which has a mass between Neptune and Saturn. 

\paragraph{GJ~229}

The system GJ~229 is located at $5.7612\pm0.0004~\pc$ \citep{Gaia2021_EDR3}. Its host star GJ~229 A is a M1 dwarf \citep{Henry2002_GJ229A_Wolf1061_spectral_type}, with a mass estimated in this work of $0.57\pm0.05~\msun$ consistent with the previous estimate from \citet{Zechmeister2009_GJ180_GJ229_GJ422_GJ433_GJ682_Mdwarfs}. By using gyrochronology, we found an age of $1.6\pm0.4~\gyr$, at $2\,\sigma$ consistent with \citet{Brandt2020_GJ229_BD} who suggested an age of $2.6\pm0.5~\gyr$ from stellar activity measurements.\\
\citet{Nakajima1995_GJ229_1planet_BD} imaged GJ~229 B at a distance of $44.3~\au$ from the star GJ~229~A. From their mass estimation of $20$--$50~\mj$ via cooling models of brown dwarfs, GJ~229~B is the first T-type brown dwarf with clear signs of methane \citep[T6 according to][]{Geballe2002_GJ229B_GJ667C_spectral_type}. 

By combining imaging, RV, and  \textit{Gaia} and \textit{Hipparcos} data, different dynamical masses are found for GJ~229~B. On one hand, \citet{Brandt2020_GJ229_BD} measured a dynamical mass of $70.4\pm4.8~\mj$, updated to $71.4\pm0.6~\mj$ by \citet{Brandt2021_GJ229B_dynamic_mass}. 
This value is significantly higher than the previous estimate and the minimal mass found by \citet{Feng2020_GJ180_GJ229_GJ422_GJ433} of $1.62~\mj$, but consistent with the inclination of $7.7^{+7.6}_{-4.4}~\degr$ found by orbital fitting \citep{Brandt2021_GJ229B_dynamic_mass}.
Nonetheless, their value is inconsistent with the value determined by using cooling models \citep[$64.8\pm0.1~\mj$,][]{Brandt2021_GJ229B_dynamic_mass}.
On the other hand, \citet{Feng2022_GJ229B} recently derived a dynamical mass of  $60.4\pm2.4~\mj$, consistent only at $4\,\sigma$ with the value derived from \citet{Brandt2021_GJ229B_dynamic_mass}. 
The brown dwarf is a bit closer than originally thought, as at~$33~\au$ \citep[e.g.,][]{Brandt2020_GJ229_BD,Brandt2021_GJ229B_dynamic_mass}.

The system hosts as well two additional planets, GJ~229~A~b first discovered by \citet{Tuomi2014_GJ180_GJ229_GJ422_GJ433_GJ682} and confirmed later by \citet{Feng2020_GJ180_GJ229_GJ422_GJ433}, who found as well the planet GJ~229~A~c. Respectively, their minimal masses are $10.0^{+3.4}_{-6.1}~\mearth$ and $7.9^{+2.4}_{-3.4}~\mearth$, while their semi-major axes are $0.90\pm0.07~\au$ and $0.34\pm0.03~\au$.
If their orbits are coplanar with the orbit of  GJ~229~B, hence face-on \citep[inclination of  $7.7^{+7.6}_{-4.4}~\degr$,][]{Brandt2021_GJ229B_dynamic_mass}, their masses would be of $73^{+67}_{-16}~\mearth$ and $57^{+17}_{-13}~\mearth$, respectively.


\paragraph{GJ~422}

The system GJ~422 is located at $12.676\pm0.005~\pc$ \citep{Gaia2021_EDR3}. By using stellar activity, we estimated for the first time the age of the system to $5.5\pm0.8~\gyr$. The host star GJ~422 corresponds to a M3.5 dwarf \citep{Hawley1996_GJ163_GJ422_GJ433_GJ667_GJ674_GJ682_GJ832_spectral_types_HD102365_binary}, with a mass estimated in this work of $0.42\pm0.05~\msun$ slightly higher but consistent with the previous estimate  from \citet{Zechmeister2009_GJ180_GJ229_GJ422_GJ433_GJ682_Mdwarfs}. \\
To date, one planet~b has been discovered but its location is debated: either at a $26$-$\dayy$ period \citep{Tuomi2014_GJ180_GJ229_GJ422_GJ433_GJ682} or at a $20$-$\dayy$ period  \citep{Feng2020_GJ180_GJ229_GJ422_GJ433}. In both cases, the minimal mass is about $10~\mearth$. We consider in this study the planet from \citet{Feng2020_GJ180_GJ229_GJ422_GJ433} ($10.4^{+3.3}_{-1.9}~\mearth$, $0.11\pm0.01~\au$) as they used more observations ($76$ measurements by combining UVES and HARPS instruments) than \citet{Tuomi2014_GJ180_GJ229_GJ422_GJ433_GJ682} ($49$, same instruments but on a shorter timeline). \\


\begin{figure}[h]
    \centering
    \includegraphics[width=\linewidth]{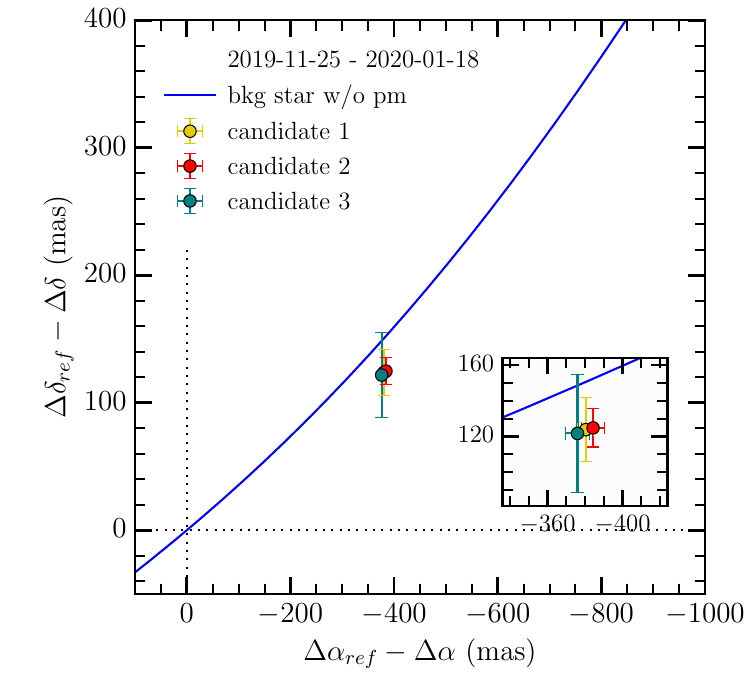}
    \caption{SPHERE/IRDIS relative astrometry differences of the offset positions of all the point-like sources detected around GJ~422 
    in both the second and third epochs (2018-01-28 and 2020-01-18).
    The expected variation of offset positions, if the companion candidates are stationary background objects are shown in blue solid line. The variation is estimated based on the parallactic and proper motions of the primary star. The five companion candidates are clearly identified as background contaminants. At the \textit{bottom-right} corner, the insert shows a zoom on the location of the candidates.
    }
    \label{fig:PMD_GJ422_2+3epochs}
\end{figure}

\paragraph{GJ~433}

The system GJ~433 is located at $9.077\pm0.002~\pc$ \citep{Gaia2021_EDR3}. By using gyrochronology, we estimated for the first time the age of the system to $500\pm150~\myr$. The host star GJ~433 corresponds to a M1.5 dwarf \citep{Hawley1996_GJ163_GJ422_GJ433_GJ667_GJ674_GJ682_GJ832_spectral_types_HD102365_binary}, with a mass estimated in this work of $0.50\pm0.05~\msun$ consistent with the previous estimate from \citet{Zechmeister2009_GJ180_GJ229_GJ422_GJ433_GJ682_Mdwarfs}. \\
GJ~433 was reported to host a low-mass planet GJ~433~b with a minimal mass of $5.8~\mearth$ at a semi-major axis of $0.058~\au$ \citep[period of $20.1~\days$,][]{Delfosse2013_GJ667_GJ433_RV_charac_planets}. They detected an additional significant signal with a much longer period of $10~\yrs$. Yet, based on the variation of activity indices on a similar timescale reported by \citet{GomesDaSilva2011_RV_GJ433_Mdwarfs_stellar_activity}, they concluded that a magnetic cycle of the star was a more likely origin. \citet{Tuomi2014_GJ180_GJ229_GJ422_GJ433_GJ682} recovered the same signals and considered the longer period signal to be a planet candidate, and called it GJ~433~c. If the planet GJ~443~c is real and the system close to edge-on, the planet~c would be a cold super-Neptune (minimal mass of $28.8^{+19.2}_{-10.5}~\mearth$) belonging to an unexplored population of Neptune-like planets. Regarding its separation of $0.5$" from its host star (corresponding to a semi-major axis of $4.7^{+1.2}_{-0.8}~\au$), the putative planet~c is probably the first realistic candidate for the direct imaging of cold Neptunes \citep{Feng2020_GJ180_GJ229_GJ422_GJ433}. In addition, \citet{Feng2020_GJ180_GJ229_GJ422_GJ433} may have discovered a third planet in the system, GJ~433~d, likely a super-earth with a minimal mass of $4.9^{+2.5}_{-1.8}~\mearth$ at a semi-major axis of $0.18^{+0.01}_{-0.02}~\au$ ($36$-$\dayy$ period). \\
%
Furthermore, \citet{Kennedy2018_GJ433_GJ649_DD} discovered an unresolved debris disk with a radius between $1$ and $30~\au$ by using Herschel.


\begin{figure}[h]
    \centering
    \includegraphics[width=\linewidth]{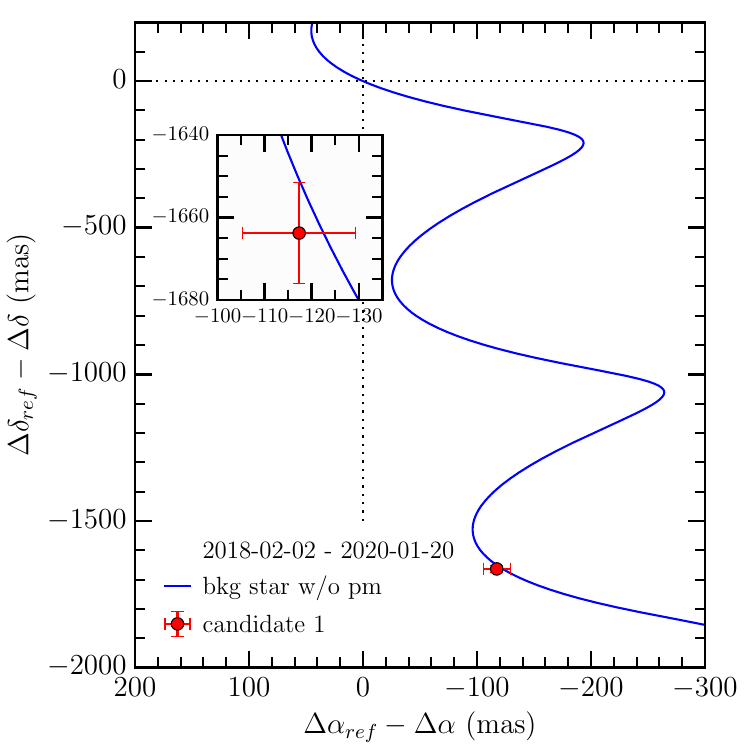}
    \caption{SPHERE/IRDIS relative astrometry differences of the offset positions of the point-like source detected around GJ~433 between the first epoch considered as reference 2018-02-02 and the second epoch 2020-01-20. (See caption from Fig. \ref{fig:PMD_GJ422_2+3epochs}) 
    }
    \label{fig:PMD_GJ433_2epochs}
\end{figure}

\paragraph{GJ~581}

The system GJ~581 is located at $6.301\pm0.001~\pc$ \citep{Gaia2021_EDR3}. We estimated the stellar age at $10\pm1~\gyr$ by using chromospheric activity which is consistent at $1\,\sigma$ with \citet{Veyette2018_GJ176_GJ581_GJ649_Wolf1061_age_chemo_kinematic} based on chemo-kinematics ($6.9^{+2.5}_{-2.3}~\gyr$) but not with the estimation based on isochrones ($4.1\pm0.3~\gyr$) from \citet{Yee2017_GJ581_FGKMstars_age}.  
Our new determination of the stellar mass ($0.35\pm0.05~\msun$) is consistent with  
\citet{AstudilloDefru2017_GJ176_GJ581_GJ674_activity_mass}. The star is a M3 star \citep{Trifonov2018_GJ176_GJ581_rv_1planet}, and its radius is $0.30\pm0.01~\rsun$  \citep{vonBraun2011_GJ581_GJ832_radius}. By looking into the atmospheric structures on the star based on observations in the visible, ultraviolet (UV), and X-ray, and especially by synthesizing integrated extreme-ultraviolet (EUV) fluxes,  \citet{Tilipman2021_GJ581_GJ832} found the star GJ~581 to be relatively inactive compared to other M dwarfs of similar spectral types (e.g., GJ~644~B, GJ~588, GJ~887, or GJ~205), with H$_2$ fluorescent emission \citep{France2013_GJ581_GJ667_GJ832_UV_H2}.
\\
Over potentially six low-mass planets discovered, three are confirmed in the system. The GJ~581~b ($15.7$-$\mearth$ minimal mass, $5.4$-$\dayy$ period) was discovered by \citet{Bonfils2005_GJ581_1planet}, while GJ~581~c ($5.0$-$\mearth$ minimal mass, $13$-$\dayy$ period) and GJ~581~d ($7.7$-$\mearth$ minimal mass, $84$-$\dayy$ period) by \citet{Udry2007_GJ581_3planets_1wrongd}, and GJ~581~e ($1.7~\mearth$-minimal mass, $3.2$-$\dayy$ period) by \citet{Mayor2009_4planets_1wrongd}. \citet{Vogt2010_GJ581_6planets} proposed the existence of two additional planets, GJ~581~f and GJ~581~g with orbital periods of $433~\days$ and $37~\days$, respectively, but their existence is not supported by \citet{Forveille2011_GJ581_only4planets_fg_wrong}, nor by \citet{Tuomi2011_GJ581_only4planets_fg_wrong}. In addition, later studies reconsidered the existence of the planet~d \citep[e.g.,][]{Baluev2013_GJ581_d_wrong_noise,Robertson2014_GJ581_d_wrong_stellar_activity,Hatzes2016_GJ581_d_wrong_stellar_activity}.
As did \citet{Trifonov2018_GJ176_GJ581_rv_1planet}, we only consider the planets GJ~581~b, GJ~581~c and GJ~581~e as confirmed and we use their derivation of the orbital values, i.e., minimal masses of $15.2\pm0.3~\mearth$, $5.7\pm0.3~\mearth$, and $1.7\pm0.2~\mearth$, and semi-major axes of $0.041\pm0.001~\au$, $0.074\pm0.001~\au$, and $0.029\pm0.001~\au$, respectively.  
The planet GJ~581~c has been more in-depth studied in the literature: its spin-orbit resonance is likely to be $3$:$2$, and the presence of liquid water has been investigated, including through 3D global climate models \citep{Leconte2013_GJ581_HD85512_3Dmodels}.

Furthermore, \citet{Lestrade2012_GJ581_DD} discovered and resolved a debris disk extending from $25~\au$ to more than $60~\au$ by using Herschel. Its inclination is estimated to be between $30$ and $70\degr$.

\paragraph{Wolf 1061 (GJ~628)}

The system Wolf~1061 (GJ~628) is the closest system to the Sun from our sample, located at $4.3078\pm0.0005~\pc$ \citep{Gaia2021_EDR3}. By using gyrochronology, we estimated an age of $7.1\pm2.0~\gyr$, consistent with the previous estimate from \citet{Veyette2018_GJ176_GJ581_GJ649_Wolf1061_age_chemo_kinematic} based on chemo-kinematics. We derive the stellar mass to a value of $0.35\pm0.05~\msun$, which is higher but in agreement with the value from \citet{Maldonado2020_GJ176_GJ433_GJ581_GJ667C_GJ674_GJ682_GJ832_Wolf1061} of $0.25\pm0.12~\msun$.
The stellar spectral type is between M3V and M4.5V spectral types \citep[e.g.,][]{Joy1974_GJ176_GJ229_GJ581_GJ649_BD06_Wolf_spectral_types,Henry2002_GJ229A_Wolf1061_spectral_type}, while the stellar radius is $0.325\pm0.012~\rsun$ \citep{Kane2017_Wolf1061_3planets_charac}. \\
%
\citet{Wright2016_Wolf1061_3planets_discovery_RV} discovered three planets Wolf~1061~b ($m\,\sin(i)=1.4\pm0.3~\mearth$, $a\sim0.04~\au$, $P\sim4.9~\days$), Wolf~1061~c ($4.3\pm0.4~\mearth$, $0.08~\au$, $18~\days$) and Wolf~1061~d ($5.21\pm0.68~\mearth$, $0.20~\au$, $67~\days$). However, \citet{AstudilloDefru2017_Wolf1061_2planets_dwrong} called into issue the existence of the planet~d, which could be caused by aliases. Hence, we consider in this paper only the planets~b and~c as confirmed.

\paragraph{GJ~649}

The system GJ~649 is located at $10.391\pm0.003~\pc$ \citep{Gaia2021_EDR3}. We derived an age of $1.40\pm0.35~\gyr$ based on gyrochronology, which is in agreement with its membership to the “old disk” following kinematics \citep{Leggett1992_GJ163_GJ649_GJ674_colors_membership_pop}
but younger than the age of $4.5^{+3.0}_{-2.0}~\gyr$ determined by \citet{Veyette2018_GJ176_GJ581_GJ649_Wolf1061_age_chemo_kinematic} following a combined chemo-kinematic approach (consistent at $2\,\sigma$). 
Our stellar mass measurement of $0.54\pm0.05~\msun$ is consistent with previous estimate from \citet{Johnson2010_GJ649_1planet} and \citet{vonBraun2014_61Vir_GJ176_GJ649_stellar_radius}. Its spectral type is M1.5 \citep{Johnson2010_GJ649_1planet}, while \citet{vonBraun2014_61Vir_GJ176_GJ649_stellar_radius} found a stellar radius equal to $0.54\pm0.02~\rsun$.\\
The system hosts one giant planet with a minimal mass of $0.328~\mj$ at a semi-major axis equal to $1.135~\au$ and eccentricity of $0.30\pm0.08$ \citep{Johnson2010_GJ649_1planet}. As the RV signal of two circular planets can mimic the signal of one eccentric planet, \citet{Wittenmyer2013_GJ649_2planets_cunconfirmed} proposed the existence of an inner smaller planet, with a minimal mass of about $9.5~\mearth$ at a semi-major axis of about $0.043~\au$ based on a statistical approach. However, this planet has not been confirmed yet. \\ 
By using Herschel, \citet{Kennedy2018_GJ433_GJ649_DD} discovered and marginally resolved a debris disk with a radius between $6$ and $30~\au$, likely to be edge-on.

\paragraph{GJ~667 C}

The system GJ~667 is a multi-stellar and planetary system located at $7.2429\pm0.0015~\pc$ \citep{Gaia2021_EDR3}. The stars GJ~667~A and GJ~667~B form a binary pair with a total dynamically derived mass of $1.27~\msun$ \citep{Soderhjelm1999_GJ667} and a semi-major axis of $13.1~\au$ \citep[period of $42.15~\yrs$,][]{Delfosse2013_GJ667_GJ433_RV_charac_planets},
around which a third star, GJ~667~C, orbits at a projected angular separation of about $32.4$", i.e., at a minimal distance of $235~\au$. 

In this work, we focused on GJ~667~C, which is a M1.5 star \citep{Hawley1996_GJ163_GJ422_GJ433_GJ667_GJ674_GJ682_GJ832_spectral_types_HD102365_binary}. We estimated its mass to $0.346\pm0.050~\msun$, consistent with previous estimates \citep{Anglada-Escude2012_GJ667C_RV_3planets, Delfosse2013_GJ667_GJ433_RV_charac_planets}. Our age determination ($6.3\pm4.0~\gyr$) via gyronochronology is in agreement with \citet{Anglada-Escude2012_GJ667C_RV_3planets}. We derived a sub-solar stellar metallicity [Fe/H] of $-0.48\pm0.10$. The stellar radius is equal to $0.30\pm0.02~\rsun$  \citep{vonBraun2013_GJ667C_stellar_radius}. In addition, \citet{France2013_GJ581_GJ667_GJ832_UV_H2} noticed moderate level of UV emission from the chromospheric and transition regions of GJ~667~C. \\
The star GJ~667~C is known to host two confirmed planets (b~and~c), while five other ones (d,~e,~f,~g and~h) have been proposed. The planets~b and~c have been discovered by \citet{Anglada-Escude2012_GJ667C_RV_3planets} with minimal masses of $5.46~\mearth$ and $4.25~\mearth$, and at periods of about $7.2~\days$ and $28.1~\days$ (semi-major axes of about $0.05~\au$ and $0.13~\au$), respectively. They were confirmed by \citet{Delfosse2013_GJ667_GJ433_RV_charac_planets}, \citet{Bonfils2013_GJ667_GJ832_RV_Mdwarves_survey}, and \citet{Robertson2014_GJ581_d_wrong_stellar_activity}. 
\citet{Anglada-Escude2013_GJ667C_3planets_habitable_zone} suggested the existence of the planet~d ($5.1^{+1.8}_{-1.7}~\mearth$, at $0.28~\au$, $92$-$\dayy$ period), planet~e ($2.7^{+1.6}_{-1.4}~\mearth$, $0.21~\au$, $62~\days$), planet~f ($2.7^{+1.4}_{-1.2}~\mearth$, $0.16~\au$, $39~\days$), planet~g ($5.1^{+1.8}_{-1.5}~\mearth$, $0.55~\au$, $256~\days$), and planet~h (tentative evidence of a $1.1^{+1.0}_{-0.9}~\mearth$, $0.09~\au$, $17~\days$).
However, \citet{Delfosse2013_GJ667_GJ433_RV_charac_planets} and \citet{Robertson2014_GJ581_d_wrong_stellar_activity} reported the RV signal associated to the planet~d as to be in reality an artifact caused by stellar rotation, harmonics, or aliases. After correcting from stellar activity, \citet{Robertson2014_GJ581_d_wrong_stellar_activity}  were not able to confirm the other planetary candidates (e,~f,~g and~h), thereby we consider them as unconfirmed.

\paragraph{GJ~674}

The system GJ~674 is the second closest system in our sample, as at $4.5528\pm0.0005~\pc$ from the Sun \citep{Gaia2021_EDR3}.  The stellar spectral type is M2.5 \citep{Hawley1996_GJ163_GJ422_GJ433_GJ667_GJ674_GJ682_GJ832_spectral_types_HD102365_binary}. 
Our stellar mass measurement of $0.40\pm0.05~\msun$ is consistent with the previous estimate from \citet{Bonfils2007_GJ674_1planet}.
We estimated an age of $2.0\pm0.5~\gyr$ based on gyrochronology, moderately higher than its previous value spanning between $0.1$ and $1.0~\gyr$ \citep{Bonfils2007_GJ674_1planet}, who considered the UVW galactic velocities corresponding to the transition between young and old disk populations \citep{Leggett1992_GJ163_GJ649_GJ674_colors_membership_pop}. Although the star is not supposed to be in its most active period, \citet{Froning2019_GJ674_UV_flare} reported a large flare and several small ones in far-ultraviolet (FUV) monitoring observations with the Hubble Space Telescope, highlighting the importance of considering the FUV emission for exoplanet atmospheric heating and chemistry.

The system GJ~674 hosts at least one planet~b of minimal mass $11.1~\mearth$ at a semi-major axis of about $0.039~\au$ \citep[$4.69$-$\dayy$ period,][]{Bonfils2007_GJ674_1planet}.
\citet{Vidotto2019_GJ674_aurora} pointed out that GJ~674~b would be a relevant planet to try to detect aurora in radio, provided that the wind of the host star is not too strong.
\begin{figure}[h]
    \centering
    \includegraphics[width=\linewidth]{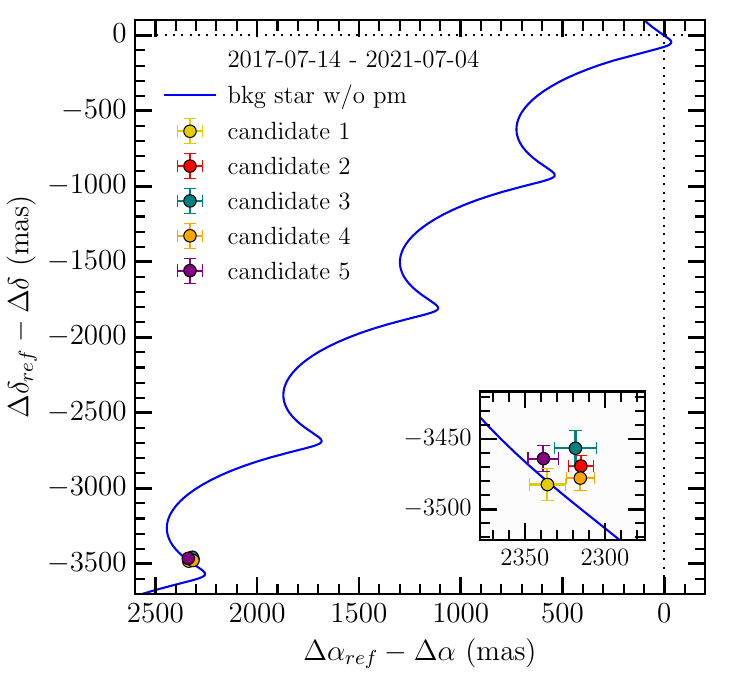}
    \caption{SPHERE/IRDIS relative astrometry differences of the offset positions of the point-like source detected around GJ~674 between the first epoch considered as reference 2017-07-14 and the second epoch 2021-07-04. (See caption from Fig. \ref{fig:PMD_GJ422_2+3epochs}) 
    }
    \label{fig:PMD_GJ674_2epochs}
\end{figure}

\paragraph{GJ~682}

The system GJ~682 (HIP~86214) located at $5.008\pm0.001~\pc$ \citep{Gaia2021_EDR3} has a M3.5 star \citep{Hawley1996_GJ163_GJ422_GJ433_GJ667_GJ674_GJ682_GJ832_spectral_types_HD102365_binary}. In our work, we estimated the stellar mass to $0.30\pm0.05~\msun$, consistent with the previous estimate from \citet{Zechmeister2009_GJ180_GJ229_GJ422_GJ433_GJ682_Mdwarfs}, but higher than  \citet{Ward-Duong2015_GJ682_binary}, estimated at $0.23~\msun$.
By using either activity or gyrochronology-based methods, we were not able to estimate a robust age for this system. Thus, in this study we consider the default age of $5\pm4~\gyr$, as no estimate is given in the literature. \citet{Ward-Duong2015_GJ682_binary} proposed a $0.14~\msun$ binary companion orbiting at a projected distance of $0.85~\au$ ($0.17"$). Since we did not re-detect it in any of our three epochs, and the field of view of GJ~682 is crowded ($10$--$26$ detections, see Table~\ref{tab:detections_per_target_summarize}, Figs.~\ref{fig:mosaic_super-earths_survey_IRDIS_cADI} and~\ref{fig:mosaic_super-earths_survey_IRDIS_ANDROMEDA}), and RV measurements did not reveal the presence of such binary companion, we consider it was a background contaminant in their 2005-05-01 epoch of observation. \\ 
%
\citet{Tuomi2014_GJ180_GJ229_GJ422_GJ433_GJ682} discovered two low-mass planets in this system, GJ~682~b and  GJ~682~c with minimal masses of $4.4^{+3.7}_{-2.4}~\mearth$ and  $8.7^{+5.8}_{-4.6}~\mearth$, at semi-major axes of $0.080^{+0.014}_{-0.004}~\au$ and $0.176^{+0.030}_{-0.003}~\au$ ($17.5$-$\dayy$ and $57$-$\dayy$ periods), respectively.
However, the planets b and c have not been confirmed in a ulterior study.  \citet{Feng2020_GJ180_GJ229_GJ422_GJ433} in their updated UCLES measurements and additional HARPS data did not identify them as planetary signal, but rather as caused by stellar activity. Thus, we considered the planets  GJ~682~b and  GJ~682~c as unconfirmed, that means the system GJ~682 might not be planetary.

\begin{figure}[h]
    \centering
    \includegraphics[width=\linewidth]{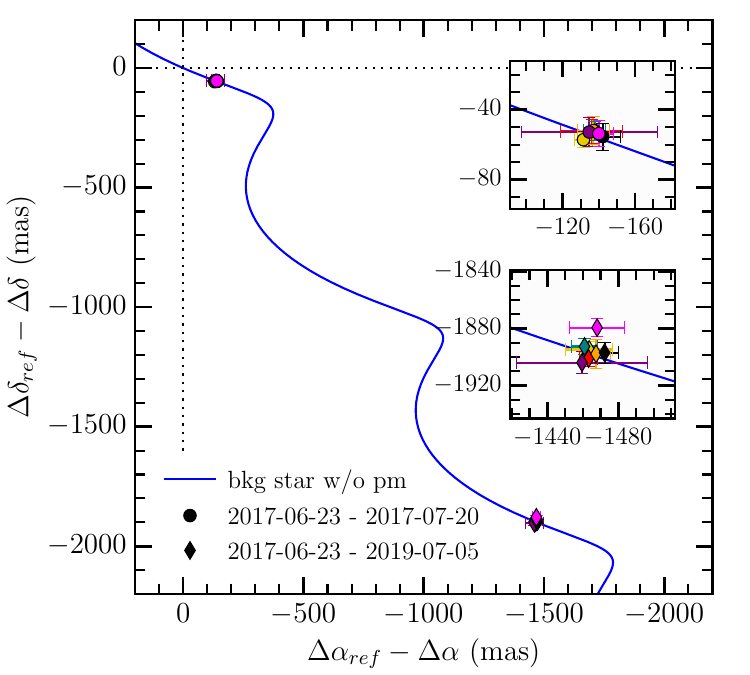} \includegraphics[width=\linewidth]{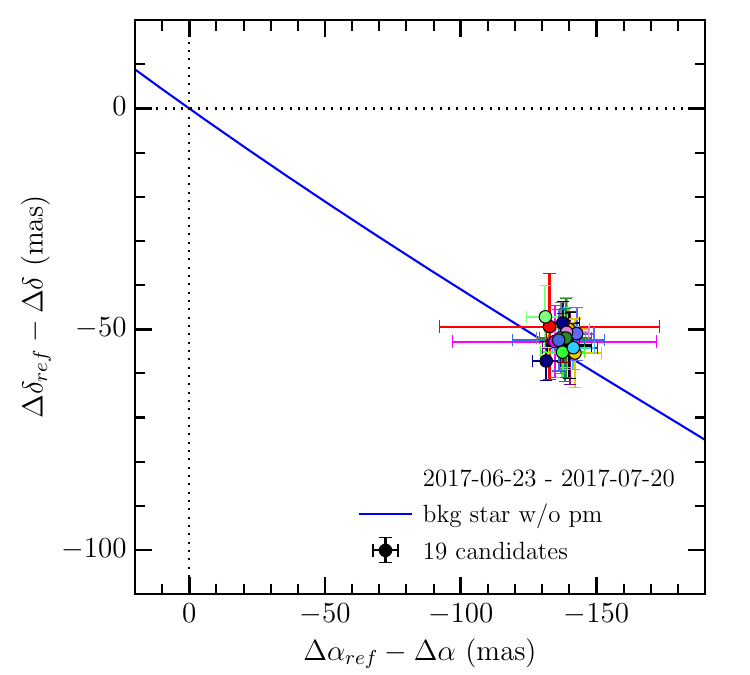}
    \caption{SPHERE/IRDIS relative astrometry differences of the offset positions of the point-like source detected around GJ~682 in the three epochs (at the \textit{top}) and  only in the first and third epochs (at the \textit{bottom}). The first epoch is considered as reference, i.e., 2017-06-22; the second epoch is 2017-06-23, and the third epoch 2019-07-05. 
     (See caption from Fig. \ref{fig:PMD_GJ422_2+3epochs})
    }
    \label{fig:PMD_GJ682_2+3epochs}
\end{figure}

\paragraph{GJ~832}

The system GJ~832 is located at $4.9671\pm0.0006~\pc$ \citep{Gaia2021_EDR3}. We estimated the stellar age at $2.4\pm0.3~\gyr$ by using gyrochronology. 
We found a stellar mass of $0.48\pm0.05~\msun$ consistent with the previous estimate from \citet{Bonfils2013_GJ667_GJ832_RV_Mdwarves_survey}. The stellar type is M2.5 \citep{Hawley1996_GJ163_GJ422_GJ433_GJ667_GJ674_GJ682_GJ832_spectral_types_HD102365_binary}, while the stellar radius is $0.50\pm0.02~\rsun$ \citep{vonBraun2011_GJ581_GJ832_radius}.
\citet{Peacock2019_GJ176_GJ832_EUV} investigated the extreme ultraviolet radiation environment around, whereas afterwards \citet{Tilipman2021_GJ581_GJ832} focused on the stellar atmospheric structures as for GJ~581. \citet{Tilipman2021_GJ581_GJ832}  found the star GJ~832 to be relatively inactive, but still more than GJ~581, and have as well a H$_2$ fluorescent emission \citep{France2013_GJ581_GJ667_GJ832_UV_H2}. \\
The system GJ~832 hosts a long-period planet named GJ~832~b. This planet~b was discovered by \citet{Bailey2009_GJ832_1planet} using  radial velocity measurements: it has a minimal mass of $0.74\pm0.06~\mj$ orbiting at a period of $3\,838\pm49~\days$ (a semi-major axis of $3.79\pm0.03~\au$) with a low eccentricity ($0.02$--$0.06$) from the most recent study \citet{Gorrini2022_stellar_activity_GJ832}. \citet{Bonfils2013_GJ667_GJ832_RV_Mdwarves_survey} confirmed the signal, and indicated a possible second signal at a $35$-$\dayy$ period, but which could yet be linked to stellar activity. This signal was considered by \citet{Wittenmyer2014_GJ832_2planets_cwrong} as the planet~c, an inner super-Earth ($5.4\pm1.0~\mearth$, $0.16\pm0.01~\au$). Recently, this planet~c have be debated by \citet{Gorrini2022_stellar_activity_GJ832} arguing for stellar activity. Hence, we only consider the planet~b as confirmed, and we removed the system GJ~832 in our statistical study of systems hosting known sub-Saturn(s).  Nonetheless, we stress the fact that the system GJ~832 is one of the four systems in our full sample with a significant proper motion anomaly signature \citep[signal-to-noise ratio equal to $14.1$,][]{Kervella2022_PMa}. This is consistent with the radial velocity measurements of the planet~b, assuming a circular orbit and an inclination of about $60\degr$ \citep{Kervella2019_PMa_inc}. The planet GJ~832~b has not yet been imaged - even though we had a detection plausible in our first epoch with SPHERE, we were not able to recover it in the best observation (see Section~\ref{sec:results_obs_DI_candidates}). 
%

\paragraph{GJ~3998}

The system GJ~3998 located at $18.18\pm0.01~\pc$ \citep{Gaia2021_EDR3} has a M1.5 star \citep{Gaidos2014_GJ180_GJ3998_Wolf1061_spectral_type}. In our work, we estimated the stellar mass to $0.55\pm0.05~\msun$, consistent with the previous estimate from \citet{Affer2016_GJ3998_2planets_mass}. 
By using gyrochronology, we estimated for the first time an age of $2.0\pm0.5~\gyr$ for this system. \\ 
\citet{Affer2016_GJ3998_2planets_mass} discovered two low-mass planets in this system, GJ~3998~b and  GJ~3998~c with minimal masses of $2.5\pm0.3~\mearth$ and  $6.3\pm0.8~\mearth$, at semi-major axes of $0.029\pm0.001~\au$ and $0.089\pm0.003~\au$ (i.e., about $2.6$-$\dayy$ and $13.7$-$\dayy$ periods), respectively.

\paragraph{HD~3651}

The system HD~3651 located at $11.108\pm0.006~\pc$ \citep{Gaia2021_EDR3} has a K0V star \citep{Gray2003_spectral_type_HD3651_HD38858}. 
By using gyrochronology and chromospheric activity, we estimated an age of $7.5\pm2.2~\gyr$ for the system, consistent with values from the literature \citep[e.g.,][]{Valenti_Fischer2005_61Vir_mass_HD3651_age, Liu2007_HD3651B_prop,Vidotto2014_age_HD3651,Bonfanti2016_HD20794_HD40307_HD69830_HD3651_HD99492_age}.
We determined a stellar mass $0.86\pm0.06~\msun$, higher than the previous estimate ($0.79\pm0.06~\msun$) from \citet{Fischer2003_HD3651_1planet_subSaturn_discovery} and a stellar metallicity [Fe/H] close to the solar one ($0.08\pm0.06$).

The system HD~3651 hosts one eccentric ($0.63\pm0.04$) Saturn-mass planet~b ($0.20~\mj$) at $0.28~\au$ (period of $62$-day) discovered by \citet{Fischer2003_HD3651_1planet_subSaturn_discovery} using radial velocity measurements from the Lick and Keck observatories. In addition, \citet{Mugrauer2006_HD3651_1BD} and \citet{Luhman2007_HD3651} independently discovered by direct imaging a wide-orbit brown dwarf ($20$--$60~\mj$, projected separation of $43"$ corresponding to $480~\au$). Its spectral type is T$7.5\pm0.5$ \citep{Liu2007_HD3651B_prop,Burgasser2007_HD3651B_prop}. \citet{Line2015_HD3651_spectro} retrieved ammonia in low-resolution near infrared spectrum for this brown dwarf, and found a good agreement between the values of the carbon-to-oxygen ratio of the brown dwarf and the host star.
\citet{Wittenmyer2013_GJ649_2planets_cunconfirmed} proposed the existence of an inner planet (minimal mass of $0.09~\mj$) at a period of $31~\days$ ($0.19~\au$) based on the same statistical approach than for GJ~649. Yet, they retracted in \citet{Wittenmeyer2007_HD3651_other_2planets}, stating that radial velocity signal from a planet with an eccentricity higher than $0.5$ cannot be explained by an architecture of two circular planets from updated simulations. It is consistent with results from \citet{Brewer2020_HD3651_EXPRESS_dynamical_stability}, who did no expect an inner planet to HD~3651 b from their radial-velocities measurements coupled to their N-body simulations, since it would be dynamically unstable.


\paragraph{HD~20794}

The system HD~20794 located at $6.041\pm0.003~\pc$ \citep{Gaia2021_EDR3} has a G8V star \citep{Pepe2011_HD20794_HD85512}. In our work, we estimated the stellar mass to $0.79\pm0.05~\msun$, lower but consistent at $1\sigma$ with the previous estimate ($0.85\pm0.02~\msun$) from \citet{Takeda2007_age}, and a metallicity [Fe/H] of $-0.36\pm0.05$.
By using chromospheric activity, we estimated an age of $6.0\pm0.3~\gyr$ for the system, younger than the previous age ($11.6\pm1.5~\gyr$) from \citet{Bonfanti2016_HD20794_HD40307_HD69830_HD3651_HD99492_age}. As HD~20794 is particularly stable over long time baselines \citep{Butler2001_stable_rv_stars_HD20794}, the system has been selected as a special star reference for radial velocity measurements, which explains the high number of observations with HARPS ($>7500$).

The system hosts several super-Earths. \citet{Pepe2011_HD20794_HD85512} reported the discovery of three planets, HD~20794~b, HD~20794~c and HD~20794~d at $0.12~\au$ ($18$-$\dayy$ period), $0.20~\au$ ($18$-$\dayy$ period) and $0.35~\au$ ($90$-$\dayy$ period), and with minimal masses of $2.7~\mearth$, $2.4~\mearth$ and $4.7~\mearth$, respectively. They are cautious regarding their planet~c, due to faint amplitude signal and as it could be linked to the stellar rotation.
\citet{Feng2017_HD20794_3planets} is suspicious as well regarding the existence of the planet~c, but points out the existence of at least one additional planet HD~20794~e  at $0.51~\au$ ($147$-$\dayy$ period), and possibly two, HD~20794~f and HD~20794~g, at $0.88~\au$ ($331$-$\dayy$ period) and $0.09~\au$ ($12$-$\dayy$ period), having minimal masses of $4.8~\mearth$, $10.3~\mearth$, and $1.0~\mearth$, respectively. In this work, we consider the planets~b,~d, and~e as confirmed detections, the others as unconfirmed.

In addition, the star hosts a cold (about $80~\mathrm{K}$) debris disk spanning between roughly $15$ and $40~\au$ \citep{Wyatt2012_61Vir_HD20794_HD38858_HD69830_HD_102365_HD136352_Herschel_submm_debris_disc,Kennedy2015_HD38858_HD20794_61Vir_HD69380_submm_debris_belt_kuiper_analog}. Hot dust was detected with VLTI/PIONIER with an excess of 1.64 ± 0.37\% \citep{Ertel2014_HD20794_Pionier} and a variability over two years \citep{Ertel2016_HD20794_Pionier_variability}. This hot dust could be located between $0.01$ and $0.1~\au$ \citep{Kennedy2015_HD38858_HD20794_61Vir_HD69380_submm_debris_belt_kuiper_analog}, with the sublimation radius estimate at $0.03~\au$ \citep{Absil2021_HD20794}.

\paragraph{HD~38858}

The system HD~38858 is located at $15.210\pm0.007~\pc$ \citep{Gaia2021_EDR3} and has a G2V star \citep{Gray2003_spectral_type_HD3651_HD38858}. We derived an age of $4.3\pm1.0~\gyr$ based on gyrochronology and chromospheric activity, in agreement with previous estimates from \citet{Takeda2007_age} and  \citet{Lovis2011_HD69830_HD85512_HD136352_HD154088_HD40307_HD38858_age}, but significantly older than the estimate from \citet{Casagrande2011_age_HD38858} of $200~\myr$ based on isochrones.
In addition, we estimated a stellar mass
of $0.91\pm0.06~\msun$, consistent with \citet{Takeda2007_age}, and a metallicity [Fe/H] of $-0.21\pm0.05$.

The system hosts at least a low-mass planet named HD~38858~b discovered by \citet{Kennedy2015_HD38858_HD20794_61Vir_HD69380_submm_debris_belt_kuiper_analog} at $0.64~\au$ ($198$-day period) and with a minimum mass of $12\pm2~\mearth$. Previously, \citet{Mayor2011_GJ38858_HD136352_HD154088_HD189567} announced the discovery of a planet with a semi-major axis of $1.04~\au$ ($407$-day period) and a minimal mass of $30.6\pm4.1~\mearth$. However, by using additional data ($96$ instead of $52$~measurements) \citet{Kennedy2015_HD38858_HD20794_61Vir_HD69380_submm_debris_belt_kuiper_analog} revealed this signal to be caused by stellar activity. 
\citet{ReadWyatt2016_HD38858_61Vir_secular_interactions} put constraints on the presence of additional low-mass planets in the system from a theoretical perspective. They reported that an eccentric ($0.2$--$0.8$) planet located between $1$ and $10~\au$ and with a minimal mass of $3$--$10~\mearth$ would be unlikely, as secular interactions with the planet HD~38858~b would cause a significant eccentricity for the planet~b that is not observed. Nonetheless, if such planet did exist, it would point for the existence of a third planet, in the vicinity of the planet  HD~38858~b, and stabilizing it.   

Furthermore, \citet{Beichman2006_HD38858_DD} discovered with \textit{Spitzer} a cool debris disk around the star, marginally resolved \citep{Krist2012_HD38858_Spitzer_DD}, and better resolved at the same wavelength ($70~\mic$) with Herschel \citet{Kennedy2015_HD38858_HD20794_61Vir_HD69380_submm_debris_belt_kuiper_analog}. These latter observations constrained the disk geometry and radial structure. 

\paragraph{HD~40307}
The system HD~40307 is located at $12.932\pm0.003~\pc$ \citep{Gaia2021_EDR3}, and hosts a K2.5V star \citep{Gray2006_spectral_type_HD69830_HD85512_HD40307_HD136352}. We derived an age of $4.4\pm2.0~\gyr$ based on chromospheric activity and gyrochronology which is consistent with e.g., \citet{Lovis2011_HD69830_HD85512_HD136352_HD154088_HD40307_HD38858_age} and \citet{Bonfanti2016_HD20794_HD40307_HD69830_HD3651_HD99492_age}. 
We estimated the stellar mass to $0.71\pm0.05~\msun$, in agreement with \citet{Takeda2007_age}, and determined a stellar sub-solar metallicity [Fe/H] of $-0.30\pm0.05$.

By using HARPS RV measurements, four to six super-Earths and/or mini-Neptunes may have been discovered in the system. In short, the three inner ones~b,~c, and~d are discovered by \citet{Mayor2009_HD40307_3planets} and the three outer ones~e,~f and~g by \citet{Tuomi2013_HD40307_6planets} who used a new analysis method. However, the existence of the planets HD~40307~e, and HD~40307~g is debated 
\citep{Diaz2016_HD40307,Unger2020_HD40307}.  
Regarding the confirmed planets  HD~40307~b,  HD~40307~c,  HD~40307~d, and HD~40307~f, \citet{Unger2020_HD40307} reported minimal masses of about $3.81~\mearth$, $6.43~\mearth$, $8.74~\mearth$, and $3.63~\mearth$, semi-major axes of about $0.05~\au$, $0.08~\au$, $0.13~\au$, and $0.25~\au$ (periods of $4.3~\days$, $9.6~\days$, $20.4~\days$, and $51.6~\days$), respectively. As for the unconfirmed planets HD~40307~e and HD~40307~g, they would have a minimal mass of $3.5\mearth$, and $7.1\mearth$, and semi-major axis of $0.19~\au$ and $0.60~\au$ (period of $34.6~\dayy$, and $198~\dayy$) following
 \citet{Tuomi2013_HD40307_6planets}.
 



Concerning their formation history, \citet{Barnes2009_HD40307_dynamic_stability} found that the planets~b, c, and d likely formed at larger radii and migrated inward into their current orbits owing to disk interactions. The dynamics of this multiple system with at that time three planets discovered has been also investigated by \citet{Papaloizou2010_HD40307_dynamic_stability}, \citet{Zhou2010_HD40307_dynamic_stability}, and \citet{YuanYuan2013_HD40307_dynamic_stability}. 
On the other hand, \citet{Brasser2014_HD40307_dynamic_stability} investigated the stability of the system based on a $6$-planet~configuration.

Furthermore, the system HD~40307 was imaged by VLT/NaCo  \citep{DietrichGinski2018_HD40307_Naco}. Nonetheless, the reported companion candidate found at a separation of $13.35$" and position angle of $146\degr$ was classified as a background object.

\paragraph{HD~69830}

The system HD~69830 is located at $12.579\pm0.006~\pc$ \citep{Gaia2021_EDR3}. We derived an age of $6.6\pm1.5~\gyr$ based on gyrochronology and chromospheric activity, which is younger than the age estimated from \citet{Takeda2007_age} of $10.4\pm2.5~\gyr$ but consistent with the one from e.g., 
\citet{Tanner2015_HD69830_star_params}. Our stellar mass estimation of $0.88\pm0.06~\msun$ is consistent with the  value from e.g., 
\citet{Tanner2015_HD69830_star_params}.  We determined a stellar metallicity [Fe/H] consistent with the solar one ($-0.03\pm0.05$). The star has a spectral type of is G8V \citep{Gray2006_spectral_type_HD69830_HD85512_HD40307_HD136352} or K0V \citet{Tanner2015_HD69830_star_params}.
From interferometric measurements with the CHARA array, \citet{Tanner2015_HD69830_star_params} estimated a physical radius of $0.906\pm0.019~\rsun$. In addition, the system HD\,69830 underwent an encounter $0.31\pm0.01~\myr$ ago at a minimal distance of $0.43^{+0.06}_{-0.02}~\pc$ with the star HD\,57901 \citep{Bertini2023_flybys_DD_HD20794_HD38858_HD69830}.

\citet{Lovis2006_HD69830_3planets} discovered three planets ~b,~c and~d with minimum masses of $10.2~\mearth$, $11.8~\mearth$ and $18.1~\mearth$ and semi-major axis of about $0.08~\au$, $0.19~\au$ and $0.63~\au$ i.e., periods of $8.7$, $31.6$
and $197~\days$, respectively. \citet{Lovis2006_HD69830_3planets}, \citet{Alibert2006_HD69830} and \citet{Lecavelier2007_HD69830b_SE_no_MN} concluded that HD~69830~b must have a mainly rocky composition such as a super-Earth, while the two outermost planets HD~69830~c and HD~69830~d should more look like super-Neptunes.
Those planets have been confirmed in later studies, e.g., \citet{Hara2022_HD69830_stellar_activity_study} who led an in-depth study of stellar activity. By using $254$~HARPS RV measurements (acquired before the change of fiber in 2015) spanning over $11.5$~years, they demonstrated that the RV periodic signals associated to the three planets have a stable amplitude, phase and period, which strengthens their planetary origin over a stellar one. Moreover, this configuration of three planets is dynamically stable even if planetary masses are shifted to Jupiter masses in case of a face-on system \citep{Ji2007_HD69830_stability_dynamics_secular_res}. \citet{Payne2009_HD69830_stability_dynamics} reported that the three planets may have migrated through the disk.

Furthermore, the system hosts a warm belt, external to the three low-mass planets known, and first constrained by \citet{Beichman2005_HD69830_DD} via SED modelling of \textit{Spitzer} observations. They show the star to harbour within $1~\au$ warm dust grains from an exozodiacal disk (exozodi), 
which would represent a total mass of about $4.6\times10^{-9}~\mearth$ by assuming only $0.25~\mic$-sized dust grains. 
The exozodi could be explained by a recent collision between planetesimals \citep{Wyatt2007_HD69830_DD,Wyatt2010_HD69830_DD,HengTremaine2010_long_lived_planetesimal_discs}. \citet{Beichman2006_HD38858_DD} pointed out that over $41$~stars hosting planets, and later on \citet{Bryden2009_HD69830} for an additional $105$~stars hosting planets, that HD~69830 is the only one among $14$~systems with a debris disk to have this IR excess exclusively at $24~\mic$. This means the system has a hot belt and insignificant amounts of cool dust. The absence of a large and cold reservoir of cometary bodies is also reported by \citet{Marshall2014_HD69830_DD_Herschel} based on Herschel data. This makes replenishing the exozodi challenging. Hence, the discovery from \citet{Lisse2007_HD69830_DD_water_grains} with \textit{Spitzer} of icy dusty grains may indicate that a recent disruptive event happened within the year of the observation. Yet, follow-up observations at higher signal-to-noise ratio still with \textit{Spitzer}  were not able to detect water ice again or any change in the emission spectrum \citep{Beichman2011_HD69830_DD_variability}. These latter indicate that the disruption of C-type asteroid(s) formed in dry and inner regions could explain the exozodi.

The status of a short-lived exozodi \citep[e.g.,][]{Beichman2005_HD69830_DD} is debated by \citet{Heng2011_HD69830_DD} which argued that the age of the system is poorly constrained and thus could be up to about $1~\gyr$. Such an age might point the hypothesis of a non-transient exozodi. In this case, the exozodi would have a mass of $3$--$4\times10^{-3}~\mearth$ i.e., from $7$ to $10$ times the one from the asteroid belt in our Solar system \citep{Krasinsky2002_mass_asteroid_belt_solar_system}. 

Complementary observations in the mid infrared with VLTI/MIDI have enabled to resolve the dust emission of the exozodi at $8$-$13~\mic$, but not with VISIR at $18.7~\mic$,  consistent with the radius of $1~\au$ derived by SED modelling \citep{Smith2009_HD69830_DD_MIDI_VISIR_MIR}. 
In addition, \citet{Beichman2011_HD69830_DD_variability} reported no variability of the amount of dust or its composition over $4$~year period with \textit{Spitzer} (upper limit of $3.3\%$ at~$1\sigma$-level).



\paragraph{HD~85512}

The system HD~85512 is located at $11.277\pm0.003~\pc$ \citep{Gaia2021_EDR3}. We derived an age of $5.1\pm1.0~\gyr$ based on gyrochronology and chromospheric activity, and a stellar mass of $0.70\pm0.05~\msun$, both consistent with e.g., \citet{Pepe2011_HD20794_HD85512}.
We found a sub-solar metallicity [Fe/H] of $-0.28\pm0.12$.
The star has a K5V spectral type, and is even more stable than HD~20794 \citep{Pepe2011_HD20794_HD85512}. 

The system is known to harbour one planet HD~85512~b of minimal mass $3.6\pm0.5~\mearth$ at $0.26\pm0.01~\au$ ($\sim58$-$\dayy$ period) on a possible circular orbit ($e=0.11\pm0.1$) and rocky \citep{Pepe2011_HD20794_HD85512}. 
The spin-orbit resonance of the planet~b is likely to be $3$:$2$ \citep{Luna2016_HD85512}.
In addition, \citet{Leconte2013_GJ581_HD85512_3Dmodels} investigated the habitability of the planet using 3D global climate models, arguing it cannot be ruled out even if the planet~b is closer than the inner edge of the classically defined habitable zone, which corresponds to a predicted runaway greenhouse transition that remains to be empirically tested \citep{Ingersoll1969,Kasting1988,Turbet2019,Turbet2020,Schlecker2023}.

\begin{figure}[h]
    \centering
    \includegraphics[width=\linewidth]{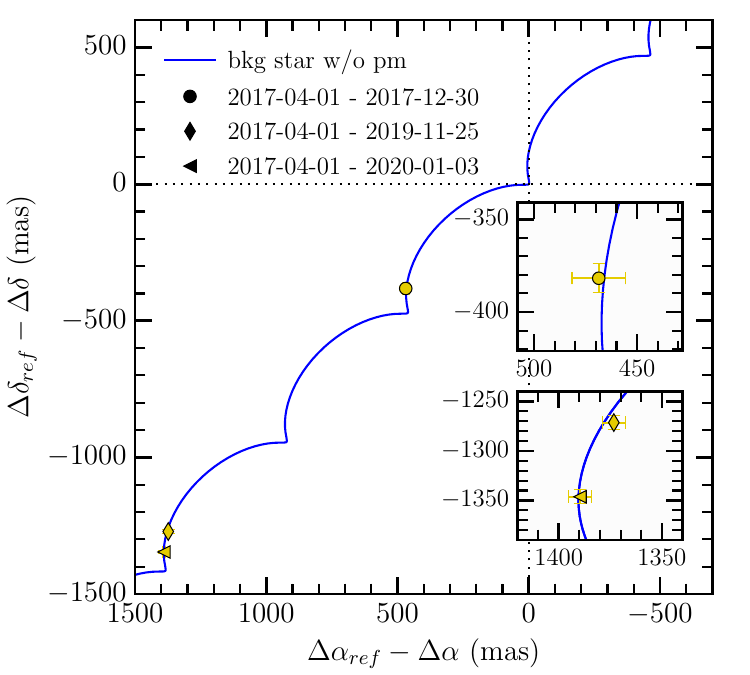}
    \caption{SPHERE/IRDIS relative astrometry differences of the offset positions of all the point-like sources detected around HD~85512 between the first epoch considered as reference 2017-04-01 and the three epochs after, 2017-12-30, 2019-11-25, and 2020-01-03. (See caption from Fig. \ref{fig:PMD_GJ422_2+3epochs})
    }
    \label{fig:PMD_HD85512_4epochs}
\end{figure}

\paragraph{HD~99492}

The system HD~99492 is located at a distance of $18.16\pm0.01~\pc$ \citep{Gaia2021_EDR3}.
We derived an age of $5.7\pm1.0~\gyr$ from gyrochronology and chromospheric activity, which is consistent with \citet{Marcy2005_HD99492_planetb} and \citet{Bonfanti2016_HD20794_HD40307_HD69830_HD3651_HD99492_age}. Its star has a mass of $0.84\pm0.05~\msun$ in agreement with \citet{Takeda2007_age} and a spectral type K2V \citep{Marcy2005_HD99492_planetb}.  We found a super-solar metallicity [Fe/H] of $0.27\pm0.09$.
The star is in a binary orbit with HD~99491, both also known as 83 Leonis B and A, respectively, separated by about $40.76$" corresponding to an average projected separation of $\sim740~\au$.

The system hosts one planet HD~99492~b of minimal mass $35.6~\mearth$ at $0.12~\au$ ($17$-$\dayy$ period) discovered by \citet{Marcy2005_HD99492_planetb}. The hypothetical outer planet HD~99492~c suggested via RV trend \citep{Marcy2005_HD99492_planetb}, is claimed by \citet{Meschiari2011_HD99492_planetc} as a giant planet of $0.36~\mj$ minimal mass orbiting at $5.4\pm0.5~\au$ (about $5000$-$\dayy$ period). However, \citet{Kane2016_HD99492_planetc_no} reported the radial velocity signal associated previously to HD~99492~c to be caused by a stellar activity cycle of about $13~\yrs$.


\paragraph{HD~102365}

The system HD~102365 (also known as HR~4523) is located at a distance of $9.320\pm0.008~\pc$ \citep{Gaia2021_EDR3}.
We derived an age of $4.3\pm0.8~\gyr$ based on chromospheric activity and gyrochronology,  younger than the age estimated from \citet{Takeda2007_age} of $9.5\pm3.0~\gyr$.  
We estimated a stellar mass of $0.81\pm0.05~\msun$, consistent with \citet{Takeda2007_age}, and a sub-solar metallicity [Fe/H] of $-0.35\pm0.03$.
The star is classified as a G3V and G5V star \citep{KeenanMcNeil1989_HD102365_stellar_type,Evans1957_HD102365_stellar_type}. In addition, the star is part of a binary system with a M4V companion located at projected separation of $322~\au$ \citep{VanBiesbrock1961_HD102365_binary,Hawley1996_GJ163_GJ422_GJ433_GJ667_GJ674_GJ682_GJ832_spectral_types_HD102365_binary}.

Based on $149$ UCLES radial velocities measurements spanning over $12~\yrs$, \citet{Tinney2011_HD102365_planetb} reported the existence of HD~102365~b of minimal mass $16.0\pm2.6~\mearth$ orbiting at $0.46\pm0.04~\au$ ($122.1\pm0.3$-$\dayy$ period) with an eccentricity of $0.34\pm0.14$. However, \citet{Zechmeister2013_HD102365_planetb_no} did not find evidence for the signal with their $62$ HARPS RV measurements spanning over $4.4~\yrs$, hence the system  HD~102365  may not be planetary.


\paragraph{61 Vir (HD~115617)}

The system 61~Vir (HD~115617) is located at $8.53\pm0.01~\pc$ \citep{Gaia2021_EDR3}. In this work we estimated the stellar age ($5.1\pm1.3~\gyr$) based on gyrochonology and stellar activity, as well as the stellar mass ($0.915\pm0.053~\msun$) and the metallicity [Fe/H] ($-0.03\pm0.04$). Our results are consistent with previous estimates from \citet{Wright2011_61Vir_age}, \citet{Vican2012_61Vir_HD3651_DD_age}, \citet{Vogt2010_61Vir_RV_3exoplanets_discovery}, \citet{Valenti_Fischer2005_61Vir_mass_HD3651_age}, and \citet{Sousa2008_sample_RV_stellar_properties_mass}.   \\
The star has a spectral type between G5V and G7V \citep[e.g.,][]{Gray2006_spectral_type_HD69830_HD85512_HD40307_HD136352} and its stellar radius is estimated to $0.987\pm0.005~\rsun$ \citep{vonBraun2014_61Vir_GJ176_GJ649_stellar_radius}. By using radial velocity measurements, \citet{Vogt2010_61Vir_RV_3exoplanets_discovery} discovered three planets~b,~c, and~d at minimal masses of $5$, $18$ and $23~\mearth$ and semi-major axes of $0.05$, $0.22$ and $0.49~\au$, respectively.  \citet{Wyatt2012_61Vir_HD20794_HD38858_HD69830_HD_102365_HD136352_Herschel_submm_debris_disc} confirmed the planets~b and~c but not the planet~d, therefore we consider in this work  61~Vir~d as unconfirmed. 

In addition, \citet{Bryden2006_61Vir_disk_discovery_spitzer} discovered with \textit{Spitzer} a debris disk surrounding the star with a fractional luminosity $L_{dust}/L_{star}$ of $2.7\cdot10^{-5}$. By using Herschel, \citet{Wyatt2012_61Vir_HD20794_HD38858_HD69830_HD_102365_HD136352_Herschel_submm_debris_disc} reported that the disk density peaks between $30$ and $100~\au$ (and even up to  $150~\au$ from ALMA observations according to \citet{Marino2017_61Vir_ALMA_general_recent_view}). \citet{Wyatt2012_61Vir_HD20794_HD38858_HD69830_HD_102365_HD136352_Herschel_submm_debris_disc} found that the disk is inclined by $77\degr$ with respect to the plane of the sky which gives a slight underestimation of $3\%$ for the planet masses by assuming the disk and the orbits of these planets to be coplanar.

\paragraph{HD~136352}

The system HD~136352 is located at $14.74\pm0.02~\pc$ \citep{Gaia2021_EDR3} and hosts a G2V star \citep{Gray2006_spectral_type_HD69830_HD85512_HD40307_HD136352}.
We derived an age of $0.57\pm0.17~\gyr$ based on chromospheric activity, which is significantly younger than the age estimated from \citet{Lovis2011_HD69830_HD85512_HD136352_HD154088_HD40307_HD38858_age} or \citet{Kane2020_HD136352_transit_age_mass}, respectively
 $4.37\pm0.5~\gyr$ and $8.3\pm3.2~\gyr$.  
 Our stellar mass estimation of $0.83\pm0.05~\msun$ is consistent with \citet{Delrez2021_HD136352_Transit_CHEOPS}.  We found a sub-solar metallicity [Fe/H] of $-0.37\pm0.03$.

\begin{figure}[h]
    \centering
    \includegraphics[width=\linewidth]{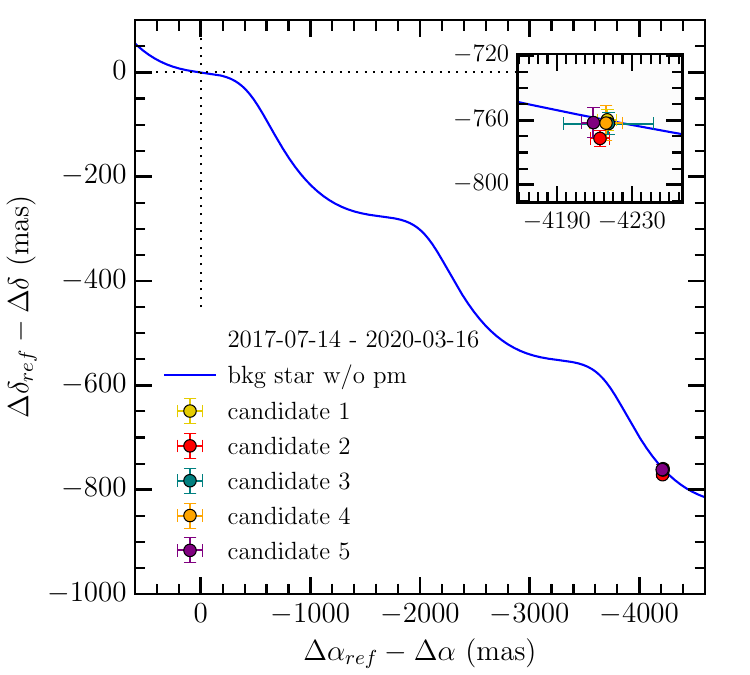}
    \caption{SPHERE/IRDIS relative astrometry differences of the offset positions of all the point-like sources detected around HD~136352 between the first epoch 2017-07-14 considered as reference and the second epoch 2020-03-16. 
    (See caption from Fig. \ref{fig:PMD_GJ422_2+3epochs})
    }
    \label{fig:PMD_HD136352_2epochs}
\end{figure}

The system hosts three planets~b,~c, and~d discovered by \citet{Mayor2011_GJ38858_HD136352_HD154088_HD189567} via RV measurements. 
\citet{Udry2019_HD136352_3planets_RV} reported minimal masses of $4.8\pm0.6~\mearth$,  $10.8\pm1.0~\mearth$, and $8.6\pm1.2~\mearth$, orbiting at about $0.09~\au$, $0.17~\au$, and $0.41~\au$ (period of about $11.6~\days$,  $27.6~\days$, and  $107.6~\days$), respectively. 
In addition, the three planets transit. The two inner ones were first detected via TESS photometry \citep{Kane2020_HD136352_transit_age_mass,Lovos2022_HD136352_Transit_TESS}, while the outer one via CHEOPS \citep{Delrez2021_HD136352_Transit_CHEOPS}.  The updated parameters derived by combining CHEOPS and TESS photometric and HARPS RV measurements do not change significantly from  \citet{Udry2019_HD136352_3planets_RV} but indicate as well an absolute mass of  $4.7\pm0.4~\mearth$,  $11.2\pm0.7~\mearth$, and $8.8\pm0.9~\mearth$, with radii of $1.66\pm0.05~\rearth$, $2.92\pm0.08~\rearth$, and  $2.56\pm0.09~\rearth$, and mean density of  $1.02\pm0.13~\densityearth$,  $0.45\pm0.05~\densityearth$, and $0.52\pm0.08~\densityearth$, respectively  \citep{Delrez2021_HD136352_Transit_CHEOPS}. 
More precisely, \citet{Delrez2021_HD136352_Transit_CHEOPS} indicated that the planet~b is a rocky mostly dry planet unlike the planets c and d that are likely to have small hydrogen-helium envelopes, and a possible water content. They reported no major eccentricity for the three planets (up to $0.25$ at $2\sigma$).


\paragraph{HD~154088}

The system HD~154088 is located at $18.273\pm0.008~\pc$ \citep{Gaia2021_EDR3}.
We derive an age of $7.1\pm0.5~\gyr$ based on chromospheric activity and a stellar mass of $0.94\pm0.05\msun$, both consistent with previous estimates from e.g., \citet{Unger2021_HD154088_HD189567}, who indicated as well the stellar spectral type to be K0IV-V. We estimate a super solar metallicity [Fe/H] of $0.25\pm0.06$. \citet{Fossati2021_HD154088_magnetic_field} reported the detection of the stellar magnetic field.

The system hosts one planet HD~154088~b with a minimal mass of $6.6~\mearth$ at $0.134~\au$ ($18.6$-$\dayy$ period) discovered by \citet{Mayor2011_GJ38858_HD136352_HD154088_HD189567} and confirmed by \citet{Unger2021_HD154088_HD189567}. The latter constrain the eccentricity below $0.19$  ($0.34$) at the  $68\%$ ($95\%$) level.

\begin{figure}[h]
    \centering
    \includegraphics[width=\linewidth]{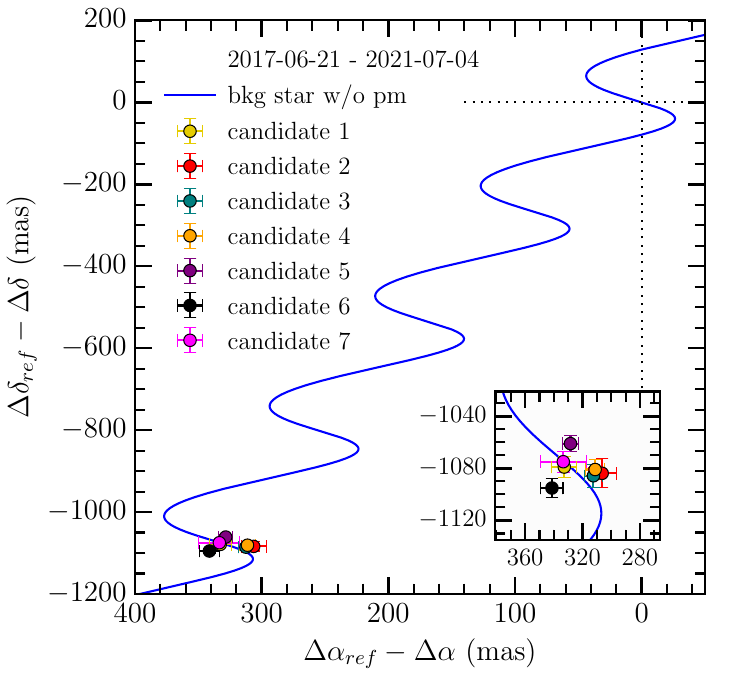}
    \caption{SPHERE/IRDIS relative astrometry differences of the offset positions of all the point-like sources detected around HD~154088 
    in the first and third epochs (2017-06-21, 2021-07-04). 
    (See caption from Fig. \ref{fig:PMD_GJ422_2+3epochs})
    }
    \label{fig:PMD_HD154088_2+3epochs}
\end{figure}

\paragraph{HD~189567}

The system HD~189567 is located at $17.93\pm0.01~\pc$ \citep{Gaia2021_EDR3}. Based on chromospheric activity, we derive an age of $4.5\pm0.5~\gyr$, younger than \citet[$9.8\pm2.1~\gyr$][]{Takeda2007_age}. We estimate a stellar mass of $0.85\pm0.05~\msun$ consistent with literature, and a metallicity [Fe/H] of $-0.27\pm0.05$. The stellar type is a G2V \citep{Gray2006_spectral_type_HD69830_HD85512_HD40307_HD136352}.

The system hosts two planets, HD~189567~b and  HD~189567~c discovered by \citet{Mayor2011_GJ38858_HD136352_HD154088_HD189567} and \citet{Unger2021_HD154088_HD189567}, respectively. \citet{Unger2021_HD154088_HD189567} derived for the planets~b and~c a minimal mass of $8.5\pm0.6~\mearth$ and  $7.0\pm0.9~\mearth$ at a semi-major axis of about $0.11~\au$ and  $0.20~\au$ (period of $14.2~\days$ and $33.7~\days$), respectively. 

\newpage

\section{Age estimation \label{app:age}}


The stellar age is an important parameter in high-contrast imaging observations to convert observed fluxes into masses of putative companion candidates, as well as the sensitivity in terms of mass of given observation. For our sample, ages from the literature were heterogeneously derived, and sometimes poorly constrained or disagreeing with each other. For these reasons, we have revisited in this work the ages of all the stars based on two different methods: gyrochronology and stellar activity.

\subsection{Gyrochronology}

Gyrochronology uses the empirical rotation period-color relation to estimate the age. This technique is calibrated by using a few age-benchmark open clusters, such as the Pleiades, the Hyades or Praesepe, can be also  used to estimate the age of individual stars, possibly complementary to other independent techniques. In the present study, considering the wide color range spanned by our targets (0.64 $<$ B$-$V $<$ 1.65 mag), we had to use three different gyrochronologic relations to infer the age.

In the color range 0.5 $<$ B$-$V $<$ 1.40 mag, corresponding to the F5--K9 spectral types,
we have used the relation:\\
\begin{equation}
    Age (Gyr) =\sqrt[1/n]{ \frac{P(d)} {a(B-V-c)^b}}\,,
\end{equation}
using  the coefficients $a=0.407$, $b=0.325$, $c=0.495$, and $n=0.566$ \citep{Mamajek08} and  $a=0.40$, $b=0.31$, $c=0.45$, and $n=0.55$  \citep{Angus15}. \\

For the M0--M1-type stars in our sample, we used the relation of \citet{Engle11}: \\
\begin{equation}
    Age (Gyr) =\sqrt[1/c]{ \frac{P (d) -a}{b}}\,,
\end{equation}
with $a=  -4.14$, $b = 15.57$ and $c = 0.688$,\\

As for the M2.5--M5.5-type stars, we used the linear relation of \citet{Engle11}:\\
\begin{equation}
    Age (Gyr) =\frac{P (d) -a}{b}\,,
\end{equation}
 with $a = 0.88$ and $b = 16.542$.\\
 
As the input rotation period for each target, we used the average value if more periods were available from the literature (see Table\,\ref{tab-age}). By exploring the existing literature on our stellar sample to gather the values of their stellar rotation periods, we found rotation periods (P) measured by using periodogram analysis of three different types of data time series: photometric  ($P_{\rm ph}$), activity index  ($P_{\rm ac}$), and radial velocity  ($P_{\rm RV}$).
In Table\,\ref{tab-data}, we list  the B$-$V color and the photometric rotation period $P_{\rm ph}$, together with its uncertainty $\sigma_P$, when known, and the corresponding references for each target.

On a total of $27$ targets, we retrieved information on the photometric rotation period $P_{\rm ph}$ for ten targets. The main reason for this low detection rate is that our targeted stars are relatively old and are expected to have very low-level photometric variability, below the current sensitivity of ground or space-based telescopes. 

We retrieved information on the rotation period $P_{\rm ac}$ from the activity index time series for $13$ targets, six of which with $P_{\rm ph}$ known, which are listed in Table\,\ref{tab-data} together with their respective references. The activity index mostly measured for the selected targets is $R^{\prime}_{HK}$. The main reason is again linked to the old age of our targeted stars, which translates into long rotation periods, combined with the scarcity of spectroscopic time series long enough to allow  a meaningful measurement of rotation. 

Finally, for seven targets we found rotation period measurements $P_{\rm RV}$ from RV time series (see Table\,\ref{tab-data}). In fact, the periodic modulation of visibility of surface magnetic inhomogeneities gives rise to quasi-periodic jitters in the RV measurements, which allow the stellar rotation period to be measured.
Out of seven targets with $P_{\rm RV}$ known,  four also have  $P_{\rm ph}$ and $P_{\rm ac}$ both known, and two have only $P_{\rm ac}$  known.

To summarize, we have the rotation period required for an estimation of the age via gyrochronology for a total of $17$ targets. We found that rotation periods measured by different methods (when available) are generally in good agreement with each other. 

\subsection{Stellar activity}

The second approach to measure the age of our targets is the use of calibrated age-activity relations. In this work, we used the relation by \citet{Mamajek08}
which is calibrated in the color range $0.5~<~\textrm{B}-\textrm{V}~<~1.40~\magg$:
\begin{equation}
\log{Age}(Gyr) = -38.053 -17.912\log{R^{\prime}_{HK}}-1.6675\log{(R^{\prime}_{HK})^2}\,,
\end{equation}
where  R$^{\prime}_{HK}$ is the chromospheric activity index.

We have explored the existing literature on the selected targets to gather measurements of magnetic activity indices. As mentioned, the most widely measured index is  the chromospheric activity index R$^{\prime}_{HK}$, which is the width of the activity sensitive Ca\,II H\&K lines normalized by the bolometric flux and corrected for a photospheric contribution.  
Except the single measurement for GJ\,3998 and GJ\,667C, all the other targets have numerous values of R$^{\prime}_{HK}$.
In Table\,\ref{tab-data}, we list for each target the R$^{\prime}_{HK}$ values and the corresponding references, and in Table\,\ref{tab-age} the average value.

 The age based on stellar activity was derived for $15$ targets, whose color is within the validity of the  \citet{Mamajek08} relation. We stress that the age of HD~136352 is likely to be underestimated by our method as our age estimation gives a rather young star of $0.57\pm0.17~\gyr$, which is discrepent by $3\sigma$ with the age from the literature ($8.2^{+3.2}_{-3.1}~\gyr$), and may not be consistent with its subsolar metallicity (see Table \ref{tab:stellar_feh} and Appendix \ref{app:metallicity}). 

Both chromospheric and gyrochronologic ages were inferred for a subsample of 7 targets (without considering GJ~581). The gyrochronologic ages inferred by \citet{Mamajek08} and \citet{Angus15} are generally in agreement within 20\%, while chromospheric and gyrochronologic ages are also in agreement within 30\%. 

\begin{figure}[h!]
    \centering
    \includegraphics[width=\linewidth]{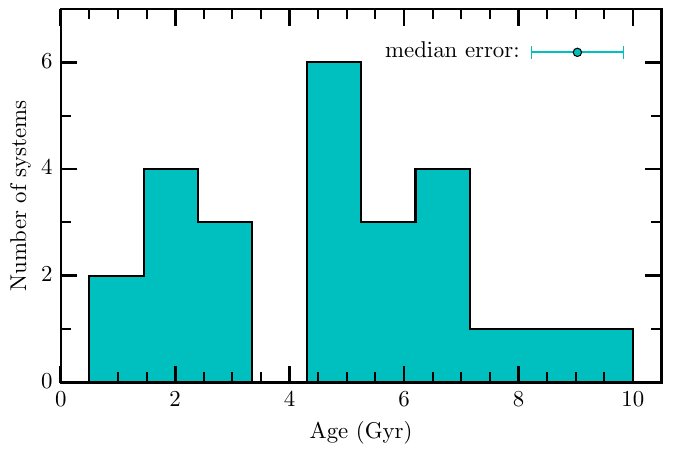}
    \caption{Distribution of the age of the stars in our sample re-determined homogeneously in this work (see values in Table~\ref{tab-age}). We indicate the median error on the age in the upper right corner.  
    }
    \label{fig:hist_age}
\end{figure}

 As final age for each system, we adopt either the gyrochronologic age, the chromospheric age, or the average value in the cases with independent realistic measurements (see Table \ref{tab-age} and Fig. \ref{fig:hist_age}). By using our methods, we are not able to estimate robust ages for GJ~180, GJ~682 or HD~136352. Regarding HD~136352, we use the age of $8.3\pm3.2~\gyr$ from \citet{Kane2020_HD136352_transit_age_mass}, while considering  GJ~180 and GJ~682 we use a default age of $5\pm4~\gyr$. Except GJ~433 ($0.50\pm0.15~\gyr$), all of our systems have ages between $1~\gyr$ and $10~\gyr$, with uncertainties up to a few gigayears.  

\onecolumn 
\begin{table*}[h!]\centering
\caption{Average rotation period and activity and corresponding activity and gyrochronologic ages (according to M: \citet{Mamajek08}; A: \citet{Angus15}; S: \citet{Engle11}. 
} \label{tab-age}
\normalsize
\begin{tabular}{lCCCCCCCC}

\hline\hline \noalign{\smallskip}
System &	\text{B}-\text{V} & <P_\text{rot}> & <R^{\prime}_{HK}> 	&	 \text{Age}_{Chromo} & \multicolumn{3}{C}{ \text{Age}_{Gyr0}} & <\text{Age}>\\
	 & \text{(mag)} & \text{(d)}  & & \text{(Gyr)} &  \multicolumn{3}{c}{ \text{(Gyr)}} \\
	 &&&&&  \text{M} &  \text{A} &  \text{S} & \text{this work} \\
	 \noalign{\smallskip}\hline\hline\noalign{\smallskip}
GJ\,163&	1.49	&50&		-5.540\pm0.084 &  &	&& 3.0 & 3.0\pm0.7 \\
GJ\,176 &	1.54& 39.5\pm1.3 & -4.955\pm0.063 & &		& & 2.3 & 2.3\pm0.6 \\
GJ\,180	&1.55& & -5.341\pm0.082 &&&&  & \\
BD-061339 &1.35& & -4.778\pm0.096 &2.9\pm1.2 & && & 2.9\pm1.2 \\
GJ\,229 &	1.48	&27.2\pm1.6	&-4.861\pm0.131& && & 1.6 & 1.6\pm0.4 \\
GJ\,422 &	1.33	&  & -5.800\pm0.044 & 5.5\pm0.8 &&&  & 5.5\pm0.8\\
GJ\,433 &	1.51&	7.94	&-5.284\pm0.192 & & && 0.50 & 0.50\pm0.15\\
GJ\,581	&1.20 & 130\pm2 & -5.459\pm0.219 & 10\pm1 & & & 7.8 & 8.9\pm1.1\\
Wolf\,1061 & 1.57&	119.1\pm1 & -5.219\pm0.209 & 	& & & 7.1 & 7.1\pm2 \\
GJ\,649 &	1.48	& 23.8\pm0.1 &-4.944\pm0.101 &  & & & 1.40 & 1.40\pm0.35\\
GJ\,667C &1.57	& 105  & -5.33 & &  & & 6.3 & 6.3\pm2 \\
GJ\,674 &	1.57	&34.1\pm1.3 & 	-5.013\pm0.086 & & & & 2.0 &  2.0\pm0.5 \\
GJ\,682	&1.65&	 & -5.363\pm0.342 & & & & \\
GJ\,832 &	1.50	& 41\pm4 & -5.215\pm0.085 & & & & 2.4 & 2.4\pm0.3 \\
GJ\,3998 &	1.51 &	32.99 	 & -4.82  & & & & 2 & 2\pm0.5 \\
HD\,3651&	0.83	&43\pm2 & -4.979\pm0.09 & 6.2\pm1.7  	&	7\pm1&8.5 & & 7.5\pm2.2 \\
HD\,20794	& 0.71	 &  &-4.971\pm0.017 & 6\pm0.3 &&& &  6\pm0.3 \\
HD\,38858	&0.64	& 24  & -4.921\pm0.043 & 5.1\pm0.8  &	4.1\pm0.5&4.3& & 4.3\pm1 \\
HD\,40307	&0.95&	34.4\pm2.4 & -4.980\pm0.144 & 	6\pm2  	&	4\pm0.4&4.8 & & 4.4\pm2 \\
HD\,69830	&0.79& 42  & -4.963\pm0.029 & 5.9\pm0.5  &	7.3\pm1&8.7& & 6.6\pm1.5 \\
HD\,85512 &	1.18	& 42\pm6 & -4.953\pm0.060 &	5.7\pm1  &	4.5\pm0.3&5.6& & 5\pm1 \\
HD\,99492 &1.02&  & -4.953\pm0.055 & 5.7\pm1.0  	&&&&5.7\pm1.0 \\
HD\,102365 & 0.67 & 24  &  -4.931\pm0.026 & 5.3\pm0.5 	&	3.6\pm0.5&4.0& & 4.3\pm0.8 \\
61\,Vir	& 0.70	&  29  & -4.973\pm0.056	& 6\pm1	&	4.7\pm0.6&5.3& & 5.1\pm1.3 \\
HD\,136352  & 0.65	&  & -4.941\pm0.044 & 0.57\pm0.17 & && & 0.57\pm0.17 \\
HD\,154088 & 0.83	 & & -5.026\pm0.027 & 7.1\pm0.5 & && & 7.1\pm0.5 \\
HD\,189567 &  0.64 &  &-4.885\pm0.028 & 4.5\pm0.5 & && & 4.5\pm0.5 \\
\noalign{\smallskip}\hline\hline
\end{tabular}
\end{table*}

\begin{longtable}[p]{l@{\hspace{.5cm}}|l@{\hspace{.5cm}}|c@{\hspace{.5cm}}c@{\hspace{.5cm}}c@{\hspace{.5cm}}|c@{\hspace{.5cm}}l@{\hspace{.5cm}}l@{\hspace{.5cm}}|l@{\hspace{.5cm}}l@{\hspace{.5cm}}l@{\hspace{.5cm}}|l@{\hspace{.5cm}}l@{\hspace{.5cm}}}
\caption{Information from the literature on the stellar rotation period and activity as measured by means of R$^{\prime}_{HK}$}\label{tab-data}\\%
\hline\hline\noalign{\smallskip} 
Name &	B$-$V & $P_{ph}$ &	$\sigma$ &	ref &	$P_{ac}$ &	$\sigma$ &	ref &	 $P_{RV}$ &	$\sigma$ &	ref &	$R^{\prime}_{HK}$ &	ref	\\
\noalign{\smallskip}\hline\hline\noalign{\smallskip}
\endfirsthead
\caption{continued.}\\%
\hline
\hline \noalign{\smallskip}
Name &	B$-$V & $P_{ph}$ &	$\sigma$ &	ref &	$P_{ac}$ &	$\sigma$ &	ref &	 $P_{RV}$ &	$\sigma$ &	ref &	$R^{\prime}_{HK}$ &	ref	\\ \noalign{\smallskip}\hline\hline\noalign{\smallskip}
\endhead
\noalign{\smallskip}\hline\hline\noalign{\smallskip}
\endfoot

\endlastfoot
GJ\,163&	1.49	&50&	---&	P8	&--- &--- &--- &--- &--- &	---&					-5.480&	P5	\\
& ---	& --- & --- &  --- &	--- &	---	& --- & --- & --- & ---&-5.600	&A7\\
\noalign{\smallskip}\hline\noalign{\smallskip}
GJ\,176&	1.54&	38.92&	---	&P3	&37&	---	&P5&	39.2&	0.2&	P2	&-4.911	&P5	\\
	& ---&	40.7	&0.4&	P6	&40.6	&---	&P7 &--- &--- &	---&			-5.00	&P2\\
	&---&	40.8	&0.1	&P4	&39.3	&0.1	&P1&--- &--- &---  &--- &--- \\
\noalign{\smallskip}\hline\noalign{\smallskip}
GJ180	&1.55&--- &--- &--- &--- &--- &--- &--- &--- & ---&										-5.283&	P5	\\
& ---	& --- & --- &  --- &	--- &	---	& --- & --- & --- & ---&-5.40	&A6\\
\noalign{\smallskip}\hline\noalign{\smallskip}
GJ229&	1.48	&27.7&	0.2	&P6	&25	&---	&P5	&29.4&	0.1&	P2	&-4.90	&A6	\\
	&---&	27.3	&0.1&	P4	&26.7&	2.4	&P2	&--- &--- &---  	&		-4.669&	P5\\
& ---	& --- & --- &  --- &	--- &	---	& --- & --- & --- & ---&-4.965	&A7\\
& ---	& --- & --- &  --- &	--- &	---	& --- & --- & --- & ---&-4.91&	P2\\
\noalign{\smallskip}\hline\noalign{\smallskip}
BD-061339 &1.35&--- &--- &--- &--- &--- &--- &--- &--- &	---&									-4.846&	A7	\\
& ---	& --- & --- &  --- &	--- &	---	& --- & --- & --- & ---&-4.710&	A17 \\
\noalign{\smallskip}\hline\noalign{\smallskip}
GJ422 &	1.33	&		--- 	&--- &--- &--- &--- &--- &--- &--- &--- &						-5.769	&P5 \\
& ---	& --- & --- &  --- &	--- &	---	& --- & --- & --- & ---&	-5.832 &	A7 \\
\noalign{\smallskip}\hline\noalign{\smallskip}
GJ433 &	1.51&	7.94&	---&	P8	& --- & --- & --- & --- & --- & --- &						-5.148&	P5	\\
& ---	& --- & --- &  --- &	--- &	---	& --- & --- & --- & ---&-5.420&	A7 \\
\noalign{\smallskip}\hline\noalign{\smallskip}
GJ581	&1.20 & --- & --- & --- &				130	&2	&	P5 	& --- &---  &	--- &	-5.776 &	P5\\
	&--- &---  &---  &---  &								132.5 &	6.3&	P1& --- & --- &	--- 	&		-5.436&	A17 \\
& ---	& --- & --- &  --- &	--- &	---	& --- & --- & --- & ---&-5.321 &	A17 \\
& ---	& --- & --- &  --- &	--- &	---	& --- & --- & --- & ---&-5.304 &	A17 \\
\noalign{\smallskip}\hline\noalign{\smallskip}
Wolf1061 & 1.57&	119&	1	&P6	&119	 &---	&P5  &---  &---  &		--- &		-5.523&	P5	\\
	&--- &	119.3&	---&	P4	& --- &--- &--- &--- &--- &--- &						-5.088&	A17\\
& ---	& --- & --- &  --- &	--- &	---	& --- & --- & --- & ---&-5.075	&A17 \\
& ---	& --- & --- &  --- &	--- &	---	& --- & --- & --- & ---&-5.190	&A17\\
\noalign{\smallskip}\hline\noalign{\smallskip}
GJ649 &	1.48	& 23.8	&0.1 &	P6	&--- &--- &--- &--- &--- &--- 	&					-4.872&	A8	\\
	&---	& 23.9	&---	&P8	&--- &--- &--- &--- &--- &--- 				&		-5.016	&A7 \\
\noalign{\smallskip}\hline\noalign{\smallskip}
GJ667C &1.57	&---  &---  &---  &---  &---  &---  &						105	&---	&P9&	-5.334&	A17		\\
\noalign{\smallskip}\hline\noalign{\smallskip}
GJ674 &	1.57	&33.29&	---	&P3	&35&	---&	P5	&36.2	&0.1&	P2	&-5.05	&A6		\\
	&---  &	35&	1&	P4&	32.9	&0.1&	P1&---  & --- & --- &				-4.885&	P5\\
	& --- &--- 		& --- &--- 	&	33.0	&0.7&	P2	&--- &--- &--- &			-5.077&	A7\\
& ---	& --- & --- &  --- &	--- &	---	& --- & --- & --- & ---&-5.04	 & P2 \\
\noalign{\smallskip}\hline\noalign{\smallskip}
GJ682	&1.65&--- 	&		--- 	&--- &--- &--- &--- &--- &--- &--- &					-4.99	&A6	\\
& ---	& --- & --- &  --- &	--- &	---	& --- & --- & --- & ---&	-5.438&	P5\\
& ---	& --- & --- &  --- &	--- &	---	& --- & --- & --- & ---&	-5.662	&A7\\
\noalign{\smallskip}\hline\noalign{\smallskip}
GJ832 &	1.50	& --- & --- & --- &			55	&---	&P5	&39.1&	0.1&	P2&	-5.23	&A6		\\
	 & --- & --- & --- & --- &				45.7	&9.3&	P1	& --- & --- & --- &			-5.10	&A4\\
	 & --- & --- & --- & --- &				39.2 &	9.4&	P2		&---  & --- & --- &				-5.182&	P5 \\
& ---	& --- & --- &  --- &	--- &	---	& --- & --- & --- & ---&-5.336&	A7\\
& ---	& --- & --- &  --- &	--- &	---	& --- & --- & --- & ---&	-5.23	&P2\\
\noalign{\smallskip}\hline\noalign{\smallskip}
GJ3998 &	1.51 &	32.99 &	--- &	P15	& --- &---  &---  &---  &---  &---  &						-4.82&	P5	\\
\noalign{\smallskip}\hline\noalign{\smallskip}
HD3651&	0.83	&41&	---	&P8	&43.4&	---&	P11	&--- &--- &	--- 	&	-4.991*&	A8	\\
		& --- &---  &---  &	---&		44&	0.4&	P10&---  &--- &---  		&	-5.020&	A9\\
		& --- & --- & --- &	---&		44&	---&	P12& --- &--- &---  			&	-4.85&	A10\\
& ---	& --- & --- &  --- &	--- &	---	& --- & --- & --- & ---&-5.02 &	A3\\
& ---	& --- & --- &  --- &	--- &	---	& --- & --- & --- & 	&-4.823&	A13\\
& ---	& --- & --- &  --- &	--- &	---	& --- & --- & --- & ---&-5.087&	A12\\
& ---	& --- & --- &  --- &	--- &	---	& --- & --- & --- & ---&-5.00&	P10\\
& ---	& --- & --- &  --- &	--- &	---	& --- & --- & --- & ---&-5.04	&A11\\
\noalign{\smallskip}\hline\noalign{\smallskip}
 HD20794	& 0.71		&--- &--- &--- &--- &--- &--- &--- &--- &--- &					-4.98	 & A10 \\
& ---	& --- & --- &  --- &	--- &	---	& --- & --- & --- & ---&-4.981* &	A1 \\
& ---	& --- & --- &  --- &	--- &	---	& --- & --- & --- & ---&-4.946*&	A2 \\
& ---	& --- & --- &  --- &	--- &	---	& --- & --- & --- & ---&-4.98	&A3 \\
\noalign{\smallskip}\hline\noalign{\smallskip}
HD38858	&0.64	&	--- &--- &	---&	24&	---	&P13	&--- &--- &--- 	&		-4.95&	A9	\\
& ---	& --- & --- &  --- &	--- &	---	& --- & --- & --- & ---&-4.918*&	A1\\
& ---	& --- & --- &  --- &	--- &	---	& --- & --- & --- & ---&-4.855*&	A2\\
& ---	& --- & --- &  --- &	--- &	---	& --- & --- & --- & ---&-4.968&	A12\\
& ---	& --- & --- &  --- &	--- &	---	& --- & --- & --- & ---&-4.89	&A14\\
& ---	& --- & --- &  --- &	--- &	---	& --- & --- & --- & ---&-4.950	&A11\\
\noalign{\smallskip}\hline\noalign{\smallskip}
HD40307	&0.95&	31.8&	3.7&	P1&	36.5&	2.3&	P2&	35.0	&0.1&	P2	&-4.949*	&A1\\
& ---	& --- & --- &  --- &	--- &	---	& --- & --- & --- & ---&-4.934*&	A2\\
& ---	& --- & --- &  --- &	--- &	---	& --- & --- & --- & ---&-4.99	&A3\\
& ---	& --- & --- &  --- &	--- &	---	& --- & --- & --- & ---&-4.83	&A4\\
& ---	& --- & --- &  --- &	--- &	---	& --- & --- & --- & ---&-4.85	&A4\\
& ---	& --- & --- &  --- &	--- &	---	& --- & --- & --- & ---&-5.259&	A5\\
& ---	& --- & --- &  --- &	--- &	---	& --- & --- & --- &---	&	-5.05	&P2\\
\noalign{\smallskip}\hline\noalign{\smallskip}
HD69830	&0.79&--- &--- &--- &				42&	---&	P13	&--- &--- &--- &			-4.95	&A9	\\
& ---	& --- & --- &  --- &	--- &	---	& --- & --- & --- & ---&-4.991	&A12\\
& ---	& --- & --- &  --- &	--- &	---	& --- & --- & --- & ---&-4.971	&A12\\
& ---	& --- & --- &  --- &	--- &	---	& --- & --- & --- & ---&-4.94	&A6\\
& ---	& --- & --- &  --- &	--- &	---	& --- & --- & --- & ---&-4.92	&A3\\
& ---	& --- & --- &  --- &	--- &	---	& --- & --- & --- & ---&-5.005*&	A15\\
& ---	& --- & --- &  --- &	--- &	---	& --- & --- & --- & ---&-4.945&	A5\\
& ---	& --- & --- &  --- &	--- &	---	& --- & --- & --- & ---&-4.98	&A11\\
\noalign{\smallskip}\hline\noalign{\smallskip}
HD85512 &	1.18	&--- &--- &--- 	&		45.9&	0.4&	P1	&34.8&	0.1	&P2	&-4.905*	&A1\\
		&--- &--- &--- &--- 	&		45.8	&5.2	&P2		&--- &--- &--- &	-4.90&	A3\\
& ---	& --- & --- &  --- &	--- &	---	& --- & --- & --- & ---&-4.96	&A4\\
& ---	& --- & --- &  --- &	--- &	---	& --- & --- & --- & ---&		-4.95	&P2\\
\noalign{\smallskip}\hline\noalign{\smallskip}
HD99492 & 1.02& --- &  --- &    --- &   --- & --- & --- & --- & ---&-4.93&--- &	A3\\
& ---	& --- & --- &  --- &	--- &	---	& --- & --- & --- & ---&-4.889	&A17\\
& ---	& --- & --- &  --- &	--- &	---	& --- & --- & --- & ---&-5.015	&A17\\
& ---	& --- & --- &  --- &	--- &	---	& --- & --- & --- & ---&-4.978	&A17\\
\noalign{\smallskip}\hline\noalign{\smallskip}
HD102365 & 0.67 &---  &---  &---  &	24	&---  &	P14	&---  &---  &---  &			-4.957 &	A12 \\
& ---	& --- & --- &  --- &	--- &	---	& --- & --- & --- & ---&-4.95 &	A16 \\
& ---	& --- & --- &  --- &	--- &	---	& --- & --- & --- & ---&-4.944* &A1 \\
& ---	& --- & --- &  --- &	--- &	---	& --- & --- & --- & ---&-4.916&	A2 \\
& ---	& --- & --- &  --- &	--- &	---	& --- & --- & --- & ---&-4.88	&A6 \\
& ---	& --- & --- &  --- &	--- &	---	& --- & --- & --- & ---&-4.93	&A3 \\
& ---	& --- & --- &  --- &	--- &	---	& --- & --- & --- & ---&-4.942&	A15 \\
\noalign{\smallskip}\hline\noalign{\smallskip}
61Vir &	0.70 & ---	& --- & --- &  29 &	--- &	P12	& --- & --- & --- & -5.001*	& A8	\\
& ---	& --- & --- &  --- &	--- &	---	& --- & --- & --- & ---&	 -5.011	& A12 \\
& ---	& --- & --- &  --- &	--- &	---	& --- & --- & --- & ---&	 	-5.04	& A9 \\
& ---	& --- & --- &  --- &	--- &	---	& --- & --- & --- & ---&	 	-5.00	& A10 \\
& ---	& --- & --- &  --- &	--- &	---	& --- & --- & --- & ---&	 	-4.87	& A16 \\
& ---	& --- & --- &  --- &	--- &	---	& --- & --- & --- & ---&	 	-4.99*& 	A1 \\
& ---	& --- & --- &  --- &	--- &	---	& --- & --- & --- & ---&	 	-4.967*	& A2 \\
& ---	& --- & --- &  --- &	--- &	---	& --- & --- & --- & ---&	 	-4.91& 	A6 \\
\noalign{\smallskip}\hline\noalign{\smallskip}
HD136352  & 0.65	&--- &--- &--- &--- &--- &--- &--- &--- &--- &									-4.91	&A16\\
& ---	& --- & --- &  --- &	--- &	---	& --- & --- & --- & ---&-5.013	&A12\\
& ---	& --- & --- &  --- &	--- &	---	& --- & --- & --- & ---&-4.949*	&A1\\
& ---	& --- & --- &  --- &	--- &	---	& --- & --- & --- & ---&-4.932*	&A2\\
& ---	& --- & --- &  --- &	--- &	---	& --- & --- & --- & ---&--4.90	&A4\\
\noalign{\smallskip}\hline\noalign{\smallskip}
HD154088 & 0.83	&---  &--- &--- &--- &--- &--- &--- &--- &--- &									-5.02	&A9	\\
& ---	& --- & --- &  --- &	--- &	---	& --- & --- & --- & ---&-5.00	& A10 \\
& ---	& --- & --- &  --- &	--- &	---	& --- & --- & --- & ---&-5.064*&	A1\\
& ---	& --- & --- &  --- &	--- &	---	& --- & --- & --- & ---&-5.02&	A6\\
\noalign{\smallskip}\hline\noalign{\smallskip}
HD189567 &  0.64		&--- &--- &--- &--- &--- &--- &--- &--- &--- &				-4.86 &	A10	\\
& ---	& --- & --- &  --- &	--- &	---	& --- & --- & --- & ---&-4.916*	&A1 \\
& ---	& --- & --- &  --- &	--- &	---	& --- & --- & --- & ---&-4.878*	&A2 \\
\noalign{\smallskip}\hline\hline\noalign{\smallskip}
\multicolumn{12}{l}{* average value}
\end{longtable}
\tablebib{References given in the columns “ref” are the following: P1 = \citet{Suarez15}, P2 = \citet{Suarez17}, P3 = \citet{KiragaStepien2007}, P4 = \citet{Suarez2016_age}, P5 = \citet{AstudilloDefru2017_GJ176_GJ581_GJ674_activity_mass}, P6 = \citet{Diez-Alonso2019_age}, P7 = \citet{Trifonov2018_GJ176_GJ581_rv_1planet}, P8 = \citet{Oelkers2018_age}, P9 = \citet{Anglada-Escude2013_GJ667C_3planets_habitable_zone}, P10 = \citet{Marsden2014_age}, P11 = \citet{Vidotto2014_age_HD3651}, P12 = \citet{Baliunas1996_age_HD3651_61Vir}, P13 = \citet{Isaacson2010_age}, P14 = \citet{Saar1997_age}, P15 = \citet{Giacobbe2020_age}, A1 = \citet{GomesdaSilva2014_age}, A2 = \citet{Meunier2017_age_FGK_stars}, A3 = \citet{Miller2015_age}, A4 = \citet{Jenkins2006_age}, A5 = \citet{Gondoin2020_age}, A6 = \citet{Hojjatpanah2019_age}, A7 = \citet{Houdebine2017_age}, A8 = \citet{Duncan1991_age}, A9 = \citet{Wright2004_age}, A10 = \citet{RochaPinto2004_age}, A11 = \citet{Katsova2011_age}, A12 = \citet{Gray2003_spectral_type_HD3651_HD38858}, A13 = \citet{Herrero2012_age}, A14 = \citet{Radick2018_age}, A15 = \citet{Zechmeister2013_HD102365_planetb_no}, A16 = \citet{Henry1996_age}, and A17 = \citet{BoroSaikia2018_age}.}
\onecolumn 
\twocolumn

\onecolumn 



\newpage

\twocolumn
\section{Stellar metallicity derivation \label{app:metallicity}}


A correlation between the presence of giant planet and the metallicity of the host star is well known \citep{FischerValenti2003_obs_metallicity}. That is why we determine homogeneously the metallicity of all the GK stars in our sample, along with their atmospheric properties.

We exploited archival high-resolution, high signal-to-noise ratio (S/N) spectra for our GK stars to derive atmospheric parameters and metallicity ([Fe/H]). Given the dominance of molecular bands, M-type dwarfs require a dedicated approach and cannot be analyzed in the same way \cite[see][and references therein]{Maldonado2020_GJ176_GJ433_GJ581_GJ667C_GJ674_GJ682_GJ832_Wolf1061}. 
Our M dwarf targets have been extensively investigated through the literature in terms of their atmospheric parameters and chemical composition. However, due to the intrinsic issues related to their analysis, different studies have provided very different results (see e.g., the exemplary case of GJ\,176 with differences of 600 K in \teff, 0.6 dex in \logg \, and 0.6 dex in metallicity [Fe/H]; \citealt{santos2013, passegger2019}). Thus, we do not discuss their composition because large systematic uncertainties plague the comparison with GK stars. 

For stars warmer than $\teff\gtrsim5\,000~\kelvin$ we obtained spectroscopic parameters by performing a differential analysis using the MOOG (\citealt{1973Sneden}; 2019 version) python wrapper $q^2$ \citep{Ramirez2014} and the \texttt{marcs} grid of model atmospheres \citep{2008gustafsson}. The Fe~{\sc i}/Fe~{\sc ii} line list comes from a compilation published in our recent works (sources for atomic parameters of the $90$ iron transitions can be found in \citealt{2020Dorazi, 2020Baratella}). The equivalent widths (EWs) have been measured with \texttt{ares} \citep{2015Sousa} and carefully checked trough the IRAF\footnote{IRAF is distributed by NOAO, which is operated by AURA under a cooperative agreement with the NSF.} task $splot$. We derived stellar parameters \teff, and \logg~by imposing excitation and ionization equilibrium, respectively; microturbulence velocities ($V_t$) were gathered by removing spurious trends between the reduced EWs of Fe~{\sc i} lines and the corresponding abundances.  
We adopted $\log(n(\text{Fe}))_\odot$ = 7.50 dex as a reference solar abundance \citep{2009asplund}.
For this sample, we also determined abundances for  Na~{\sc i} (non-LTE corrections have been applied following \citealt{2011lind}), Mg~{\sc i}, Si~{\sc i}, Ca~{\sc i}, Ti~{\sc i}, Ti~{\sc ii}, Cr~{\sc i}, Cr~{\sc ii}, and Ni~{\sc i}, adopting the same model atmosphere and code that is described above. As for Fe, we assumed solar reference abundances from \cite{2009asplund}.
Results are reported in Table \ref{tab:elements}.

Errors on the stellar parameters, metallicity [Fe/H], and elemental abundances have been calculated in the standard way: they include internal errors due to the EW measurements and the propagation of uncertainties related to the atmospheric parameters. We refer the reader to our previous investigations, e.g., \citet{2020Dorazi} and \citet{2020Baratella}.
Regarding the four cooler stars of this GK-type sub-sample (namely GJ~667C, HD~40307, HD~85512, and BD-061339), we instead fixed atmospheric parameters from photometry: \teff~are from \textit{Gaia} (DR2 or eDR3 depending on availability for each target) and 2MASS~JHK colors using the relationship by \cite{2021Casagrande}. We exploited the \texttt{colte} python program developed and maintained by L. Casagrande\footnote{\url{https://github.com/casaluca/colte}} in order to obtain average \teff~with weighted errors calculated through Monte Carlo method. Surface gravities have been calculated starting from those \teff, masses as given in Table \ref{tab:target_properties}, \textit{Gaia} parallaxes and bolometric correction for K band photometry (\citealt{2014Casagrande}, \citeyear{2018Casagrande}). We then measured EWs of neutral iron lines and found the [Fe/H] values as reported in Table~\ref{tab:stellar_feh}. 

\begin{figure}[h!]
    \centering
    \includegraphics[width=\linewidth]{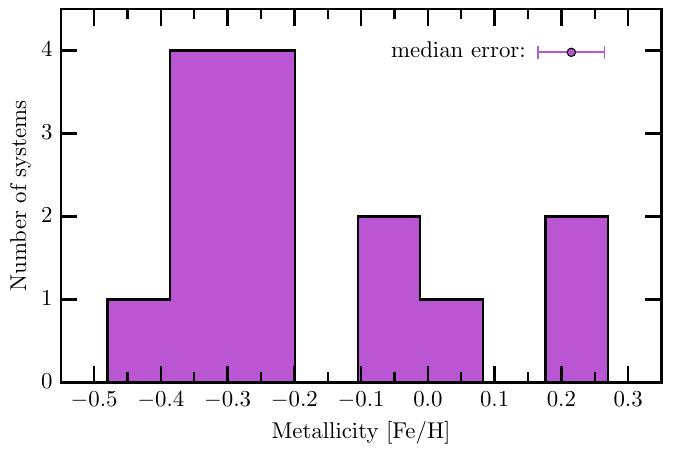}
    \caption{Distribution of the metallicity of the GK host stars in our sample that we estimated in this work (see values in Table~\ref{tab:stellar_feh}). We indicate the median error on the metallicity in the upper right corner. 
    }
    \label{fig:hist_metallicity}
\end{figure}

Overall, we note that in our sample only three GK host stars have a super-solar metallicity: HD~99492, HD~154088, and HD~3651, the latter one only at a~$2\,\sigma$~level (see Table~\ref{tab:stellar_feh} and Fig.~\ref{fig:hist_metallicity}). We also notice that two of them (HD~99492 and HD~3651) are in a wide binary system. 
On the contrary, $9$ GK host stars have a significant ($>4\,\sigma$) sub-solar metallicity. 

\onecolumn


\begin{table*}[htbp]
\centering
\caption{Atmospheric parameters and metallicity [Fe/H] for the GK star sample. 
}
\begin{tabular}{lCCCCcC}
\hline\hline\noalign{\smallskip}
Star & T_{\rm eff} & \log g & V_t & \text{[Fe/H]} & \text{Spectrum} & S/N \\
       &   \text{(K)}      & \text{(dex)} &  \text{( km s}^{-1}\text{)} & \text{(dex)} & & \text{(per pixel)}\\
\noalign{\smallskip}\hline\hline\noalign{\smallskip}       
HD~3651 & 5274\pm107 & 4.42\pm0.20 	& 0.87\pm 0.25 & +0.08\pm 0.06 & HARPN & 280\\
HD~20794 & 5501\pm66 & 4.49\pm0.15  &  0.88\pm0.11 & -0.36\pm 0.05 & HARPS & 350\\
HD~38858 & 5788\pm75 & 4.56\pm0.11 	&  0.95\pm0.11 & -0.21\pm0.05 & HARPS & 400 \\
HD~69830 & 5458\pm72 & 4.55\pm0.12 	&  0.76\pm0.15 & -0.03\pm 0.05 & HARPS & 380\\
HD~99492 & 5562\pm120 & 4.58\pm0.19 &  1.06\pm20  &  +0.27\pm0.09 & HARPS & 250 \\
HD~102365&  5608\pm50 & 4.27\pm0.12 &  0.91\pm0.07 & -0.35\pm 0.03 & HARPS & 350\\
61 Vir & 5557\pm48 & 4.40\pm0.12 	&  0.82\pm0.10 & -0.03\pm 0.04&  HARPS & 300\\
HD~136352 & 5655\pm48 & 4.24\pm0.12 &  0.88\pm0.06 & -0.37\pm0.03 & HARPS & 345\\
HD~154088&  5416\pm70 & 4.46\pm0.20 &  0.96\pm0.20 & +0.25\pm0.06 & HARPS & 312\\
HD~189567&  5715\pm45 & 4.36\pm0.13 &  0.92\pm0.07 & -0.27\pm0.03 & HARPS & 330 \\
\noalign{\smallskip}  \hline \noalign{\smallskip}
BD-061339 & 4061\pm52 & 4.68\pm0.05 & 0.79\pm0.12 &-0.20\pm0.10 &  HARPS & 60 \\
GJ~667\,C & 4939\pm41 & 4.50\pm0.10  & 0.90\pm0.10 &-0.48\pm0.10 & HARPS & 290 \\
HD~40307& 4871\pm65 & 4.61\pm0.05 & 0.89\pm0.10 & -0.30\pm0.05 & ESPRESSO & 1000 \\
HD~85512 & 4389\pm115 & 4.48\pm0.10 &0.75\pm0.10 & -0.28\pm0.12 & HARPS & 195 \\
\hline
\hline
\noalign{\smallskip}
\label{tab:stellar_feh}
\end{tabular}
\end{table*}

\begin{table*}[h!]
\centering
\caption{Elemental abundances with corresponding errors for the G-type stars in our sample. 
}\label{tab:elements}
\begin{tabular}{lCCCCCCC}
\hline\hline\noalign{\smallskip}
Star & \text{[Na/Fe]}_{\rm NLTE} & \text{[Mg/Fe]} & \text{[Si/Fe]} & \text{[Ca/Fe]} & \text{[Ti/Fe]} & \text{[Cr/Fe]} & \text{[Ni/Fe]} \\
\hline\hline\noalign{\smallskip}
HD~3651 & 0.11\pm0.05 &  0.10\pm0.07 & 0.00\pm0.05 &  0.04\pm0.06 & 0.04\pm0.05 & 0.11\pm0.08 & 0.10\pm0.08\\   
HD~20794 & 0.06\pm0.05& 0.25\pm0.04 &  0.14\pm0.05	& 0.18\pm0.07 &  0.28\pm0.05 & 0.04\pm0.05 & 0.08\pm0.04 \\	 
HD~69830 & 0.03\pm0.05 & -0.01\pm0.04 &   -0.01\pm0.04 & 0.00\pm0.05  & 0.05\pm0.06 & 0.05\pm0.06 & 0.04\pm0.05\\	 
HD~38858 & -0.01\pm0.05 & -0.03\pm0.06 &  -0.03\pm0.07 & 0.03\pm0.08 & 0.04\pm0.04 & 0.04\pm0.08 & -0.01\pm0.08 \\
HD~99492 & 0.12\pm0.04 & -0.03\pm0.06 &   0.06\pm0.06 &  -0.04\pm0.07 &0.05\pm0.06 & 0.08\pm0.06 & 0.10\pm0.06	\\
HD~102365 & 0.08\pm0.05 &0.18\pm0.04 &0.11\pm0.05 &  0.12\pm0.16 & 0.13\pm0.05 & 0.01\pm0.03 & 0.06\pm0.07 \\ 
61 Vir & 0.06\pm0.05 & -0.01\pm0.06 &  -0.01\pm0.05 &  0.03\pm0.05 &  0.06\pm0.05 & 0.00\pm0.05 & 0.03\pm0.06 \\ 
HD~136352 & 0.13\pm0.06& 0.24\pm0.05 &  0.13\pm0.06 & 0.14\pm0.10& 0.17\pm0.06  & 0.05\pm0.04 & 0.05\pm0.06\\
HD~154088 & 0.15\pm0.06  & 0.07\pm0.06 &  0.12\pm0.04 & 0.04\pm0.05 & 0.07\pm0.08 & 0.16\pm0.07 & 0.10\pm0.10 \\
HD~189567 & 0.08\pm0.09 & 0.05\pm0.04 &  0.05\pm0.04 & 0.06\pm0.10 & 0.08\pm0.10 & 0.01\pm0.05 & 0.01\pm0.07\\
\noalign{\smallskip}\hline\hline
\end{tabular}
\end{table*}

\newpage 

\onecolumn
\section{Log of the VLT/SPHERE observations \label{app:log_obs}}
\begin{figure*}[h]
    \centering \normalsize
    \includegraphics[width=\linewidth]{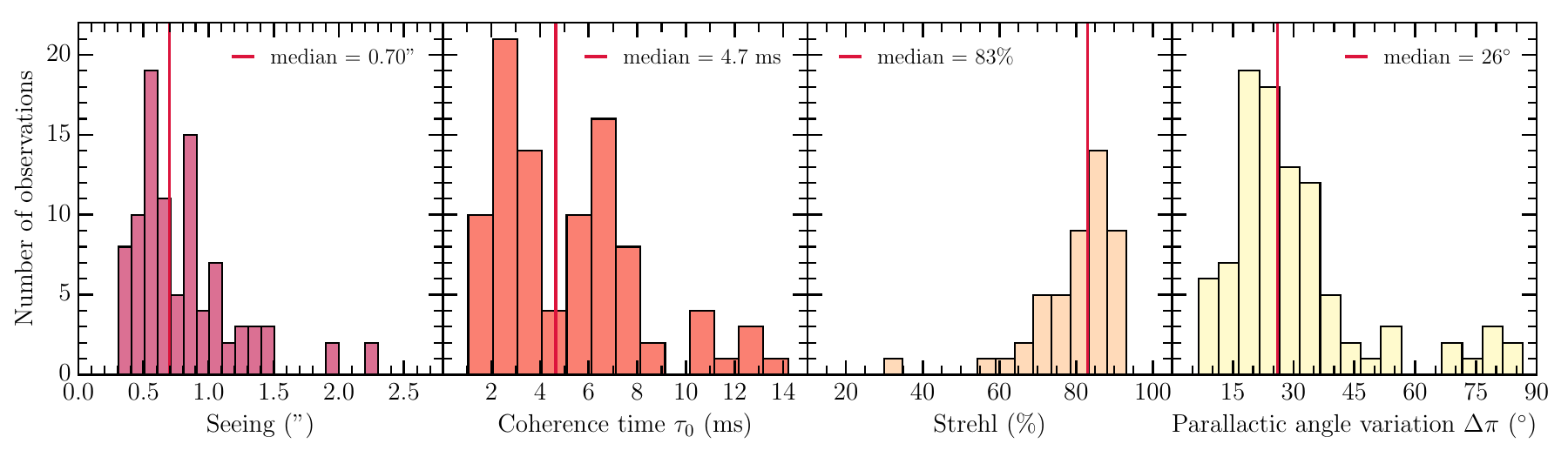}
    \vspace{-0.2cm}
    \caption{Distribution of the observing conditions: seeing, coherence time of the adaptive optic system, strehl, and parallactic angle variation. }
    \label{fig:hist_obslog}
\end{figure*}

Table \ref{tab:obslog} lists the $94$ observations with VLT/SPHERE, including the observational conditions, represented in Fig.~\ref{fig:hist_obslog} too. 

\vspace{-0.2cm}


\begin{small}
\begin{longtable}[h]
{ccccCCcCCCcC}
\caption{Log of the $94$ observations with the instrument VLT/SPHERE.} \label{tab:obslog}\\%
\hline\hline\noalign{\smallskip} 
System & Date & Instru. & Filter & \text{N}_\text{exp} & \text{DIT} & Grade & \Delta \pi & \epsilon & \tau_0 &  \textnormal{Sr} & \textnormal{airmass} \\
& & & & & \text{(s)} & & \text{(\degr)} & \text{(")} & \text{(ms)}  & \textnormal{(\%)} & \\
\noalign{\smallskip}\hline\hline\noalign{\smallskip}
\endfirsthead
\caption{continued.}\\%
\hline
\hline \noalign{\smallskip} 
System & Date & Instru. & Filter & \text{N}_\text{exp} & \text{DIT} & Grade & \Delta \pi & \epsilon & \tau_0 &  \textnormal{Sr} & \textnormal{airmass} \\
& & & & & \text{(s)} & & \text{(\degr)} & \text{(")} & \text{(ms)}  & \textnormal{(\%)} & \\
\hline\hline\noalign{\smallskip}
\endhead
\noalign{\smallskip}\hline\hline\noalign{\smallskip}
\endfoot
\endlastfoot 
GJ\,163 & 2018-10-12 & IRDIS & H23 & 32 & 96 & B & 23.7 & 0.68 & 2.4 & 30 & 1.14 \\
GJ\,163 & 2018-10-12 & IFS & YJ & 32 & 96 & B & 23.8 & 0.70 & 2.4 & -- & 1.14 \\
GJ\,176& 2018-10-08 & IRDIS & H23 & 64 & 48 & A & 16.9 & 0.90 & 3.7 & 77 & 1.38 \\
GJ\,176& 2018-10-08 & IFS & YJ & 32 & 96 & A & 16.6 & 0.88 & 3.8 & -- & 1.39 \\
GJ\,180 & 2018-10-09 & IRDIS & H23 & 32 & 96 & A & 78.3 & 0.77 & 3.2 & 73 & 1.01 \\
GJ\,180 & 2018-10-09 & IFS & YJ & 32 & 96 & A & 78.4 & 0.84 & 2.5 & -- & 1.01 \\
BD-061339 & 2018-11-24 & IRDIS & H23 & 32 & 96 & A & 31.3 & 0.59 & 12.0 & 84 & 1.06 \\
BD-061339 & 2018-11-24 & IFS & YJ & 32 & 96 & A & 31.7 & 0.53 & 14.2 & -- & 1.07 \\
GJ\,229 & 2018-11-27 & IRDIS & H23 & 48 & 64 & A & 9.0 & 0.59 & 4.4 & 87 & 1.03 \\
GJ\,229 & 2018-11-27 & IFS & YJ & 32 & 96 & A & 9.2 & 0.63 & 4.0 & -- & 1.03 \\
GJ\,422 & 2018-01-28 & IRDIS & H23 & 30 & 96 & A & 19.4 & 0.38 & 10.2 & 75 & 1.20 \\
GJ\,422 & 2018-01-28 & IFS & YJ & 30 & 96 & A & 19.6 & 0.38 & 10.2 & -- & 1.20 \\
GJ\,422 & 2019-11-25 & IRDIS & K12 & 24 & 96 & C & 8.4 & 0.57 & 5.8 & 84 & 1.79 \\
GJ\,422 & 2019-11-25 & IFS & YJH & 24 & 96 & C & 8.4 & 0.65 & 3.9 & -- & 1.79 \\
GJ\,422 & 2020-01-18 & IRDIS & K12 & 24 & 96 & A & 15.4 & 0.47 & 10.7 & 87 & 1.20 \\
GJ\,422 & 2020-01-18 & IFS & YJH & 24 & 96 & A & 15.4 & 0.44 & 8.7 & -- & 1.20 \\
GJ\,433 & 2018-02-02 & IRDIS & H23 & 30 & 96 & A & 68.0 & 1.01 & 6.7 & 84 & 1.01 \\
GJ\,433 & 2018-02-02 & IFS & YJ & 30 & 96 & A & 68.4 & 1.01 & 6.7 & -- & 1.01 \\
GJ\,433 & 2019-06-08 & IRDIS & K12 & 15 & 64 & C & 23.9 & 0.86 & 3.7 & 82 & 1.01 \\
GJ\,433 & 2019-06-08 & IFS & YJH & 10 & 96 & C & 23.3 & 0.86 & 3.7 & -- & 1.01 \\
GJ\,433 & 2020-01-20 & IRDIS & K12 & 15 & 64 & C & 21.6 & 0.36 & 12.3 & 91 & 1.01 \\
GJ\,433 & 2020-01-20 & IFS & YJH & 10 & 96 & C & 20.6 & 0.36 & 12.3 & -- & 1.01 \\
GJ\,581 & 2017-07-21 & IRDIS & H23 & 30 & 96 & A & 35.3 & 0.55 & 6.9 & 81 & 1.05 \\
GJ\,581 & 2017-07-21 & IFS & YJ & 30 & 96 & A & 35.6 & 0.56 & 6.4 & -- & 1.05 \\
Wolf\,1061 & 2017-07-28 & IRDIS & H23 & 120 & 24 & A & 34.9 & 0.82 & 4.0 & 79 & 1.04 \\
Wolf\,1061 & 2017-07-28 & IFS & YJ & 30 & 96 & A & 34.1 & 0.78 & 4.5 & -- & 1.04 \\
GJ\,649 & 2017-08-20 & IRDIS & H23 & 60 & 48 & B & 14.1 & 0.84 & 5.6 & 81 & 1.57 \\
GJ\,649 & 2017-08-20 & IFS & YJ & 30 & 96 & B & 13.9 & 0.84 & 5.6 & -- & 1.57 \\
GJ\,667 & 2017-06-27 & IRDIS & H23 & 30 & 96 & A & 53.6 & 1.36 & 2.6 & 66 & 1.02 \\
GJ\,667 & 2017-06-27 & IFS & YJ & 30 & 96 & A & 54.2 & 1.36 & 2.6 & -- & 1.02 \\
GJ\,674 & 2017-07-14 & IRDIS & H23 & 30 & 96 & A & 27.2 & 0.67 & 3.2 & 73 & 1.08 \\
GJ\,674 & 2017-07-14 & IFS & YJ & 30 & 96 & A & 27.4 & 0.67 & 3.2 & -- & 1.08 \\
GJ\,674 & 2021-07-04 & IRDIS & K12 & 68 & 32 & A & 21.2 & 0.54 & 5.8 & 85 & 1.08 \\
GJ\,674 & 2021-07-04 & IFS & YJH & 23 & 96 & A & 21.2 & 0.54 & 5.8 & -- & 1.08 \\
GJ\,682 & 2017-06-23 & IRDIS & H23 & 60 & 48 & C & 31.5 & 1.95 & 1.4 & 58 & 1.07 \\
GJ\,682 & 2017-06-23 & IFS & YJ & 30 & 96 & C & 31.0 & 1.95 & 1.4 & -- & 1.07 \\
GJ\,682 & 2017-07-20 & IRDIS & H23 & 30 & 96 & B & 30.7 & 0.55 & 6.5 & 72 & 1.07 \\
GJ\,682 & 2017-07-20 & IFS & YJ & 30 & 96 & B & 31.0 & 0.55 & 6.5 & -- & 1.07 \\
GJ\,682 & 2019-07-05 & IRDIS & K12 & 24 & 96 & B & 25.0 & 0.54 & 6.5 & 88 & 1.06 \\
GJ\,682 & 2019-07-05 & IFS & YJH & 24 & 96 & B & 25.1 & 0.52 & 6.5 & -- & 1.06 \\
GJ\,832 & 2017-05-27 & IRDIS & H23 & 46 & 64 & A & 26.7 & 1.06 & 2.5 & 78 & 1.10 \\
GJ\,832 & 2017-05-27 & IFS & YJ & 30 & 96 & A & 25.9 & 0.97 & 2.9 & -- & 1.10 \\
GJ\,832 & 2017-06-21 & IRDIS & H23 & 184 & 16 & C & 27.7 & 2.30 & 1.1 & 61 & 1.10 \\
GJ\,832 & 2017-06-21 & IFS & YJ & 90 & 32 & C & 26.9 & 2.30 & 1.4 & -- & 1.10 \\
GJ\,832 & 2017-06-27 & IRDIS & H23 & 46 & 64 & C & 26.7 & 1.50 & 1.9 & 71 & 1.10 \\
GJ\,832 & 2017-06-27 & IFS & YJ & 30 & 96 & C & 25.9 & 1.50 & 1.9 & -- & 1.10 \\
GJ\,832 & 2017-07-03 & IRDIS & H23 & 46 & 64 & B & 25.9 & 1.42 & 1.8 & 75 & 1.10 \\
GJ\,832 & 2017-07-03 & IFS & YJ & 30 & 96 & B & 25.0 & 1.30 & 1.9 & -- & 1.10 \\
GJ\,832 & 2022-07-29 & IRDIS & H23 & 95 & 64 & A & 55.4 & 0.53 & 7.6 & 83 & 1.10 \\
GJ\,832 & 2022-07-29 & IFS & YJ & 56 & 96 & A & 47.3 & 0.56 & 7.2 & -- & 1.10 \\
GJ\,3998 & 2017-05-14 & IRDIS & H23 & 30 & 96 & B & 18.2 & 1.29 & 2.5 & 67 & 1.23 \\
GJ\,3998 & 2017-05-14 & IFS & YJ & 30 & 96 & B & 18.3 & 1.37 & 1.9 & -- & 1.23 \\
GJ\,3998 & 2019-07-27 & IRDIS & K12 & 24 & 96 & B & 14.4 & 0.59 & 4.8 & 85 & 1.23 \\
GJ\,3998 & 2019-07-27 & IFS & YJH & 24 & 96 & B & 14.4 & 0.68 & 3.6 & -- & 1.24 \\
HD\,3651 & 2017-07-11 & IRDIS & H23 & 332 & 10 & A & 18.8 & 1.22 & 2.0 & 78 & 1.45 \\
HD\,3651 & 2017-07-11 & IFS & YJ & 66 & 48 & A & 16.8 & 1.08 & 2.4 & -- & 1.45 \\
HD\,20794 & 2017-09-03 & IRDIS & H23 & 736 & 3 & C & 34.4 & 0.45 & 7.2 & 89 & 1.06 \\
HD\,20794 & 2017-09-03 & IFS & YJ & 406 & 6 & C & 37.5 & 0.47 & 6.8 & -- & 1.06 \\
HD\,20794 & 2017-09-04 & IRDIS & H23 & 552 & 6 & B & 44.9 & 0.88 & 2.2 & 87 & 1.06 \\
HD\,20794 & 2017-09-04 & IFS & YJ & 261 & 12 & B & 42.3 & 0.90 & 2.1 & -- & 1.06 \\
HD\,38858 & 2018-11-23 & IRDIS & H23 & 48 & 64 & A & 26.2 & 0.31 & 6.6 & 89 & 1.09 \\
HD\,38858 & 2018-11-23 & IFS & YJ & 32 & 96 & A & 26.1 & 0.32 & 7.7 & -- & 1.09 \\
HD\,40307 & 2018-12-22 & IRDIS & H23 & 48 & 64 & A & 18.2 & 0.48 & 8.7 & 89 & 1.27 \\
HD\,40307 & 2018-12-22 & IFS & YJ & 32 & 96 & A & 18.1 & 0.52 & 8.1 & -- & 1.27 \\
HD\,69830 & 2018-12-18 & IRDIS & H23 & 96 & 32 & A & 37.2 & 0.64 & 5.8 & 84 & 1.04 \\
HD\,69830 & 2018-12-18 & IFS & YJ & 32 & 96 & A & 36.7 & 0.70 & 5.5 & -- & 1.04 \\
HD\,85512 & 2017-04-01 & IRDIS & H23 & 60 & 48 & C & 32.8 & 1.04 & 2.4 & 80 & 1.06 \\
HD\,85512 & 2017-04-01 & IFS & YJ & 30 & 96 & C & 32.3 & 1.04 & 2.4 & -- & 1.06 \\
HD\,85512 & 2017-12-30 & IRDIS & H23 & 60 & 48 & B & 32.8 & 0.46 & 6.5 & 86 & 1.06 \\
HD\,85512 & 2017-12-30 & IFS & YJ & 30 & 96 & B & 32.2 & 0.46 & 6.5 & -- & 1.06 \\
HD\,85512 & 2019-11-25 & IRDIS & K12 & 136 & 16 & C & 7.0 & 0.39 & 7.9 & 91 & 1.55 \\
HD\,85512 & 2019-11-25 & IFS & YJH & 34 & 64 & C & 6.5 & 0.38 & 7.7 & -- & 1.56 \\
HD\,85512 & 2020-01-03 & IRDIS & K12 & 102 & 32 & B & 78.4 & 0.74 & 12.6 & 93 & 1.13 \\
HD\,85512 & 2020-01-03 & IFS & YJH & 34 & 64 & B & 24.6 & 0.82 & 10.3 & -- & 1.06 \\
HD\,99492 & 2018-02-03 & IRDIS & H23 & 30 & 96 & A & 22.7 & 0.94 & 5.8 & 85 & 1.13 \\
HD\,99492 & 2018-02-03 & IFS & YJ & 30 & 96 & A & 22.9 & 0.92 & 6.3 & -- & 1.13 \\
HD\,99492 & 2020-03-16 & IRDIS & K12 & 24 & 96 & B & 17.5 & 0.80 & 4.8 & 87 & 1.13 \\
HD\,99492 & 2020-03-16 & IFS & YJH & 24 & 96 & B & 17.6 & 1.06 & 3.4 & -- & 1.14 \\
HD\,102365 & 2018-04-07 & IRDIS & H23 & 136 & 24 & A & 40.1 & 0.44 & 7.7 & 91 & 1.05 \\
HD\,102365 & 2018-04-07 & IFS & YJ & 135 & 24 & A & 40.9 & 0.46 & 7.1 & -- & 1.05 \\
61\,Vir & 2017-06-28 & IRDIS & H23 & 208 & 16 & A & 76.1 & 0.83 & 2.6 & 87 & 1.01 \\
61\,Vir & 2017-06-28 & IFS & YJ & 203 & 16 & A & 15.8 & 0.78 & 2.9 & -- & 1.01 \\
HD\,136352 & 2017-07-14 & IRDIS & H23 & 70 & 48 & A & 31.2 & 0.82 & 2.9 & 80 & 1.10 \\
HD\,136352 & 2017-07-14 & IFS & YJ & 68 & 48 & A & 30.8 & 0.88 & 2.8 & -- & 1.10 \\
HD\,136352 & 2020-03-16 & IRDIS & K12 & 84 & 24 & A & 19.3 & 0.65 & 6.1 & 91 & 1.09 \\
HD\,136352 & 2020-03-16 & IFS & YJH & 64 & 32 & A & 19.7 & 0.64 & 6.1 & -- & 1.10 \\
HD\,154088 & 2017-06-21 & IRDIS & H23 & 90 & 32 & A & 21.6 & 1.20 & 2.6 & 73 & 1.03 \\
HD\,154088 & 2017-06-21 & IFS & YJ & 90 & 32 & A & 22.1 & 1.20 & 2.4 & -- & 1.03 \\
HD\,154088 & 2019-07-22 & IRDIS & K12 & 56 & 32 & C & 86.6 & 0.50 & 3.4 & 80 & 1.00 \\
HD\,154088 & 2019-07-22 & IFS & YJH & 28 & 64 & C & 84.1 & 0.51 & 3.5 & -- & 1.00 \\
HD\,154088 & 2021-07-04 & IRDIS & K12 & 28 & 32 & A & 34.6 & 0.55 & 6.6 & 91 & 1.00 \\
HD\,154088 & 2021-07-04 & IFS & YJH & 14 & 64 & A & 34.6 & 0.55 & 6.6 & -- & 1.00 \\
HD\,189567 & 2017-05-29 & IRDIS & H23 & 368 & 8 & A & 18.2 & 0.90 & 2.5 & 83 & 1.37 \\
HD\,189567 & 2017-05-29 & IFS & YJ & 60 & 12 & A & 18.1 & 0.93 & 2.5 & -- & 1.37 \\
\noalign{\smallskip}\hline\hline
\end{longtable}
\end{small}


\section{Reduced observations}

\begin{figure*}[h!]
    \centering
    \includegraphics[width=0.95\linewidth]{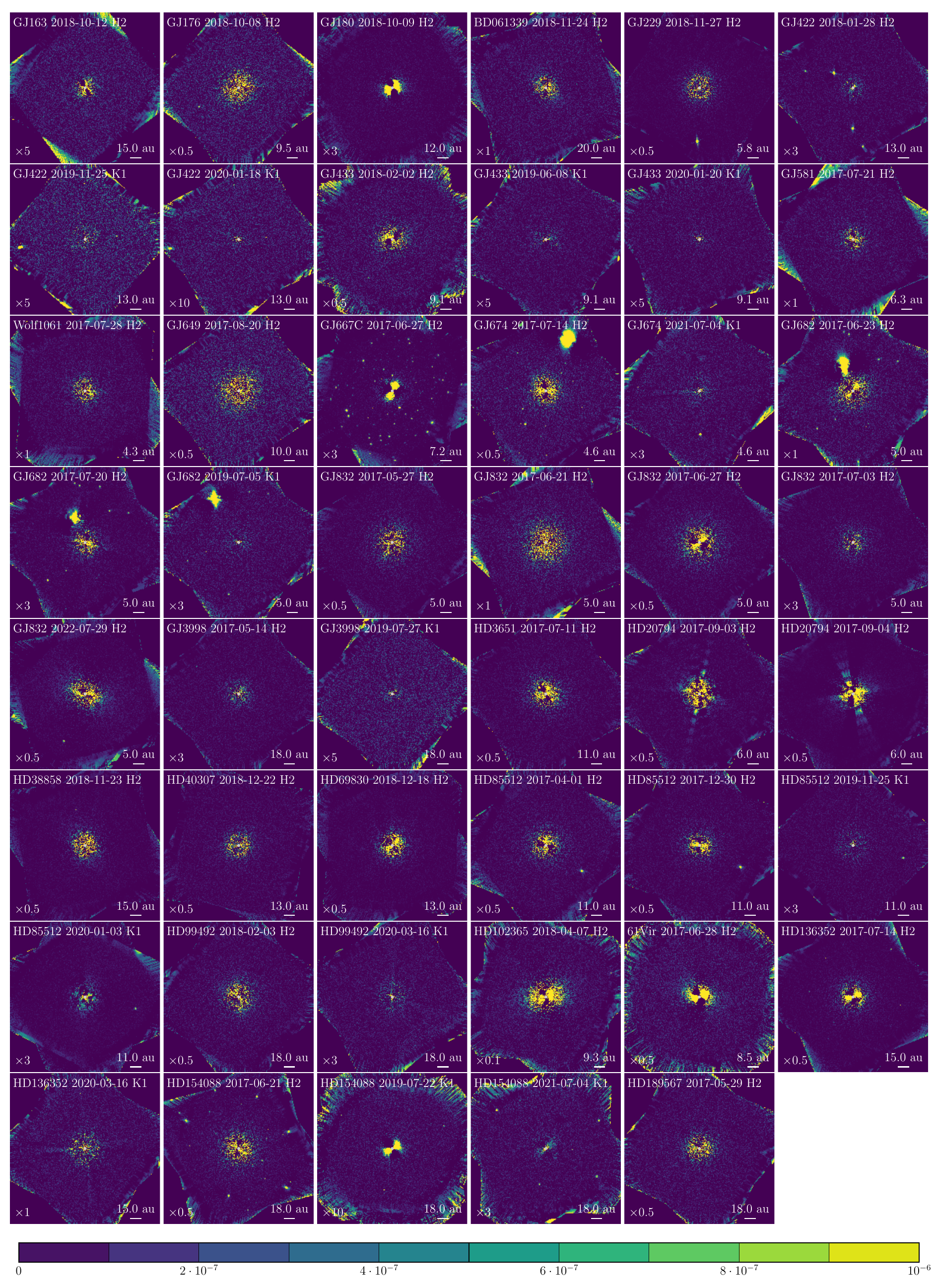}
    \caption{
    Reduced images from SPHERE-IRDIS for the whole survey for SpeCal-cADI reductions. The factor in the bottom left hand corner of each figure indicates the number by which to multiply the color bar to have the corresponding contrast values.
    }
    \label{fig:mosaic_super-earths_survey_IRDIS_cADI}
\end{figure*}

\begin{figure*}[p]
    \centering
    \includegraphics[width=0.95\linewidth]{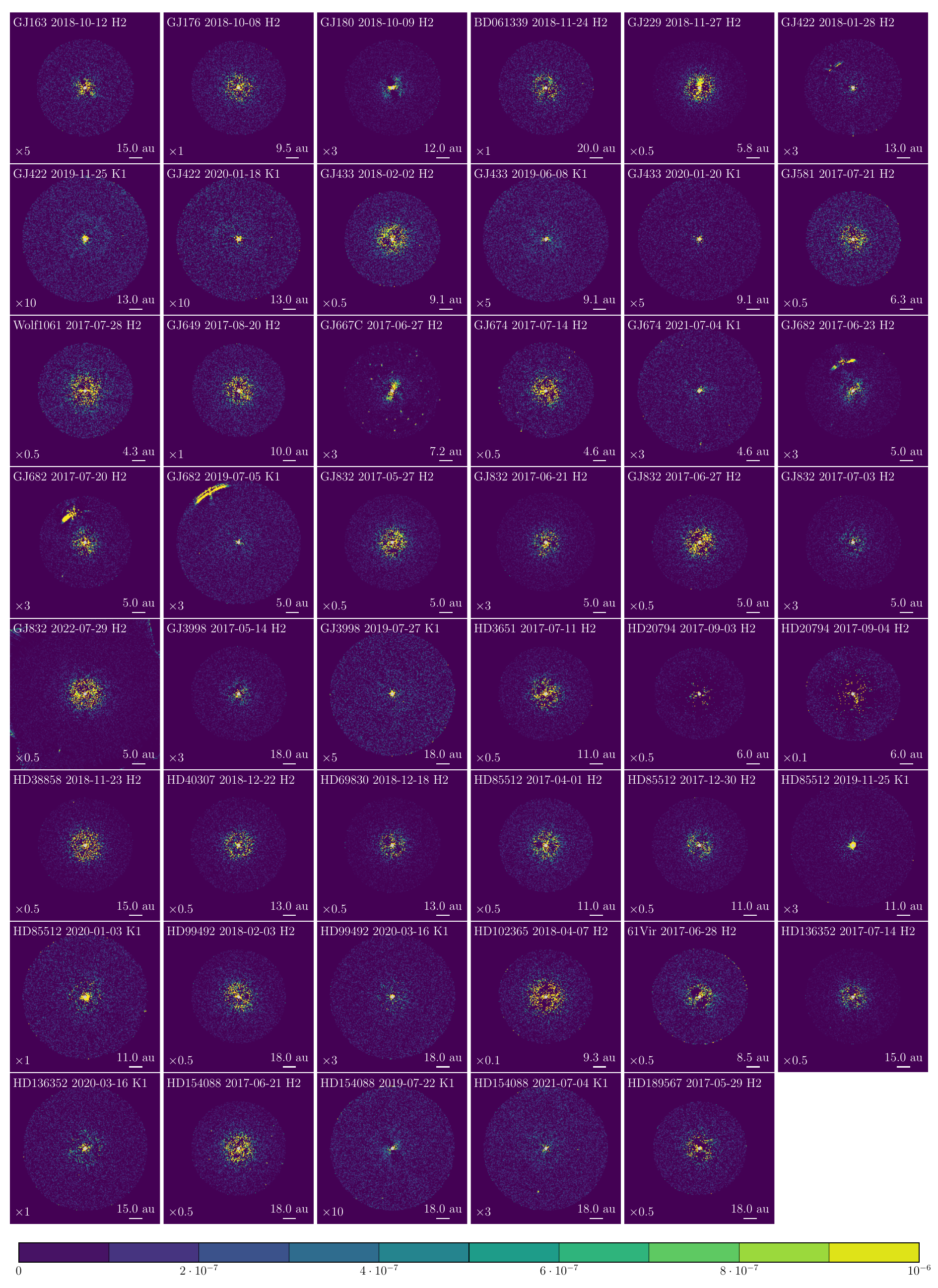}
    \caption{
    Reduced images from SPHERE-IRDIS for the whole survey for SpeCal-TLOCI ADI reductions.}
    \label{fig:mosaic_super-earths_survey_IRDIS_TLOCI}
\end{figure*}

\begin{figure*}[p]
    \centering
    \includegraphics[width=0.95\linewidth]{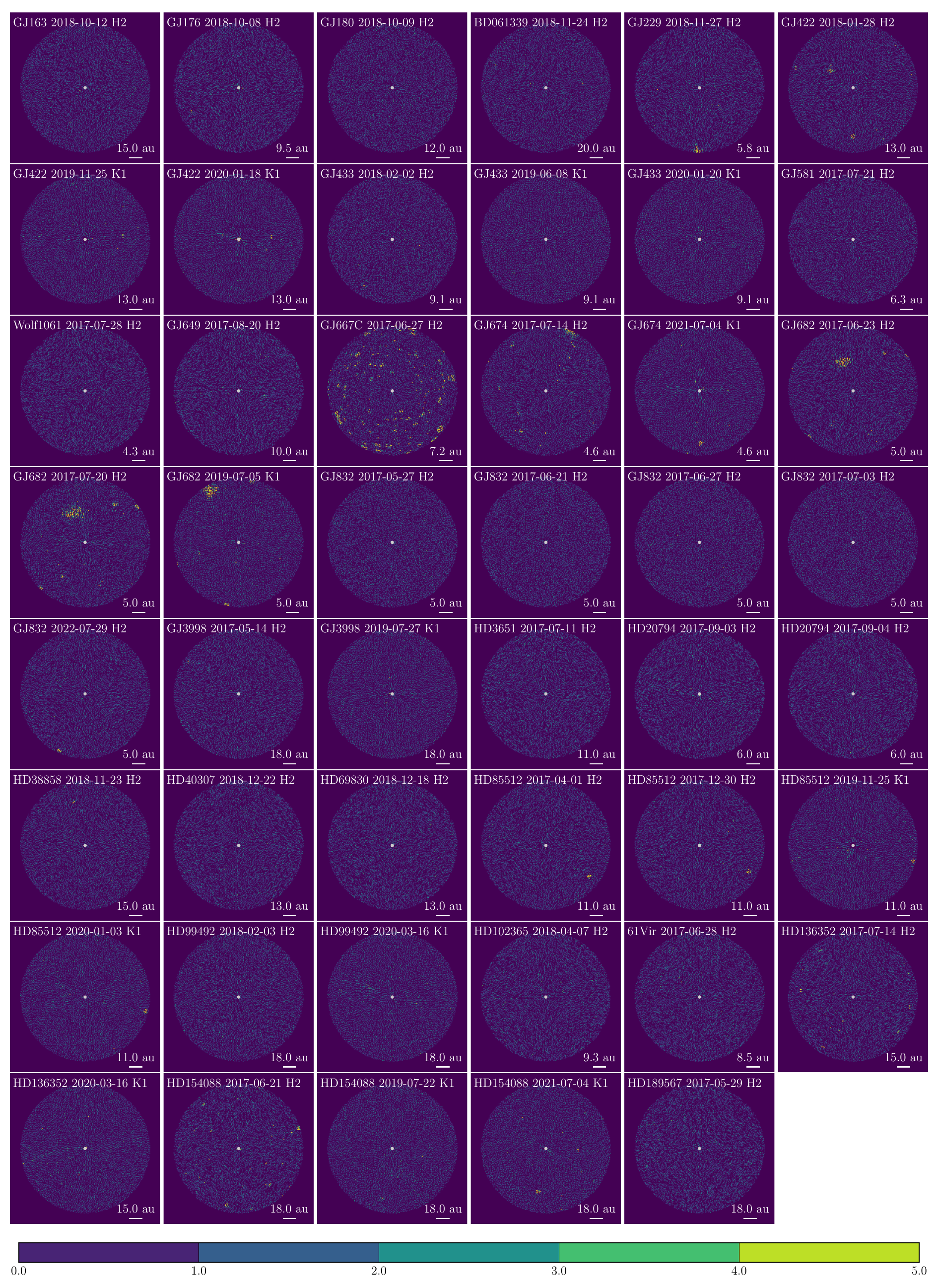}
    \caption{
    Reduced images from SPHERE-IRDIS for the whole survey for ANDROMEDA reductions.}
    \label{fig:mosaic_super-earths_survey_IRDIS_ANDROMEDA}
\end{figure*}

\begin{figure*}[p]
    \centering
    \includegraphics[width=0.95\linewidth]{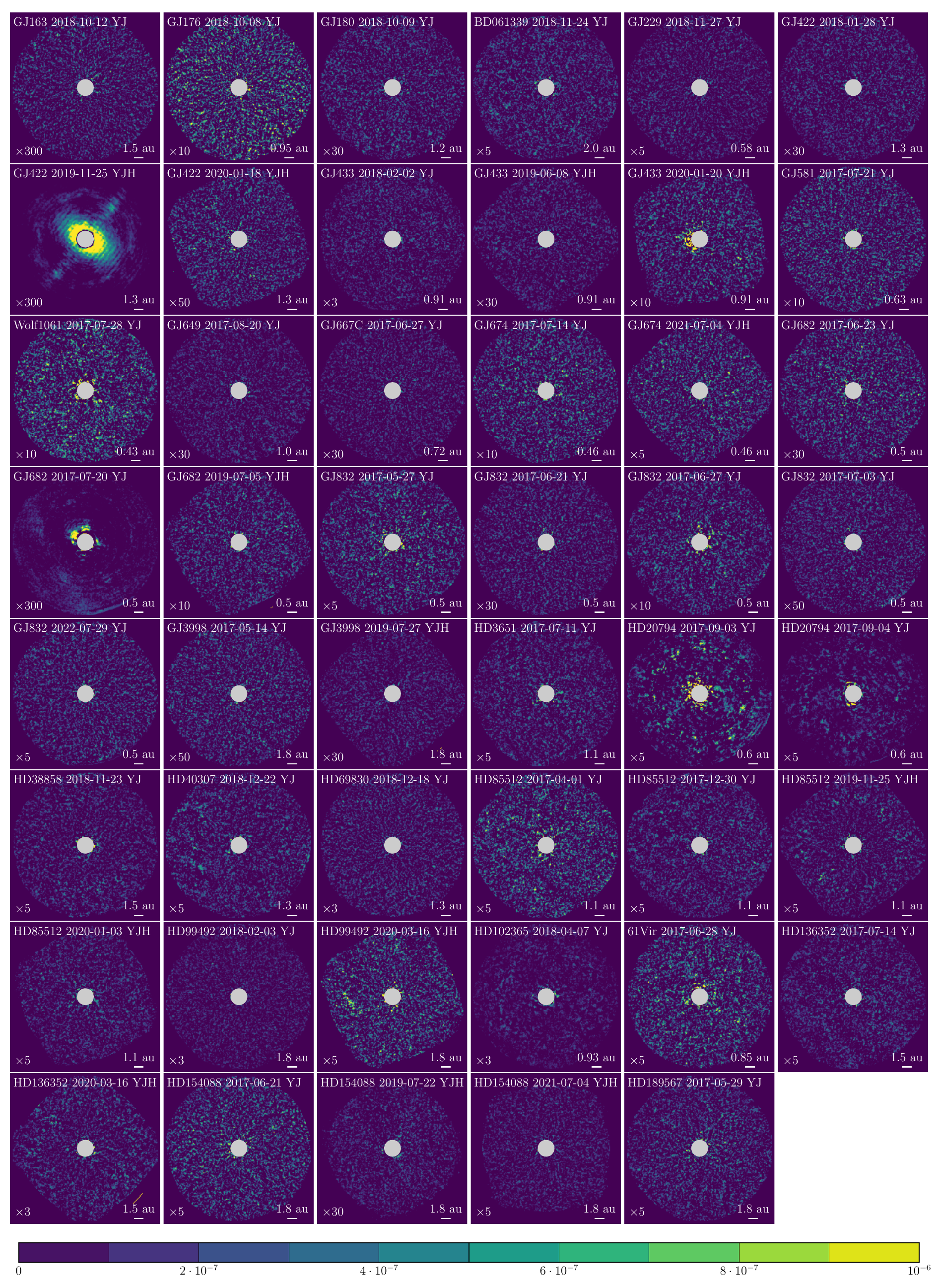}
    \caption{
    Reduced images from SPHERE-IFS for the whole survey for SpeCal-PCA-ADSI reductions.}
    \label{fig:mosaic_super-earths_survey_IFS_PCAPad}
\end{figure*}

\begin{figure*}[p]
    \centering
    \includegraphics[width=0.95\linewidth]{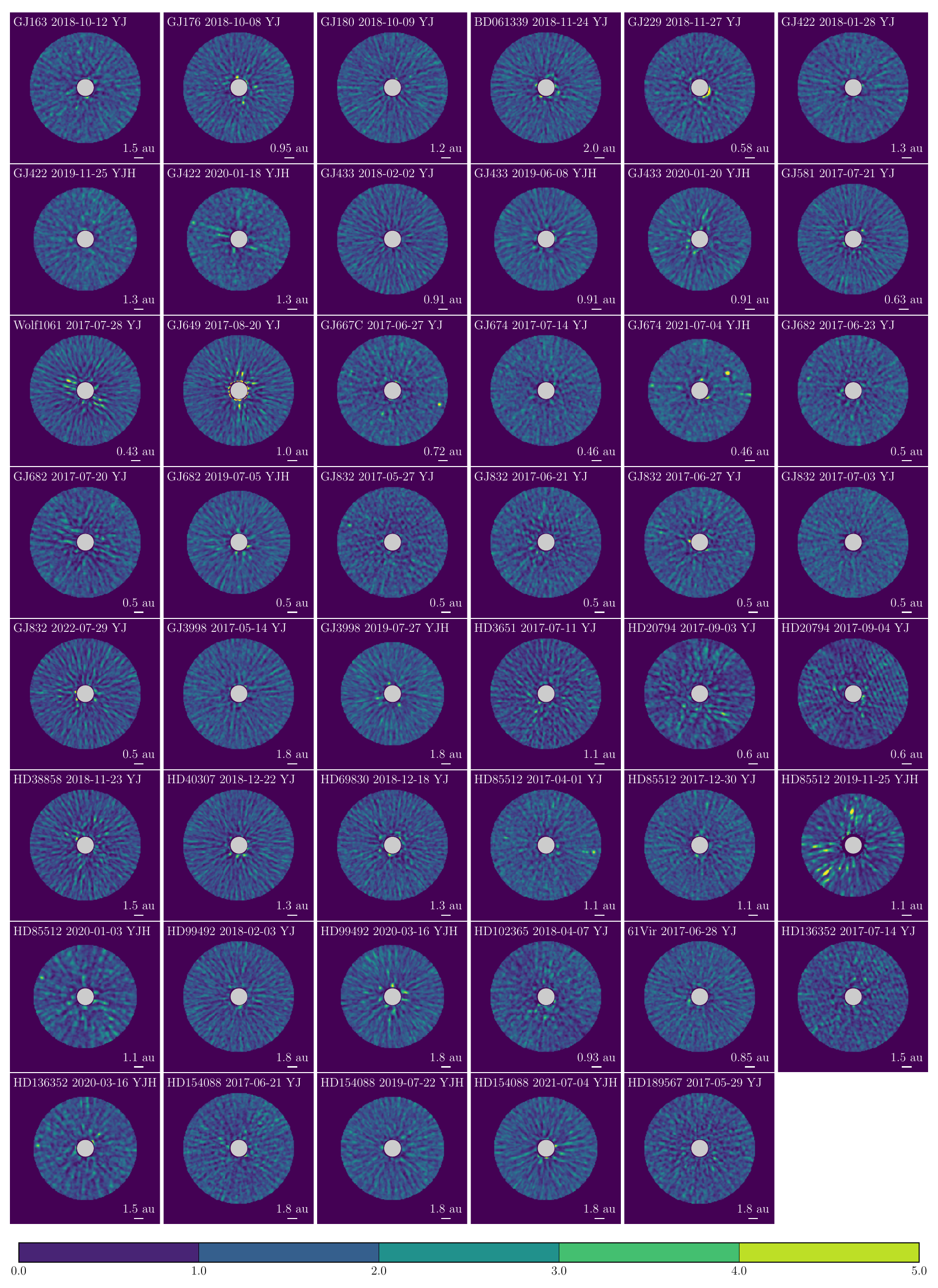}
    \caption{
    Reduced images from SPHERE-IFS for the whole survey for ANDROMEDA-ASDI reductions.}
    \label{fig:mosaic_super-earths_survey_IFS_ANDROMEDA}
\end{figure*}

Figures \ref{fig:mosaic_super-earths_survey_IRDIS_cADI}--\ref{fig:mosaic_super-earths_survey_IRDIS_ANDROMEDA} show the reduced images via SpeCal-cADI, SpeCal-TLOCI, ANDROMEDA ADI algorithms for all the IRDIS observations. In addition, Figs.  \ref{fig:mosaic_super-earths_survey_IFS_PCAPad} and \ref{fig:mosaic_super-earths_survey_IFS_ANDROMEDA} show the reduced images via SpeCal-PCA ASDI and ANDROMEDA ASDI algorithms for all the IFS observations.

 \newpage

 \newpage

 \newpage

 \newpage

\newpage

\onecolumn

\section{Companion candidates from ANDROMEDA and SpeCal processing  \label{app:diag_detections_algo}}

\begin{figure*}[h!]
    \centering \includegraphics[width=0.48\linewidth]{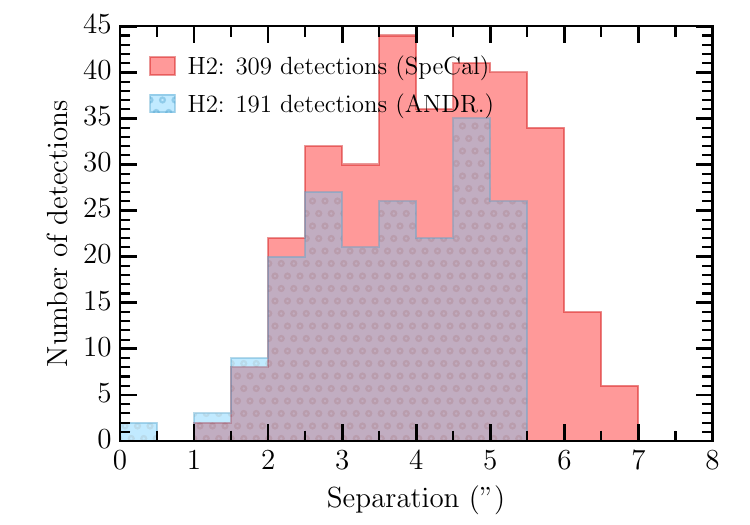} \includegraphics[width=0.48\linewidth]{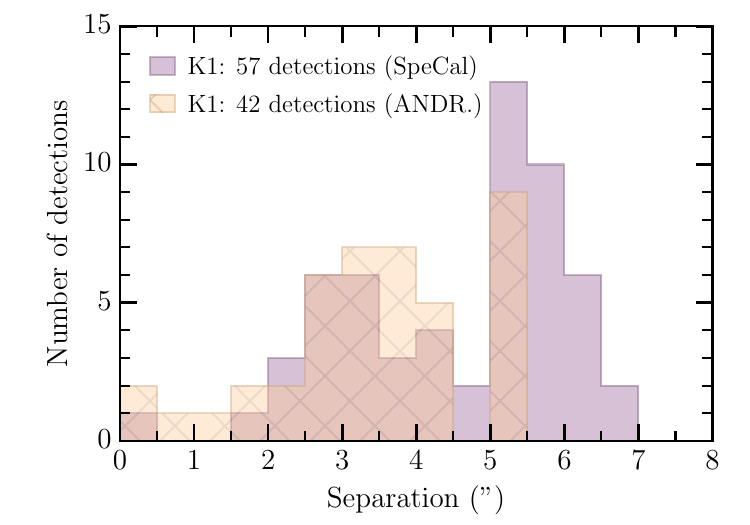}
    
    \centering \includegraphics[width=0.48\linewidth]{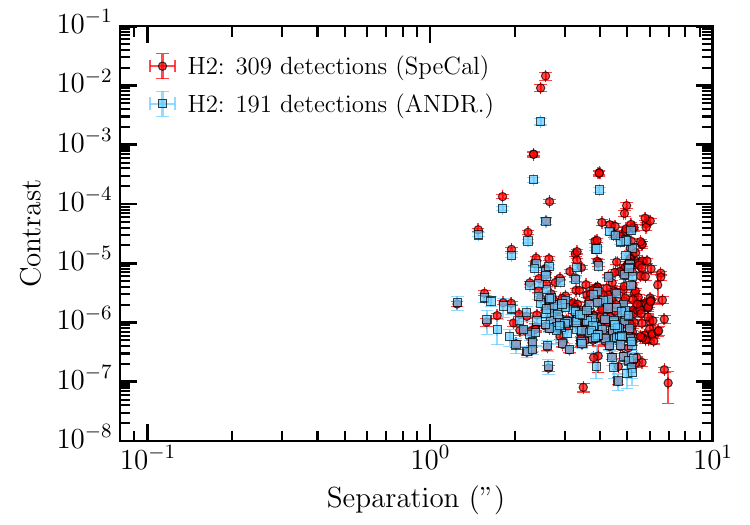} \includegraphics[width=0.48\linewidth]{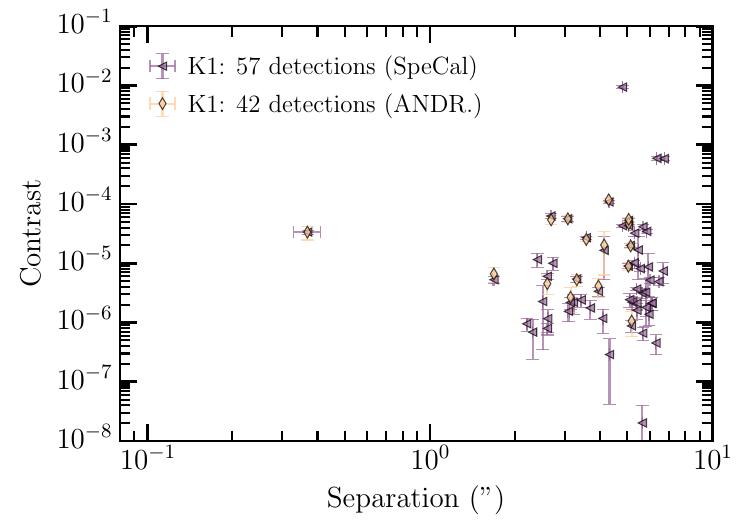}
    
    \caption{At the \textit{top}, distribution of the separations of all the point-source detections from SPHERE-IRDIS in the H2-band (\textit{on the left}) and the K1-band (\textit{on the right}). At the \textit{bottom}, contrasts and separations of all the point-source detections from SPHERE-IRDIS in the H2-band (\textit{on the left}) and the K1-band (\textit{on the right}).
    The post-processing algorithms used are SpeCal-cADI and ANDROMEDA-ADI.
    }
    \label{fig:hist_plot_all_candidates}
\end{figure*}

Figure \ref{fig:hist_plot_all_candidates} reports the distribution of the detections  for SPHERE-IRDIS observations with the ANDROMEDA-ADI or SpeCal-cADI processing, as well as the separations and contrasts of the detections.
We can see that below $2"$, ANDROMEDA processing results in additional detections compared to SpeCal-cADI, which is expected \citep[e.g.,][]{Cantalloube2015_andromeda}.

In addition, we note that concerning bright ($>10^{-4}$ or $10^{-5}$) candidates, there is a bias between ANDROMEDA and SpeCal photometry: the same candidate can have a difference up to $0.3~\magg$ depending on which processing is used to retrieve the photometry. This photometric bias correlates with the apparent magnitude of the candidate. The brighter is the detection, the larger is the bias. There is no visible correlation with the separation. In our observations, this effect is only visible in the H2 band, corresponding to observations with bright candidates. 
By investigating this effect on the known brown dwarf GJ~229~B for which the photometry is reported in the literature, we found that SpeCal-cADI (and Specal-non-ADI) was consistent with the values from \citet{Nakajima1995_GJ229_1planet_BD} and \citet{Faherty2012_GJ229B_magnitude}, contrary to ANDROMEDA.

\section{Input parameters for Gaia and Hipparcos results via the \texttt{GaiaPMEX} tool}

Table~\ref{tab:paramPMEX} summarizes the parameters used for running the tool \texttt{GaiaPMEX} (Kiefer et al. (in prep.) for the systems GJ\,832 and GJ\,229, the only two that led to a positive detection from \textit{Gaia} and \textit{Hipparcos} data (see~\ref{sec:results_gaiapmex})

\begin{table*}[]
\caption{Parameters used with the \texttt{GaiaPMEX} tool for the systems GJ\,229 and GJ\,832.}
\label{tab:paramPMEX}\centering
\begin{tabular}{lccc}
\noalign{\smallskip}\hline \hline  \noalign{\smallskip}
Parameter & Unit & GJ\,229 & GJ\,832 \\
\noalign{\smallskip}\hline \hline  \noalign{\smallskip}
\multicolumn{4}{c}{\textit{Aliases}} \\ 
HIP                                                                        &                         & 29295                    & 106440                   \\
Gaia DR3                                                                   &                         & 2940856402123426176 & 6562924609150908416 \\
 \noalign{\smallskip}\hline   \noalign{\smallskip}
\multicolumn{4}{c}{\textit{Main parameters}} \\ 
$M_\star$                                                                  & $\mathrm{M_{\odot}}$    & 0.570                        & 0.480                        \\
$\sigma_{M\star}$                                                          & $\mathrm{M_{\odot}}$    & 0.049975713178077            & 0.0498015394994533           \\
V                                                                          & $\mathrm{}$             & 8.12                     & 8.67                     \\
RA                                                                         &                         & 06:10:34.4589                & 21:33:33.9004                \\
DEC                                                                        &                         & -21:52:04.163                & -49:00:45.468                \\
 \noalign{\smallskip}\hline  \noalign{\smallskip}
\multicolumn{4}{c}{\textit{Gaia parameters}} \\ 
\texttt{ra}                                               & deg  & 92.644                       & 323.391                      \\
\texttt{ra\_error}                                        & deg  & 0.008                        & 0.018                        \\
\texttt{dec}                                              & deg  & -21.868                      & -49.013                      \\
\texttt{dec\_error}                                       & deg  & 0.014                        & 0.014                        \\
\texttt{pmra}                                             & $\mathrm{mas\,yr^{-1}}$ & -135.692                     & -45.917                      \\
\texttt{pmra\_error}                                      & $\mathrm{mas\,yr^{-1}}$ & 0.011                        & 0.023                        \\
\texttt{pmdec}                                            & $\mathrm{mas\,yr^{-1}}$ & -719.178                     & -816.875                     \\
\texttt{pmdec\_error}                                     & $\mathrm{mas\,yr^{-1}}$ & 0.017                        & 0.018                        \\
\texttt{parallax}                                         & $\mathrm{mas}$          & 173.57                       & 201.33                       \\
\texttt{parallax\_error}                                  & $\mathrm{mas}$          & 0.02                         & 0.02                         \\
\texttt{phot\_g\_mean\_mag}                               & $\mathrm{}$             & 7.31                         & 7.74                         \\
\texttt{bp\_rp}                                           & $\mathrm{}$             & 2.074                        & 2.240                        \\
\texttt{astrometric\_matched\_transits}                   & $\mathrm{}$             & 82                           & 47                           \\
\texttt{astrometric\_n\_good\_obs\_al}                    & $\mathrm{}$             & 723                          & 414                          \\
\texttt{astrometric\_params\_solved}                      & $\mathrm{}$             & 31                           & 31                           \\
\texttt{astrometric\_excess\_noise} or AEN$_\mathrm{obs}$ & $\mathrm{mas}$          & 0.123                        & 0.160                        \\
\texttt{ruwe} or RUWE$_\mathrm{obs}$                      & $\mathrm{}$             & 1.013                        & 1.097                        \\
 \noalign{\smallskip}\hline \noalign{\smallskip}
\multicolumn{4}{c}{\textit{Hipparcos parameters}} \\ 
e$_\mathrm{RA} \cos \mathrm{DEC}$                                          & $\mathrm{mas}$          & 0.381                        & 0.420                        \\
e$_\mathrm{DEC}$                                                           & $\mathrm{mas}$          & 0.670                        & 0.600                        \\
$\sigma_\mathrm{pos}$                                                      & $\mathrm{mas}$          & 0.771                        & 0.732                        \\
 \noalign{\smallskip}\hline   \noalign{\smallskip}
\multicolumn{4}{c}{\textit{Kervella et al. (2022) parameters}} \\ 
pmRAH2EG3b                                                                 & $\mathrm{mas\,yr^{-1}}$ & -136.406                     & -46.046                      \\
e\_pmRAH2EG3b                                                              & $\mathrm{mas\,yr^{-1}}$ & 0.015                        & 0.018                        \\
pmDEH2EG3b                                                                 & $\mathrm{mas\,yr^{-1}}$ & -714.974                     & -816.289                     \\
e\_pmDEH2EG3b                                                              & $\mathrm{mas\,yr^{-1}}$ & 0.019                        & 0.020                        \\
PMaRAH2EG3b                                                                & $\mathrm{mas\,yr^{-1}}$ & 0.712                        & 0.140                        \\
e\_PMaRAH2EG3b                                                             & $\mathrm{mas\,yr^{-1}}$ & 0.018                        & 0.029                        \\
PMaDEH2EG3b                                                                & $\mathrm{mas\,yr^{-1}}$ & -4.224                       & -0.547                       \\
e\_PMaDEH2EG3b                                                             & $\mathrm{mas\,yr^{-1}}$ & 0.025                        & 0.027                        \\
$\Vert \mathrm{PMa} \Vert$ (or PMa$_\mathrm{obs}$)                          & $\mathrm{mas\,yr^{-1}}$ & 4.284                        & 0.565                        \\
$\sigma_\mathrm{PMa,obs}$                                                  & $\mathrm{mas\,yr^{-1}}$ & 0.025                        & 0.027                        \\
 \noalign{\smallskip}\hline   \noalign{\smallskip}
\multicolumn{4}{c}{\textit{Noise parameters and simulations of single systems}} \\ 
$\sigma_\mathrm{AL}$                                                       & $\mathrm{mas}$          & 0.083                        & 0.086                        \\
$\sigma_\mathrm{attitude}$                                                 & $\mathrm{mas}$          & 0.064                        & 0.082                        \\
$\sigma_\mathrm{jitter}$                                                   & $\mathrm{mas}$          & 0.129                        & 0.135                        \\
AEN$_\mathrm{simu}$ as single                                              & $\mathrm{mas}$          & 0.123                        & 0.124                        \\
$\sigma_\mathrm{AEN,simu,single}$                                          & $\mathrm{mas}$          & 0.015                        & 0.021                        \\
PMa$_\mathrm{simu}$ as single                                              & $\mathrm{mas\,yr^{-1}}$ & 0.060                        & 0.059                        \\
$\sigma_\mathrm{PMa,simu,single}$                                          & $\mathrm{mas\,yr^{-1}}$ & 0.034                        & 0.033                        \\
AEN$_\mathrm{obs}$ significance                                            & $\mathrm{}$             & 0.007                        & 1.696                        \\
PMa$_\mathrm{obs}$ significance                                            & $\mathrm{}$             & $>$3.82                      & $>$3.82                      \\
\noalign{\smallskip}\hline \hline  \noalign{\smallskip}
\end{tabular}
\end{table*}

\newpage

\onecolumn

\section{Comparison of detection method sensitivity on our survey  \label{app:detlim_pedagogical}}

\begin{figure*}[h!]
    \centering
    \includegraphics[width=0.48\linewidth]{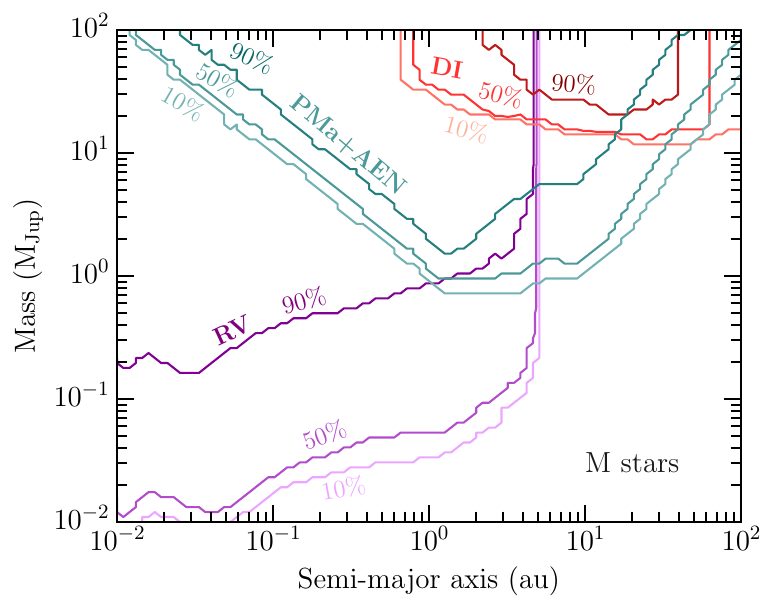}
    \includegraphics[width=0.48\linewidth]{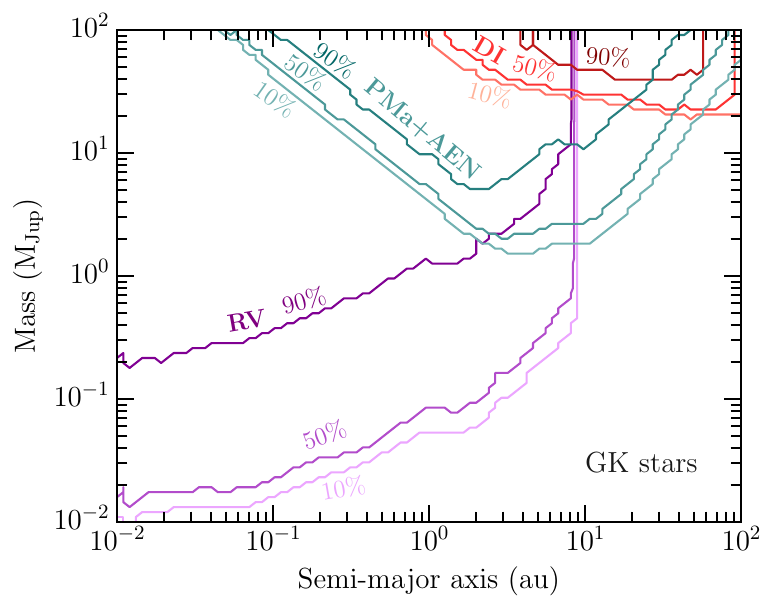}
    \caption{Comparison of the sensitivity of the direct imaging (DI, red), proper motion anomaly and \textit{Gaia} astrometric  excess noise (PMa+AEN, blue) and radial velocity (RV, purple) detections methods on our survey. We took the median of the detection probabilities at $10\%$, $50\%$, and $90\%$ the M-stars (\textit{left}) and the GK-stars (\textit{right}). }
    \label{fig:detlim_pedagogical}
\end{figure*}

Detection methods of exoplanet suffer from observational biases. Combining different detection techniques enable to have a more global view on planetary systems (see~Fig.~\ref{fig:detlim_pedagogical}), even though observers are still blind to low-mass exoplanets at wide orbits (e.g., an exoplanet of $1~\mearth$ planet orbiting at $20~\au$).

\begin{figure*}[h!]
    \centering
    \includegraphics[width=\linewidth]{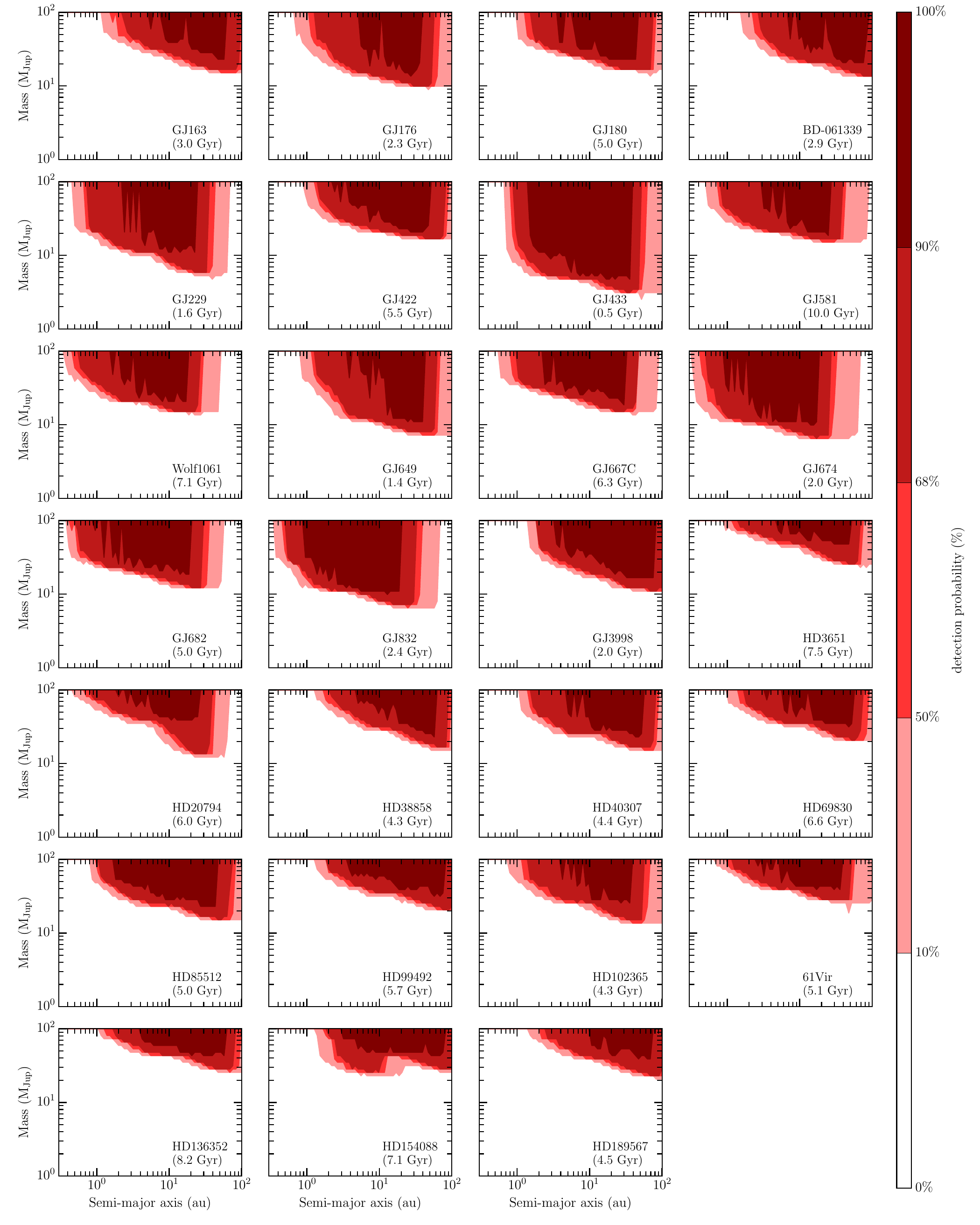}
    \caption{Detection limits from direct imaging observations (SPHERE-IRDIS and SPHERE-IFS) for each system of the survey obtained with the \texttt{MESS3} tool.  
    }
    \label{fig:mosaic_detlims_cond2003_DIonly}
\end{figure*}

Figure~\ref{fig:mosaic_detlims_cond2003_DIonly} shows sensitivity from high-contrast imaging SPHERE-IRDIS and SPHERE-IFS observations for each system of our sample (see Section~\ref{sec:results_obs_detlims}). For a few mature systems, high-contrast imaging observations are sensitive to companions of $\lesssim10~\mj$ beyond a few astronomical units (e.g.,~GJ\,433)  or dozen of astronomical units (e.g.,~GJ\,229, GJ\,649, GJ\,674, GJ\,832).

\end{appendix}

\end{document}